\documentclass{pasj01}

\begin{document} 
\Received{}
\Accepted{}

\title{AGN Jet Power, Formation of X-ray
Cavities, and FR~I/II Dichotomy in Galaxy Clusters}

\author{Yutaka \textsc{Fujita},\altaffilmark{1}}
\altaffiltext{1}{Theoretical Astrophysics, Department of Earth and
Space Science, Graduate School of Science, Osaka University, 1-1
Machikaneyama-cho, \\ Toyonaka, Osaka 560-0043}
\email{fujita@vega.ess.sci.osaka-u.ac.jp}

\author{Nozomu \textsc{Kawakatu},\altaffilmark{2}}
\altaffiltext{2}{Faculty of Natural Sciences, National Institute of
Technology, Kure College, 2-2-11 Agaminami, Kure, Hiroshima, 737-8506}
\author{Isaac \textsc{Shlosman}\altaffilmark{3,1}}
\altaffiltext{3}{Department of Physics and Astronomy, University of
Kentucky, Lexington, KY 40506-0055, USA}


\maketitle

\begin{abstract}
We investigate the ability of jets in active galactic nuclei (AGNs) to
break out of the ambient gas with sufficiently large advance velocities.
Using observationally estimated jet power, we analyze 28 bright
elliptical galaxies in nearby galaxy clusters. Because the gas density
profiles in the innermost regions of galaxies have not been resolved so
far, we consider two extreme cases for temperature and density
profiles. We also follow two types of evolution for the jet cocoons:
being driven by the pressure inside the cocoon (Fanaroff-Riley [FR]\,I
type), and being driven by the jet momentum (FR\,II type). Our main
result is that regardless of the assumed form of density profiles, jets
with observed powers of $\lesssim 10^{44}\rm\: erg\: s^{-1}$ are not
powerful enough to evolve as FR\,II sources. Instead, they evolve as
FR\,I sources and appear to be decelerated below the buoyant velocities
of the cocoons when jets were propagating through the central dense
regions of the host galaxies. This explains the reason why FR\,I sources
are more frequent than FR\,II sources in clusters. Furthermore, we
predict the sizes of X-ray cavities from the observed jet powers and
compare them with the observed ones --- they are consistent within a
factor of two if the FR\,I type evolution is realized. Finally, we find
that the jets with a power $\gtrsim 10^{44}\rm\: erg\: s^{-1}$ are less
affected by the ambient medium, and some of them, but not all, could
serve as precursors of the FR\,II sources.
\end{abstract}

\section{Introduction}
\label{sec:intro}

Active galactic nuclei (AGNs) in the centers of elliptical galaxies
often produce relativistic jets. These jets propagate through the hot
interstellar medium (ISM) of host galaxies, and are enveloped by cocoons
(e.g., \cite{beg89a}). Recent X-ray observations have revealed that
brightest cluster galaxies (BCGs) are often associated with X-ray
cavities that are thought to be relics of these cocoons. Their sizes
range from a few kpc to a few hundred kpc (e.g.,
\cite{bir04a,nul05a,nul05b,mcn05a}), and many of them are located at
$r\sim 20$~kpc from the galactic centers \citep{bir04a}.

The time-averaged jet power can be estimated by measuring the volume of
cavities and the timescale of their buoyant rise (e.g.,
\cite{bir04a,raf06a,rus13b}).  In order to create large cocoons or
cavities, the jets have to break out of the central dense region of the
host galaxies. Jet evolution at this breakout phase can affect their
subsequent evolution, because they are expected to be decelerated by
interaction with the ambient medium (e.g.,
\cite{dey97a,ode98a,car02,per03}), and their advance speed can determine
whether they evolve into Fanaroff-Riley~I (FR\,I) or FR\,II types
(e.g. \cite{kaw08a}).

Previous studies of the relation between AGN jet power and X-ray
cavities have not addressed the issue of whether jets are actually
capable of forming the observed cavities (e.g.,
\cite{all06a,raf06a,bal08a,mcn11a}, see also \cite{kaw08a}). In this
work we investigate whether jets in BCGs have enough power during the
breakout phase to form the observed large X-ray cavities using available
data on these cavities and the ambient gas. We use the cosmological
parameters $\Omega_{\rm m0}=0.3$, $\Omega_{\rm \Lambda 0}=0.7$, and
$h=0.7$.  Unless otherwise noted, errors are the 1~$\sigma$ values.

\section{Models}
\label{sec:method}

\subsection{Breakout Jets}
\label{sec:breakout}

We assume that a pair of relativistic jets generated around a
supermassive black hole (SMBH) in the galactic center advance into the
ambient medium, forming a cocoon that envelops them (e.g.,
\cite{beg89a}). While most of the galaxies in our sample are known as
FR\,I sources, a few are FR\,IIs and FR\,II-like sources (e.g. Cygnus~A,
see table~\ref{tab:pot}). Moreover, the FR\,Is might have evolved as
FR\,IIs when they were young. Thus, we consider both FR\,I and II type
evolution for the cocoons.  Figure~\ref{fig:cocoon} shows schematic
evolution of a cocoon or cavity. Note that while the cocoon evolution is
controlled by the jets (phase A$\rightarrow$B), the cavity evolution is
controlled by buoyancy (phase B$\rightarrow$D). Below, we discuss
evolution during phases A and B. Effects of the evolution in phases B
and D are addressed in section~\ref{sec:FRI} and
appendix~\ref{sec:app_cavity}. We do not consider the effects of
recurrent jet activities.\footnote{Hydra~A in our sample has a
cavity-in-cavity structure \citep{wis07a}. However, the outermost cavity
dominates the inner cavities in formation energy and contributes mostly
to the average jet power. Thus, we focus on the outermost cavity.}

\subsubsection{FR\,I type evolution}
\label{sec:FRImodel}

\begin{figure}
 \begin{center}
  \includegraphics[width=84mm]{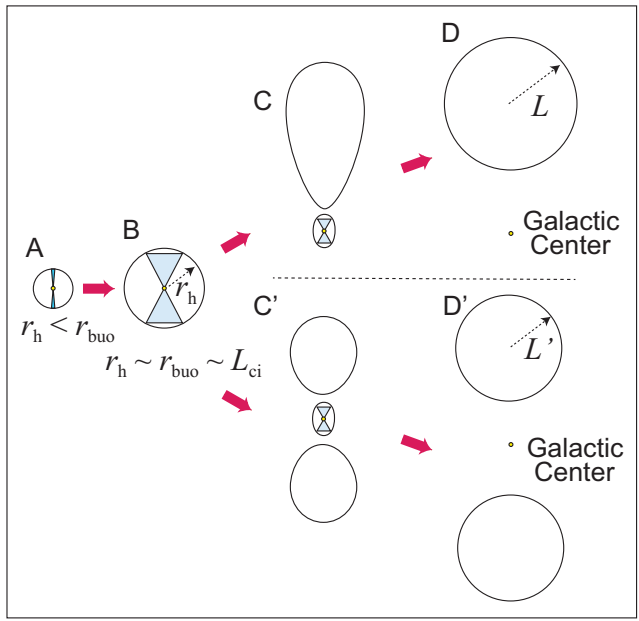}
 \end{center}
\caption{Schematic figure for the evolution of a cocoon (cavity). While
sequence A$\rightarrow$B$\rightarrow$C$\rightarrow$D is the case when
only one cavity is formed due to asymmetry of the environment, etc.
Sequence A$\rightarrow$B$\rightarrow$C'$\rightarrow$D' is the case where
two cavities are formed.  Phase A: Cocoon expansion is driven by jets
($r_{\rm h}<r_{\rm buo}$). Phase B (B') : Buoyancy starts to dominate
the jets for the cocoon expansion and the cocoon starts to rise by
buoyancy as a cavity ($r_{\rm h}\sim r_{\rm buo}\sim L_{\rm ci}$). Phase
C (C'): Cavity is detached from the jets, and rises in the hot gas by
buoyancy. Phase D (D'): Cavity is observed as an X-ray cavity. For more
details see the text.}  \label{fig:cocoon}
\end{figure}

Evolution of young cocoons is driven by the jets inside them
(phase A in figure~\ref{fig:cocoon}). If the cocoon expands due to
high pressure inside it, we call the evolution FR\,I type, which is
governed by an energy conservation law:
\begin{equation}
\label{eq:FRI}
 P_{\rm j} 
= \frac{1}{\gamma_{\rm c}-1} 
\frac{dp(r_{\rm h})}{dr_{\rm h}}v_{\rm I}(r_{\rm h})
\frac{4\pi r_{\rm h}^3}{3}
+
\frac{\gamma_{\rm c}}{\gamma_{\rm c}-1} 4\pi r_{\rm h}^2 v_{\rm
I}(r_{\rm h}) p(r_{\rm h}) 
\:,
\end{equation}
where $r_{\rm h}$ is the cocoon radius or the distance to the jet head
from the galactic center, $\gamma_{\rm c} (=4/3)$ is the adiabatic index
of the relativistic gas in the cocoon, $v_{\rm I}$ is the advance
velocity of the jet or the expansion velocity of the cocoon, $p$ is the
pressure of the hot gas outside the cocoon, and $P_{\rm j}$ is the power
of the twin jets. We
assume that $P_{\rm j}$ is constant with time. The first term on the
right hand side of equation~(\ref{eq:FRI}), denoted as $\xi(r)$, is
generally much smaller than the second term, because the pressure
gradient is modest in the central region of our sample
galaxies. Previous studies often ignored it (e.g., \cite{chu00a}), but
it is kept here because we quantitatively consider the pressure
profile of the ambient medium.

Jets must have an advance velocity that is larger than the buoyant
velocity of the cocoon, in order to create a large cavity whose size is
comparable to or larger than that of the host galaxy. This is because
when the buoyant velocity is larger, the cocoon detaches from the jet,
before it grows substantially. The buoyant velocity is comparable to or
a factor of a few smaller than the sound velocity of the ambient hot gas
(e.g., \cite{bir04a}). Thus, the condition for the formation of a large
cavity is
\begin{equation}
\label{eq:condI}
P_{\rm j} > \xi(r) + 
\frac{\gamma_{\rm c}}{\gamma_{\rm c}-1} 
4\pi r^2 f_{\rm 1} c_{\rm s}(r) p(r) \:,
\end{equation}
where $r$ is the distance from the galactic center which must be sufficiently 
large. The sound speed in the ambient gas, $c_{\rm s}=\sqrt{\gamma
k_{\rm B} T/(\mu m_{\rm p})}$, is a function of gas temperature $T$,
where $\gamma\, (=5/3)$ is the adiabatic index of the gas, $\mu\,
(=0.6)$ is the mean molecular weight, and $m_{\rm p}$ is the proton
mass. The buoyant velocity is given by $f_{\rm 1} c_{\rm s}$, where $f_1
(\lesssim 1)$ --- the reduction factor can be constrained from
observations (section 3).  When the left hand side of
equation~(\ref{eq:condI}) equals the right hand side, the jets no longer
drive the cocoon expansion --- the cocoon starts to rise in the hot gas
by buoyancy and forms a cavity (phases B$\rightarrow$C$\rightarrow$D in
figure~\ref{fig:cocoon}).  We assume that AGN activity creates only
one cavity. If it creates two cavities (phases
B$\rightarrow$C'$\rightarrow$D' in figure~\ref{fig:cocoon}), the size of
the cavities decreases only by a factor of $2^{1/3}\approx 1.26$ ($L$
and $L'$ in figure~\ref{fig:cocoon}), which does not affect the
following discussion.

\subsubsection{FR\,II type evolution}
\label{sec:FRIImodel}

If a cocoon expansion is driven by the jet momentum,
in phase A (figure~\ref{fig:cocoon}), we call the evolution the FR\,II
type.  The momentum flux of the jet is balanced with the ram pressure of
the ambient gas over the cross-sectional area at the head of the cocoon,
$(A_{\rm h})$:
\begin{equation}
\label{eq:FRII}
 P_{\rm j}/(2c) = \rho (r_{\rm h})v_{\rm h}^2(r_{\rm h})
A_{\rm h}(r_{\rm h})\:,
\end{equation}
where $\rho$ is the density of the ambient medium, $v_{\rm h}$ is the
advance velocity of the cocoon head or the velocity of the hot spot at
the end of the jet, and $r_{\rm h}$ is the distance from the galactic
center or the SMBH to the hot spot. Note that the velocity of the hot
spot ($v_{\rm h}\ll c$) decreases gradually with time, while the
velocity of the jet material is $c$. Assuming a constant jet
power, the condition for the formation of a large cavity is
\begin{equation}
\label{eq:condII}
 P_{\rm j}/(2c) > \rho (r) f_1^2 c_{\rm s}(r)^2 A_{\rm h}(r)\:.
\end{equation}
When the left and right hand sides become nearly equal, the evolution is
controlled by the buoyancy, as is the case of the FR~I type (sequence
B$\rightarrow$C$\rightarrow$D or sequence
B$\rightarrow$C'$\rightarrow$D' in figure~\ref{fig:cocoon}). It is
difficult to estimate the area of the cocoon head, $A_{\rm h}$, in the
condition~(\ref{eq:condII}). However, it can be related to the size of
the hot spot, $r_{\rm HS}$, and we use the values obtained by
\citet{kaw08a}. They compiled observational data for the sizes of the
hot spots in 117 radio sources, and found that it is simply represented
by a broken power-law of
\begin{equation}
\label{eq:rHS}
 r_{\rm HS}(r_{\rm h})\approx 0.3\: (r_{\rm
h}/{\rm 1~kpc})^a\: \rm kpc\:,
\end{equation}
where $a=1.34\pm 0.24$ for $r_{\rm h}<1$~kpc and $a=0.44\pm 0.08$ for
$r_h>1$~kpc. Following \citet{kaw09a}, we assume that
\begin{equation}
\label{eq:Ah}
 A_{\rm h}(r_{\rm h})=f_2 \pi r_{\rm HS}^2(r_{\rm h})\:.
\end{equation}
Although the value of factor $f_2$ can be estimated to lie in the range
of $10<f_2<100$ (e.g., \cite{kaw08a,kaw09a}), we fix it conservatively
at the lower limit, $f_2=10$, allowing the jet to break out with less
power. Equation~(\ref{eq:rHS}) is mainly based on the observations of
compact symmetric objects (CSOs) and medium-size symmetric objects
(MSOs) for $r_{\rm h}\lesssim 10$~kpc. Most of our sample galaxies are
FR\,Is and we study their jet evolution at $r_{\rm h}\lesssim
10$~kpc. Thus, we implicitly assume that the CSOs and MSOs are the
precursors of FR\,Is as well as of FR\,IIs (e.g., \cite{ode98a}). Note
that a jet is not likely to extend substantially beyond the central
region of the host galaxy in its lifetime, $\lesssim 10^8$~yr (e.g.,
\cite{bir08a}), if it is heavily decelerated by the ambient medium as we
show below.

\subsection{Density and Temperature Profiles of Ambient Medium}
\label{sec:den}

As a next step, we construct a model for the density and temperature
profiles of the hot galactic gas under conditions~(\ref{eq:condI})
and~(\ref{eq:condII}). We assume that the gas distribution is
spherically-symmetric for the sake of simplicity. Even with the superb
angular resolution of {\it Chandra X-ray Observatory}, it is difficult
to resolve the central region of a galaxy on a scale of the Bondi radius
$r_{\rm B}$. Thus, we are required to extrapolate the density and
temperature inward from the innermost measurement radius, $r_{\rm in}$,
to the Bondi radius. Previously, the extrapolation was often made by
assuming a power-law density profile and a constant temperature
\citep{all06a,bal08a}. However, it is not certain whether such an
assumption is justified. Hence, we consider two models for the profiles
that are physically motivated, and represent the two extremes that
encompass the real density profiles --- these are used as test cases.

\subsubsection{Low-temperature model}

In this model, the hot gas is assumed to be in pressure equilibrium:
\begin{equation}
\label{eq:hydroeq1}
 -\frac{dp}{dr} = \rho g\:,
\end{equation}
where $g(r)$ is the gravitational acceleration including three
components, i.e., $g=g_\bullet + g_{\rm gal} + g_{\rm cl}$, where
$g_\bullet$ is the SMBH contribution, $g_{\rm gal}$ is the galaxy
contribution, and $g_{\rm cl}$ is the cluster contribution.  The gas
temperature around the galaxy center is comparable to the virial
temperature of the host galaxy. For given boundary conditions that are
consistent with observations, equation~(\ref{eq:hydroeq1}) can be
integrated and the pressure profile $p(r)$ and the density profile
$\rho(r)$ can be obtained. Using these profiles, the Bondi accretion
rate $\dot{M}_{\rm B}$ can be estimated. The details are deferred to
appendix~\ref{sec:app_low}.

\subsubsection{Isentropic model}

\begin{figure*}
  \begin{center}
    \begin{tabular}{c}
      \begin{minipage}{0.5\hsize}
        \begin{center}
          \includegraphics[width=84mm]{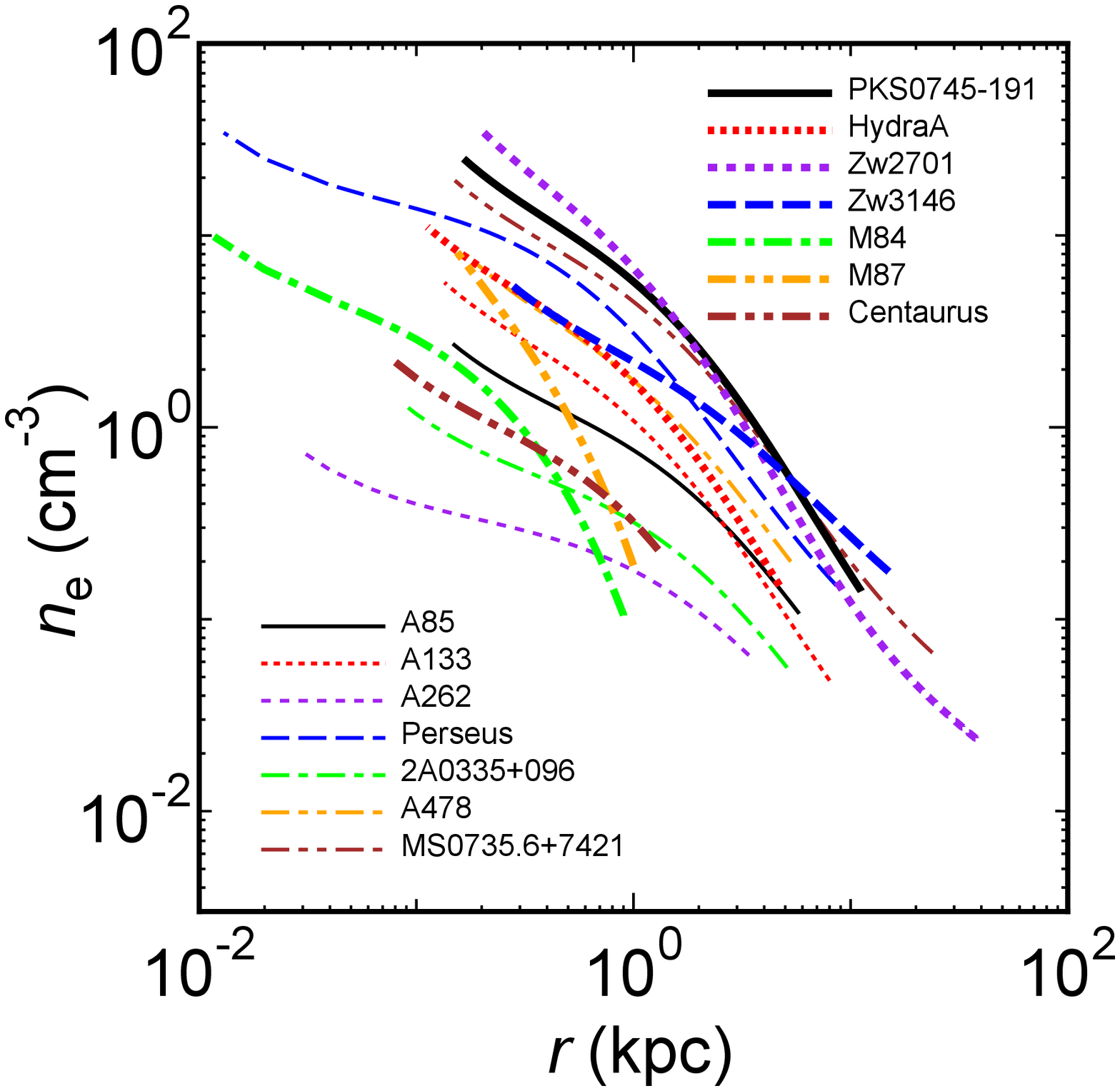}
        \end{center}
      \end{minipage}
      \begin{minipage}{0.5\hsize}
        \begin{center}
          \includegraphics[width=84mm]{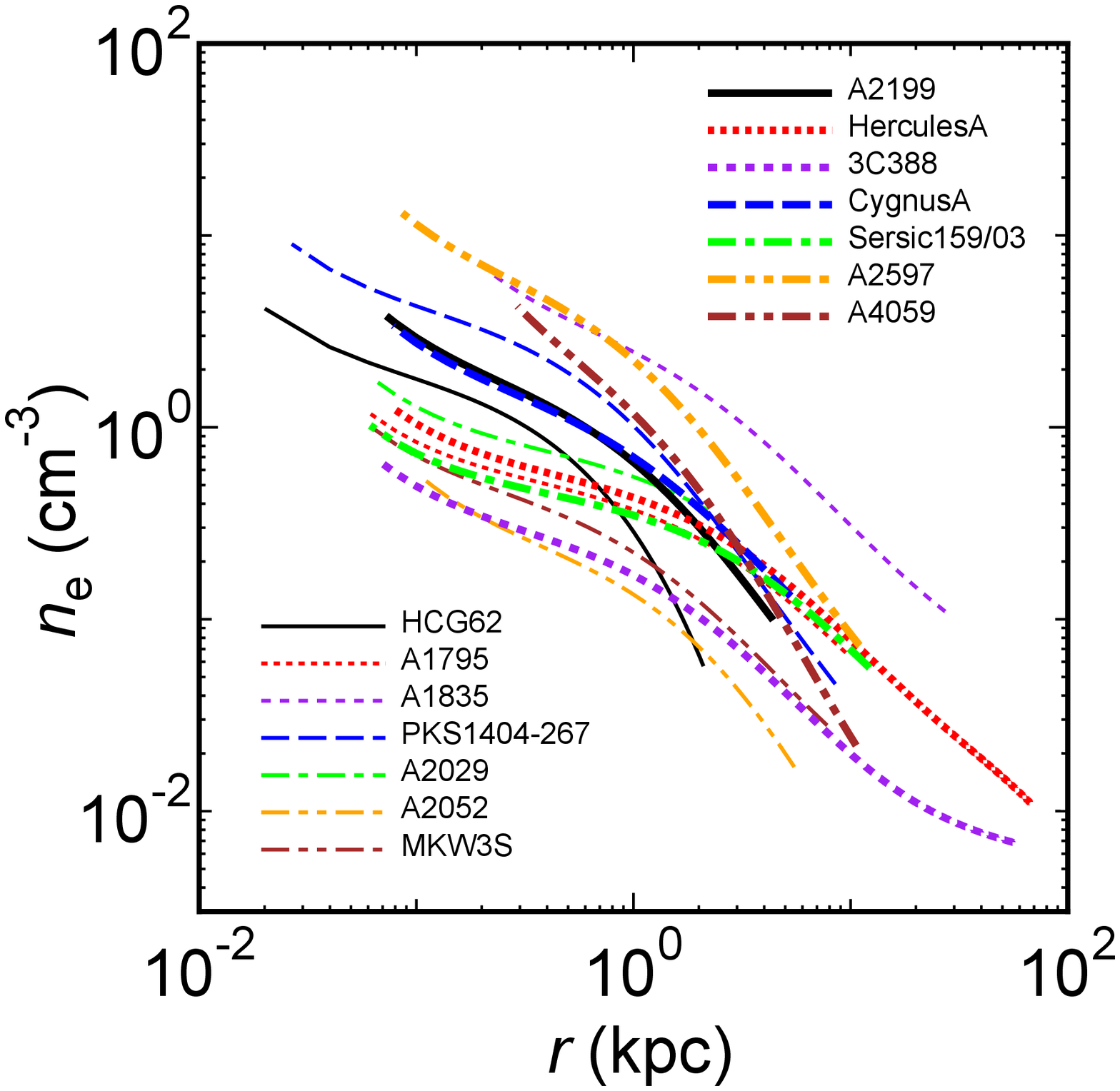}
        \end{center}
      \end{minipage}
    \end{tabular}
     \begin{tabular}{c}
      \begin{minipage}{0.5\hsize}
        \begin{center}
          \includegraphics[width=84mm]{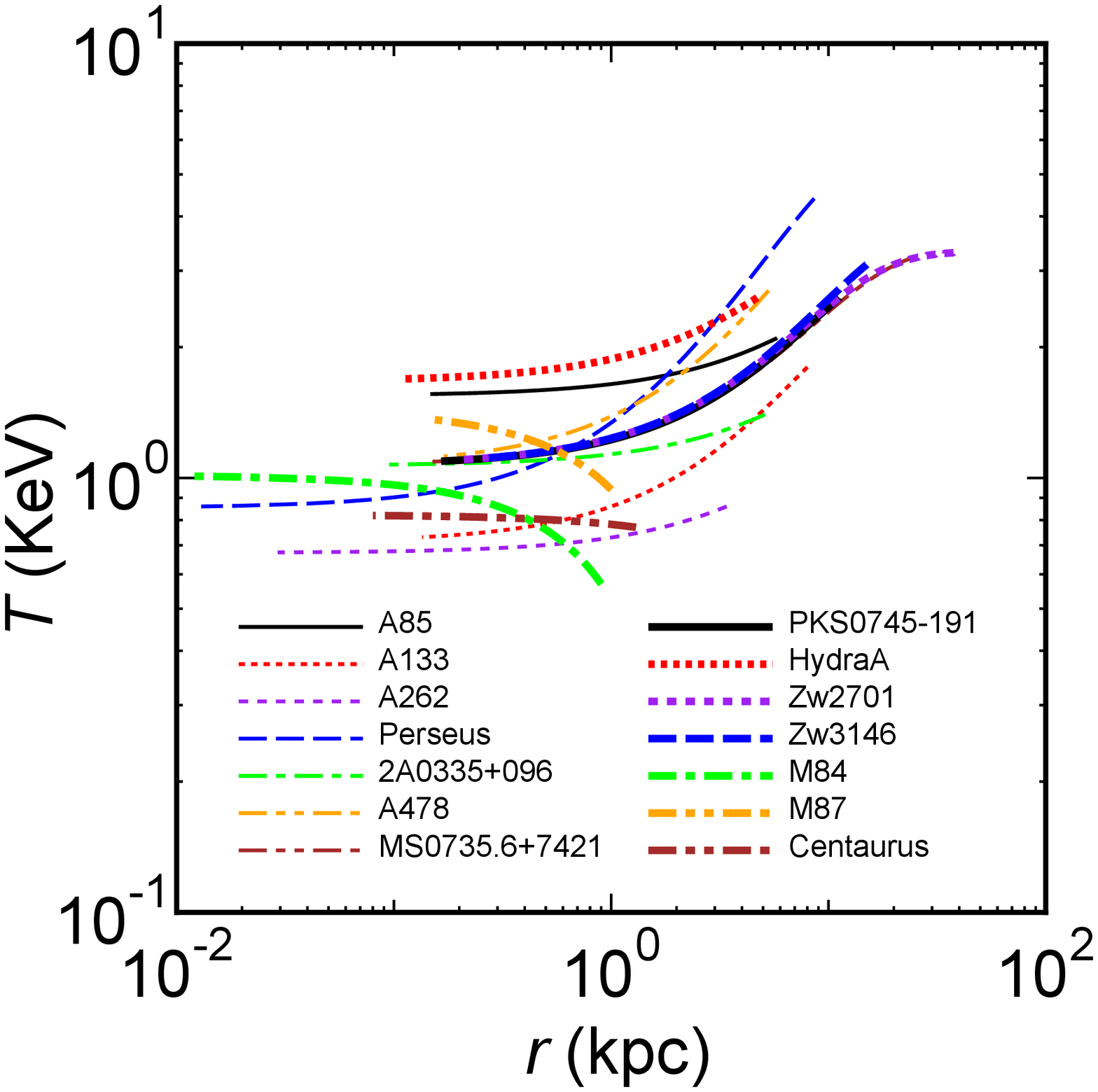}
        \end{center}
      \end{minipage}
      \begin{minipage}{0.5\hsize}
        \begin{center}
          \includegraphics[width=84mm]{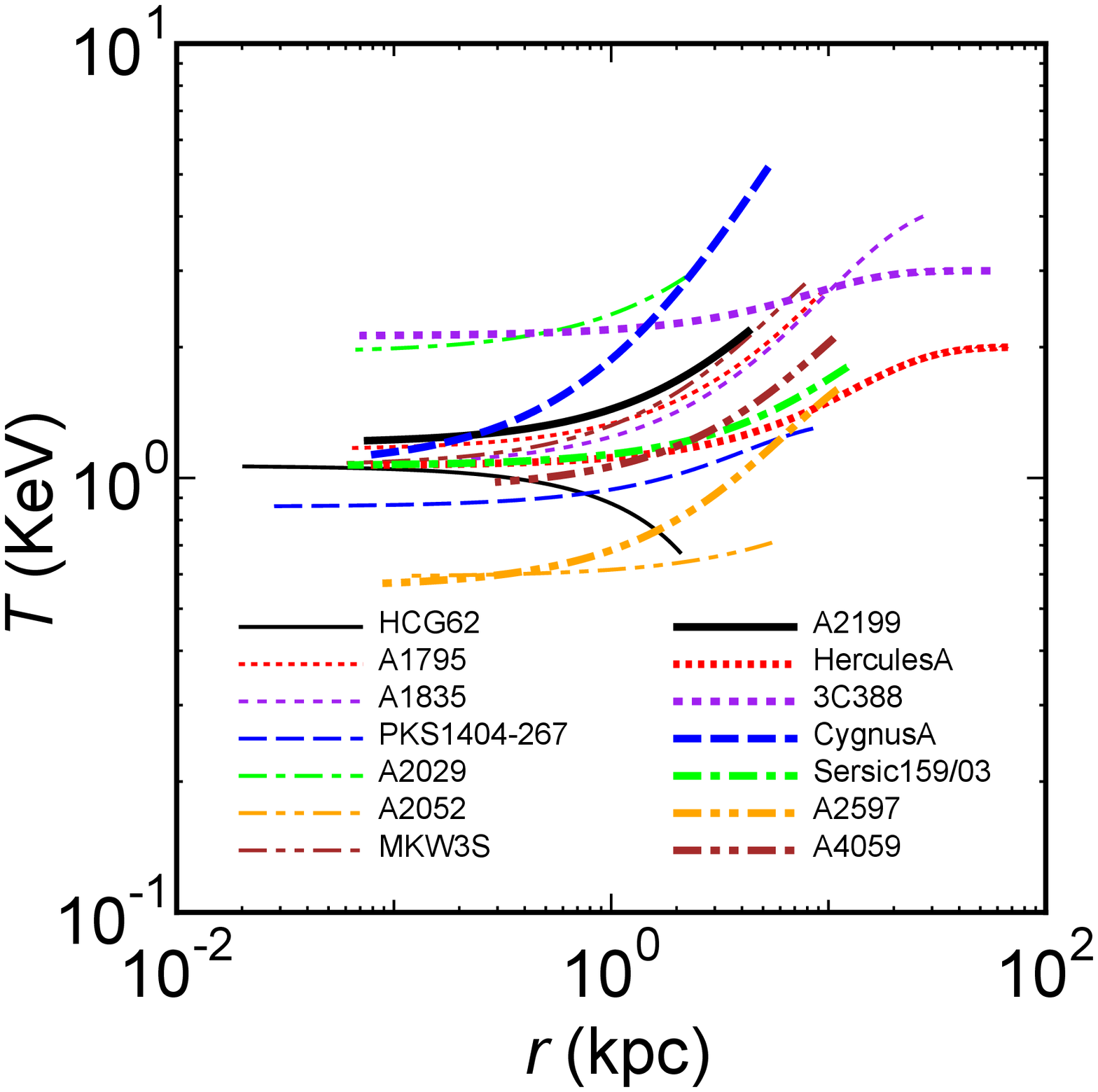}
        \end{center}
      \end{minipage}
    \end{tabular}
  \end{center}
 \caption{Most probable density and temperature profiles based on the
 low-temperature model. The right and left ends of each curve
 correspond to $r_{\rm in}$ and $r_{\rm B}$, respectively.}
 \label{fig:nT}
\end{figure*}

\begin{figure*}
  \begin{center}
    \begin{tabular}{c}
      \begin{minipage}{0.5\hsize}
        \begin{center}
          \includegraphics[width=84mm]{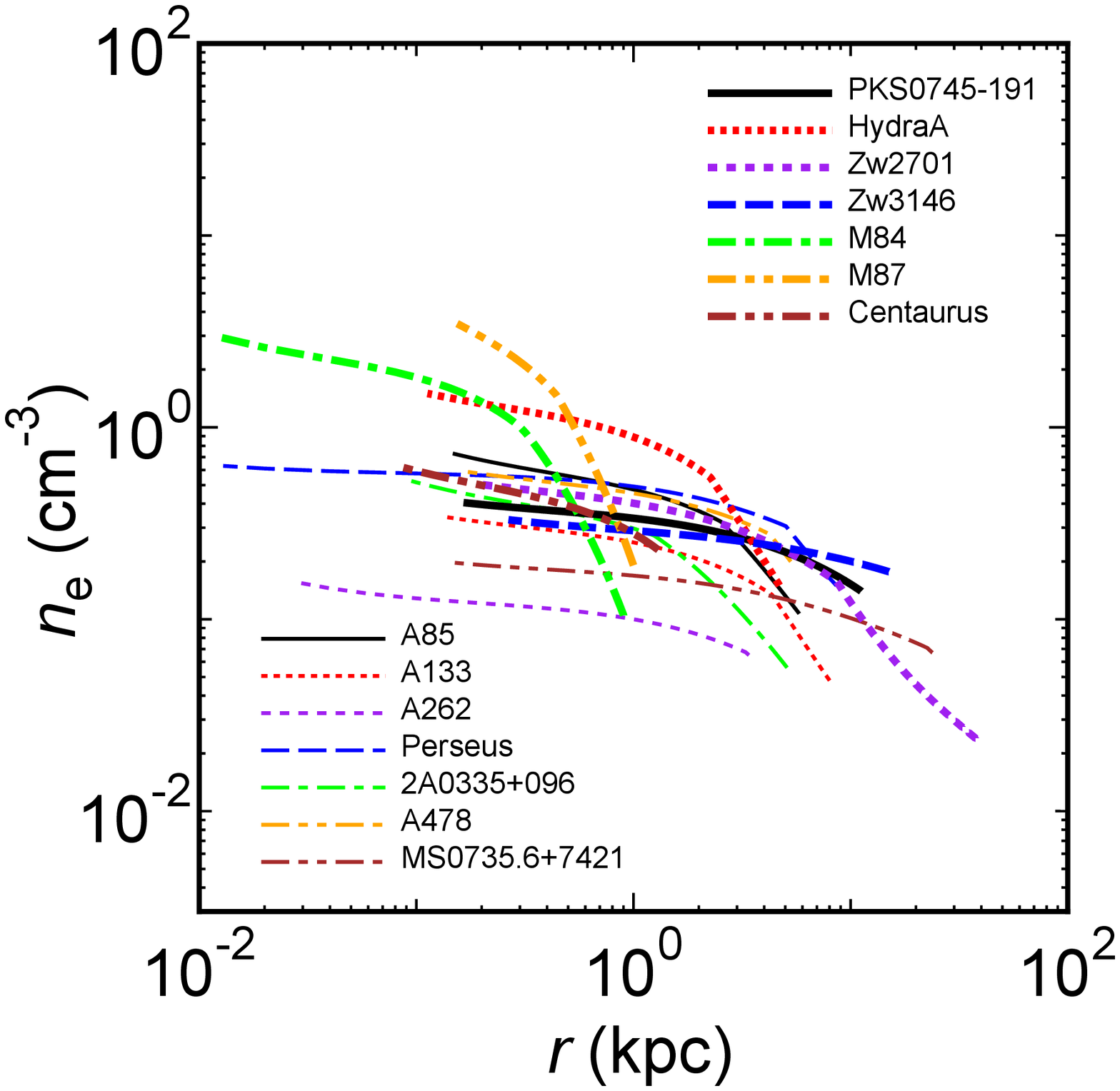}
        \end{center}
      \end{minipage}
      \begin{minipage}{0.5\hsize}
        \begin{center}
          \includegraphics[width=84mm]{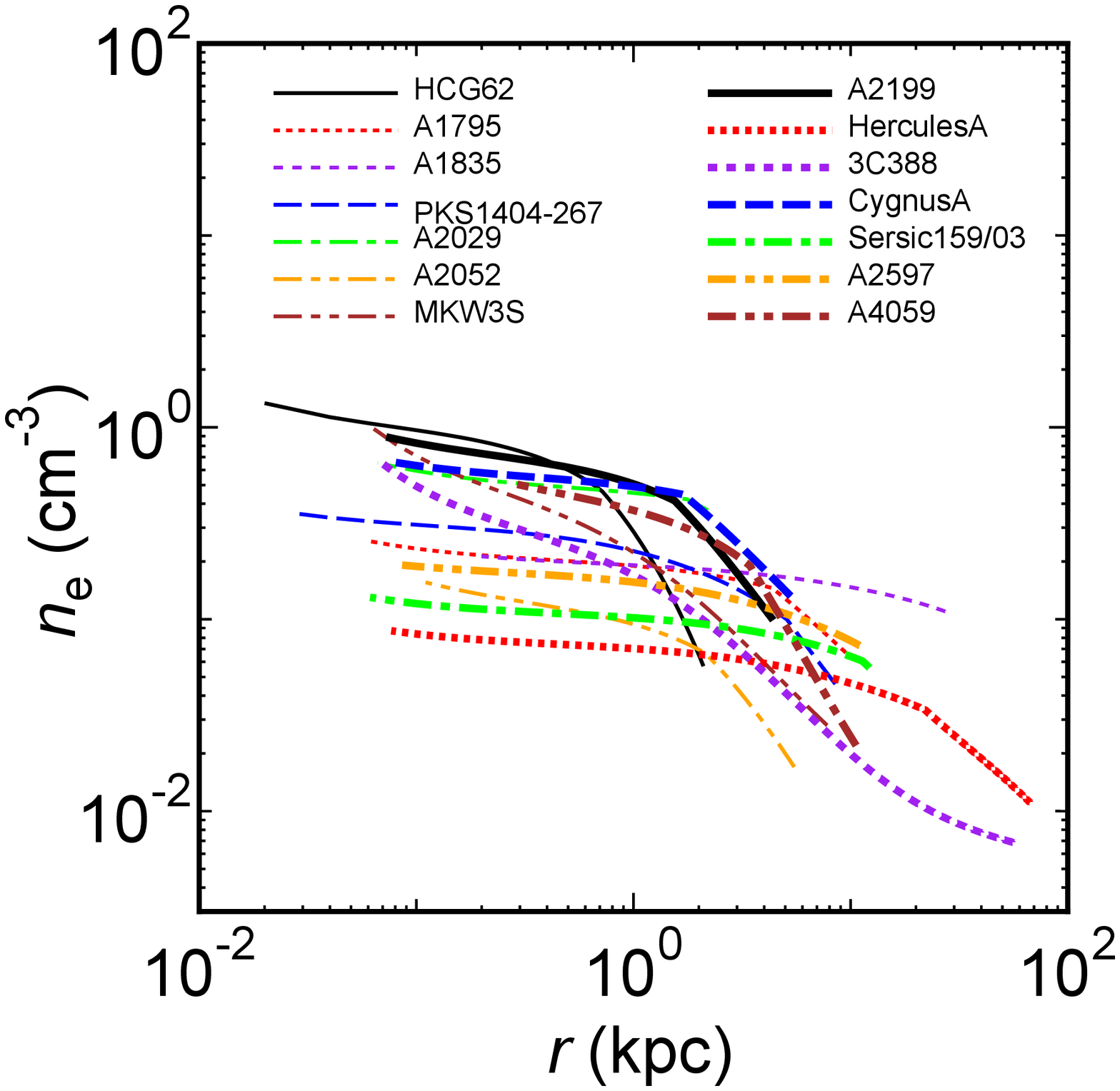}
        \end{center}
      \end{minipage}
    \end{tabular}
     \begin{tabular}{c}
      \begin{minipage}{0.5\hsize}
        \begin{center}
          \includegraphics[width=84mm]{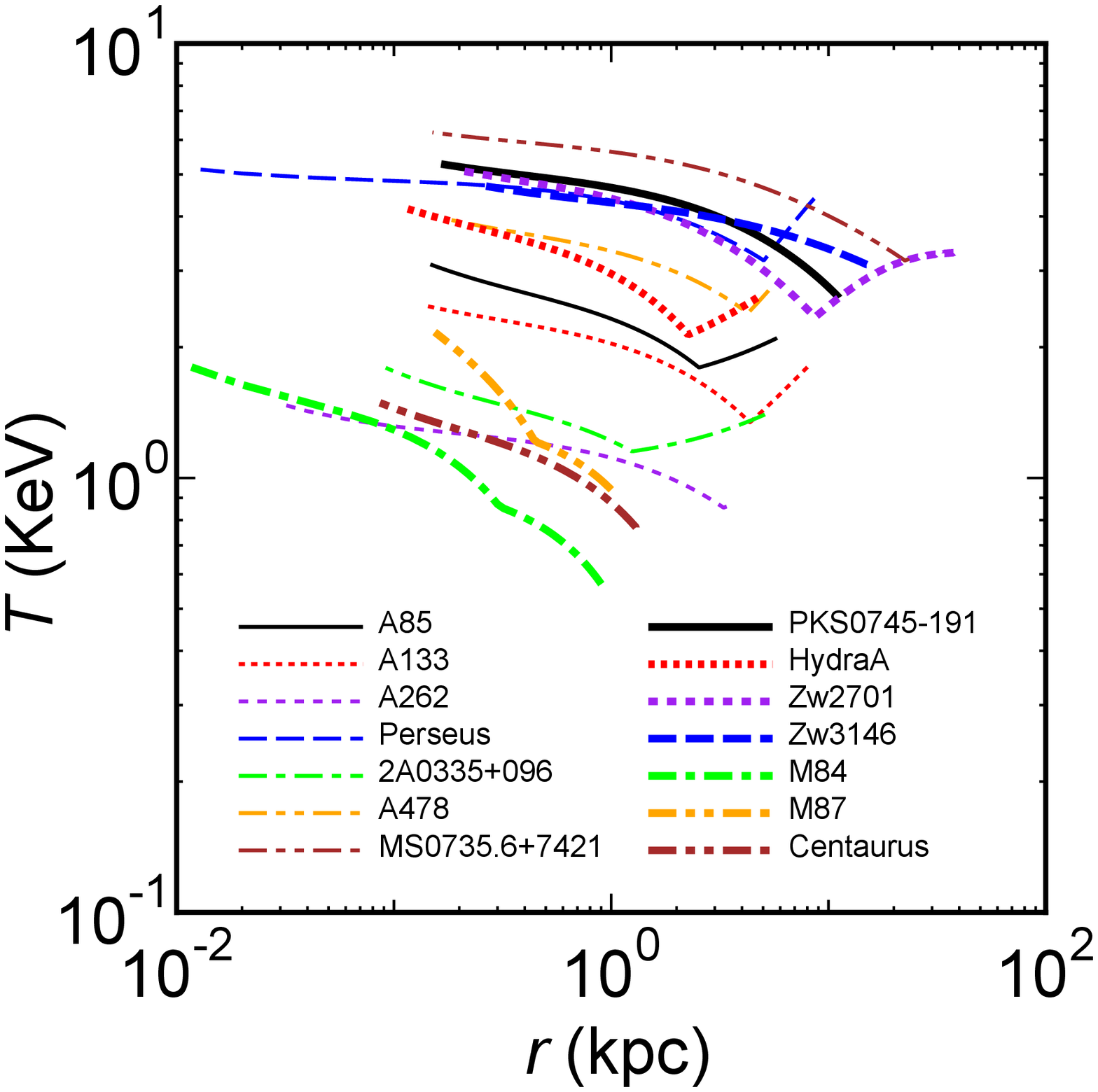}
        \end{center}
      \end{minipage}
      \begin{minipage}{0.5\hsize}
        \begin{center}
          \includegraphics[width=84mm]{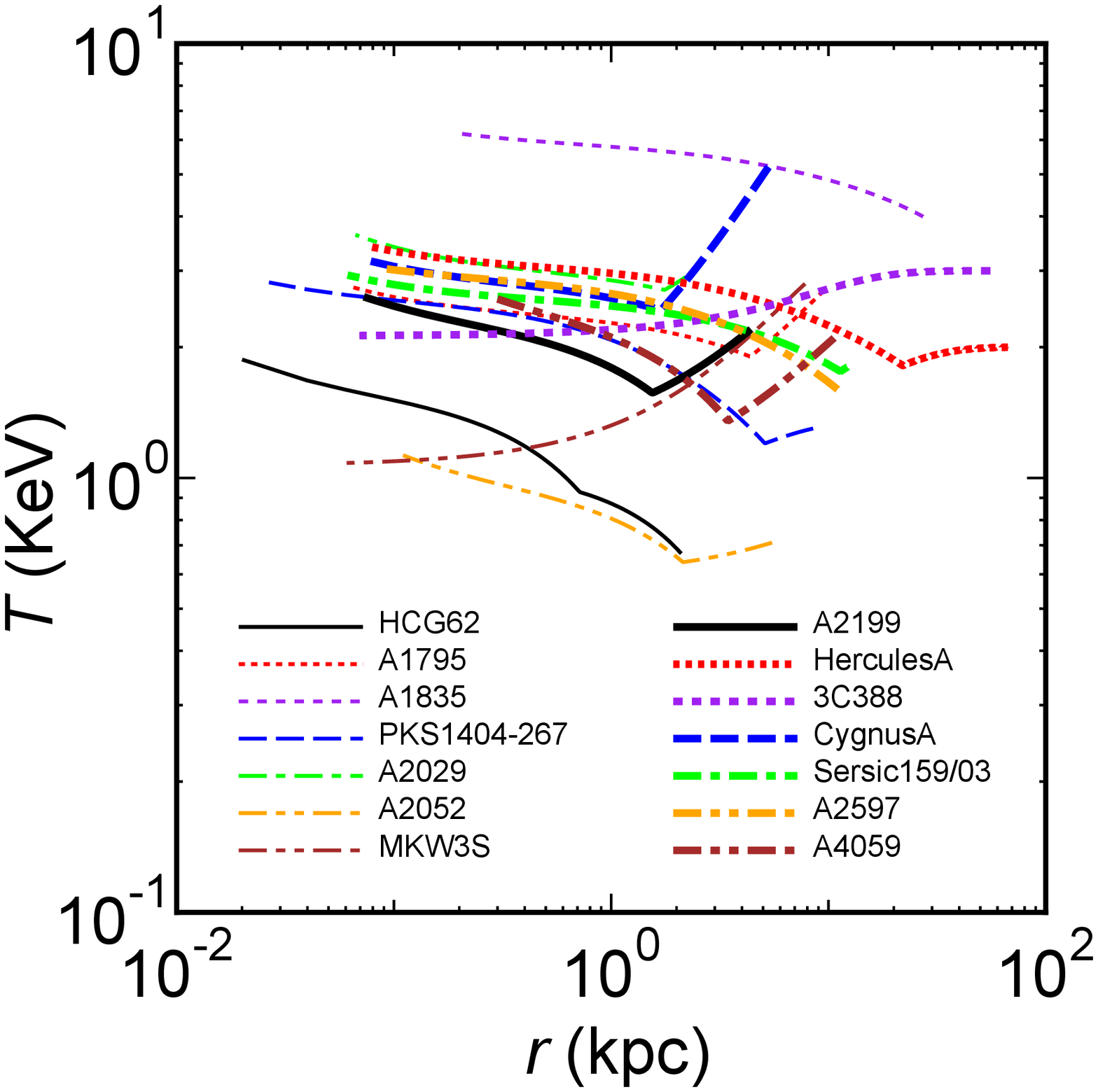}
        \end{center}
      \end{minipage}
    \end{tabular}
  \end{center}
 \caption{Most probable density and temperature profiles based on the
 isentropic model. The right and left ends of each curve correspond
 to $r_{\rm in}$ and $r_{\rm B}$ for the low-temperature model,
 respectively.}  \label{fig:nTe}
\end{figure*}

The low-temperature model gives fairly high densities at the center of
the galaxies (see figure~\ref{fig:nT}), which results in a short cooling
time of the gas, $t_{\rm cool}$. Recent numerical simulations have shown
that thermal instabilities may develop if the condition $t_{\rm
cool}/t_{\rm ff}\lesssim 10$ is satisfied, where $t_{\rm ff} =
(2r/g)^{1/2}$ is the free-fall time (e.g.,
\cite{gas12c,mcc12a,sha12a,mee15a}). Under these conditions, a
substantial fraction of the hot gas may turn into cold gas, and the
density of the remaining hot gas, which occupies most of the volume, can
decrease significantly. The hot gas creates an entropy core at the
center of the galaxy (e.g. \cite{gas13a}). In fact, \citet{fuj16a}
have shown that this is plausibly the situation in the center of
NGC~1275 in the Perseus cluster --- namely, most of the volume
in the central region ($\lesssim 10$~pc) is occupied by tenuous gas
($\lesssim 1\:\rm cm^{-3}$; see also figure~\ref{fig:nTe}).

However, the evolution of the cold component, which can form as a result
of a thermal instability, is expected to differ from that of the hot
one. Its accretion rate can be much higher than the Bondi accretion rate
(see section 5), but analyzing its properties is beyond the scope of
this paper.

Thus, we consider a model in which the density and temperature profiles
are the same as those in the low-temperature model for $r>r_{\rm s}$,
where $r_{\rm s}$ is the radius outside which $t_{\rm cool}/t_{\rm ff}>
10$. For $r<r_{\rm s}$, the entropy of the remaining hot gas is
constant, and thus the relation between the pressure and the density is
given by $p\propto \rho^\gamma$ ($\gamma=5/3$). Moreover, we assume that
the hot gas is in pressure equilibrium (equation~\ref{eq:hydroeq1}). The
cooling time is given by
\begin{equation}
 t_{\rm cool}=\frac{1.5\: n k_{\rm B} T}
{n_{\rm i}n_{\rm e}\Lambda(T,Z)}\:,
\end{equation}
where $n_{\rm i}$ is the ion density. The cooling function $\Lambda$
depends on the temperature $T$ and metal abundance $Z$:
\begin{eqnarray}
 \Lambda(T,Z)&=&2.41\times 10^{-27}
\left[0.8+0.1\left(\frac{Z}{Z_\odot}\right)\right]
\left(\frac{T}{\rm K}\right)^{0.5}\nonumber\\
& &+ 1.39\times 10^{-16}
\left[0.02+0.1\left(\frac{Z}{Z_\odot}\right)^{0.8}\right]\nonumber\\
& &\times\left(\frac{T}{\rm K}\right)^{-1.0}\rm\: erg\: cm^3\:.
\end{eqnarray}
This function approximates the one derived by \citet{sut93a} for
$T\gtrsim 10^5$~K and $Z\lesssim 1\: Z_\odot$ \citep{fuj13a}. We fix the
abundance at $Z=0.5\: Z_\odot$. In this isentropic model, we do not
discuss the Bondi accretion of the hot gas, because the accretion of the
cold gas is expected to dominate. In the following, we consider the
properties of this model between $r=r_{\rm B}$ and $r_{\rm in}$, where
$r_{\rm B}$ is the Bondi radius for the low-temperature model. The
choice of the inner boundary does not affect the results.

 The actual density profiles highly plausibly lie between those
predicted by the low-temperature model and those predicted by the
isentropic model, because thermal instabilities are ignored in the
former and the constant entropy profile is an extreme assumption in the
latter. In other words, from the physical point of view, it is unlikely
that the actual profiles lie outside the two extreme profiles considered
here.

\begin{figure*}
  \begin{center}
    \begin{tabular}{c}
      \begin{minipage}{0.5\hsize}
        \begin{center}
          \includegraphics[width=84mm]{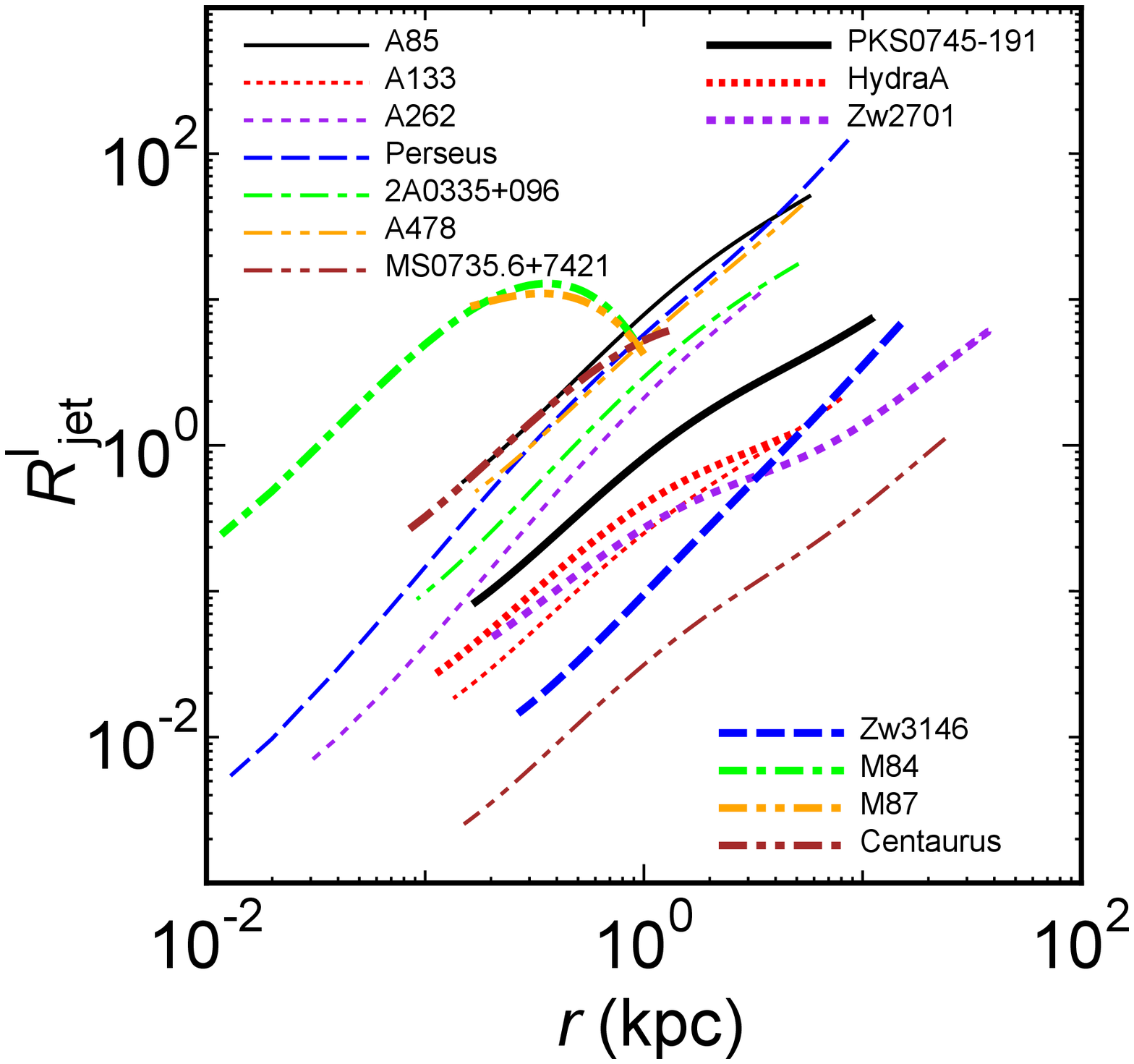}
        \end{center}
      \end{minipage}
      \begin{minipage}{0.5\hsize}
        \begin{center}
          \includegraphics[width=84mm]{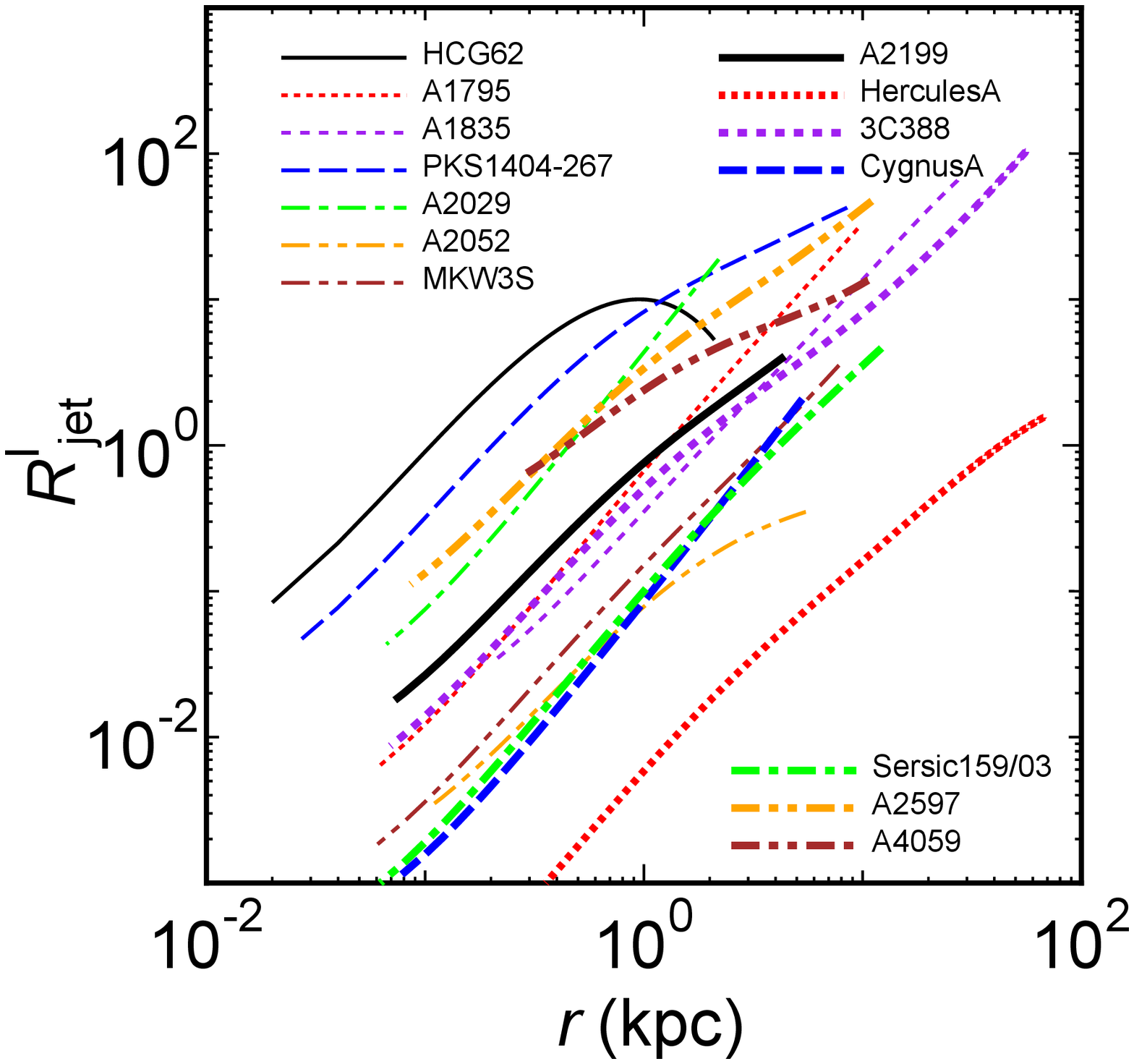}
        \end{center}
      \end{minipage}
    \end{tabular}
  \end{center}
 \caption{Most probable profiles of ${\cal R}^{\rm I}_{\rm jet}$ for
 the FR\,I low-temperature model.}  \label{fig:RI}
\end{figure*}

\begin{figure*}
  \begin{center}
    \begin{tabular}{c}
      \begin{minipage}{0.5\hsize}
        \begin{center}
          \includegraphics[width=84mm]{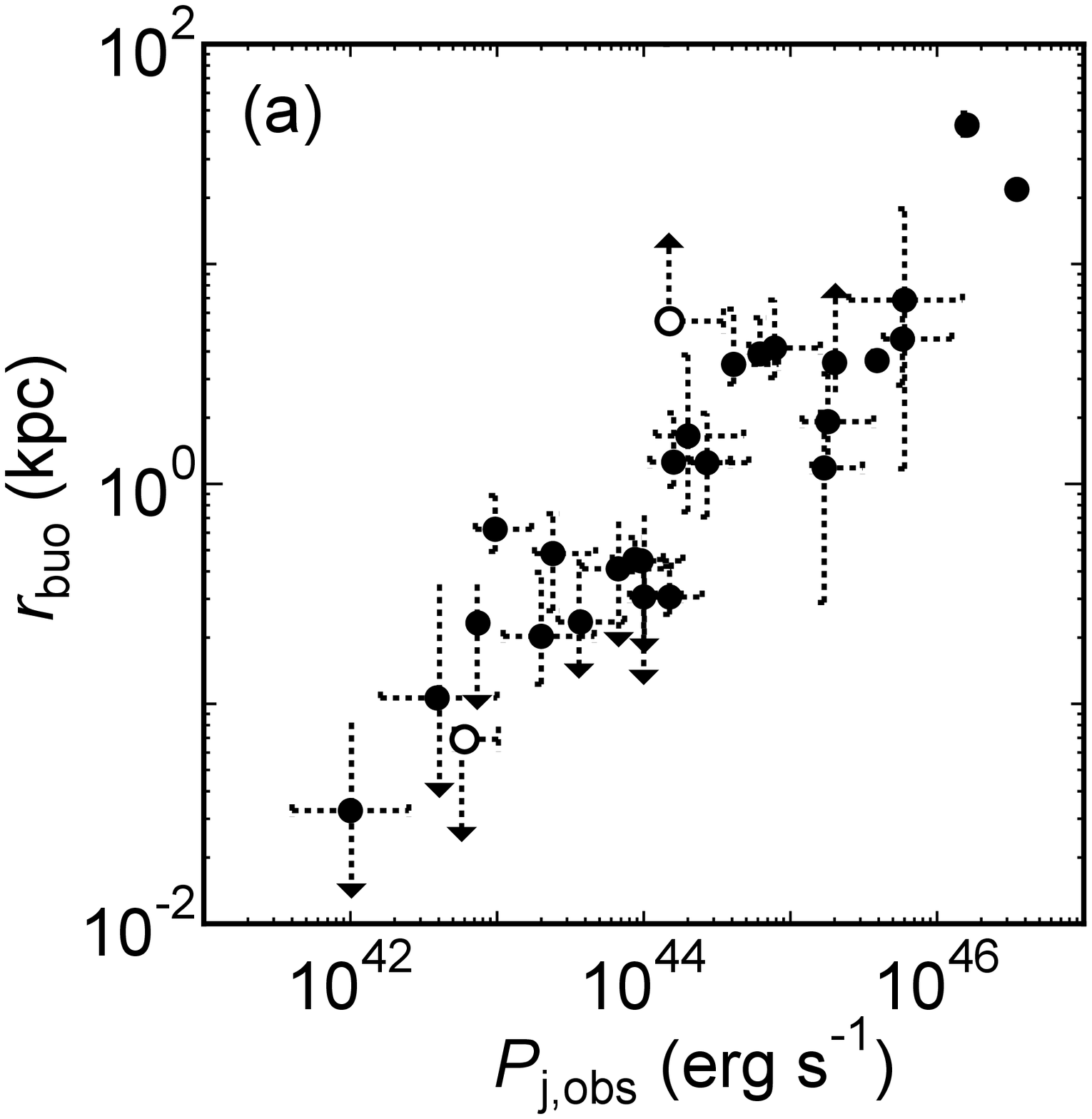}
        \end{center}
      \end{minipage}
      \begin{minipage}{0.5\hsize}
        \begin{center}
          \includegraphics[width=84mm]{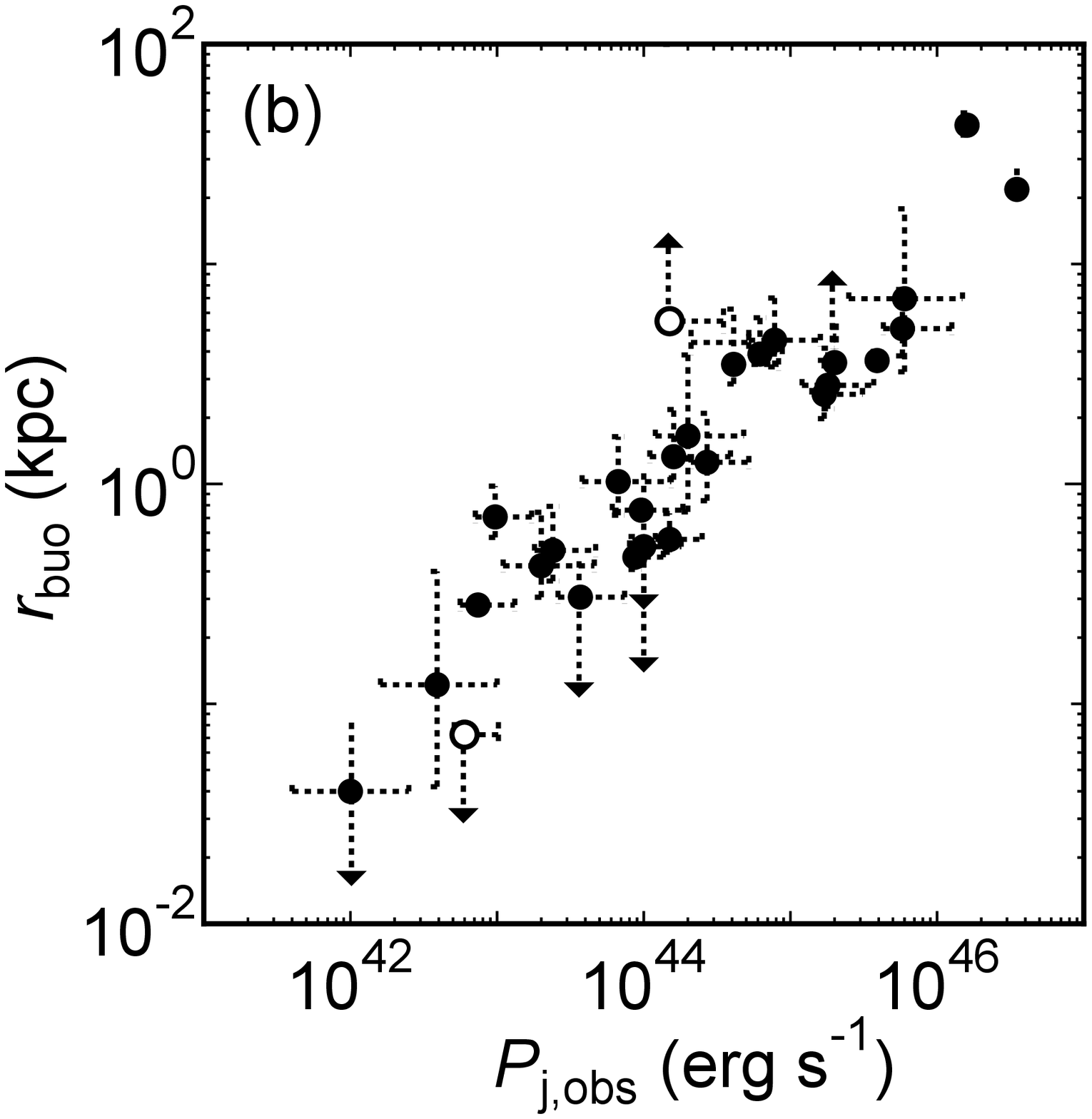}
        \end{center}
      \end{minipage}
    \end{tabular}
  \end{center}
 \caption{(a) Relation between the jet power, $P_{\rm j,obs}$, and
 $r_{\rm buo}$ for the FR\,I low-temperature model with $P_{\rm j}=P_{\rm
j,obs}$. Filled circles show the most probable values. Arrows indicate
that lower or upper limits are not determined. Open circles show upper
or lower limits. If $r_{\rm buo}>r_{\rm in}$ for all realizations
($r_{\rm buo}=+\infty$ in table~\ref{tab:jetI}), we show $r_{\rm in}$
as the lower limit. (b) Same as (a) but for the FR~I isentropic model.}
\label{fig:PrbuoI}
\end{figure*}

\begin{figure*}
  \begin{center}
   \includegraphics[width=84mm]{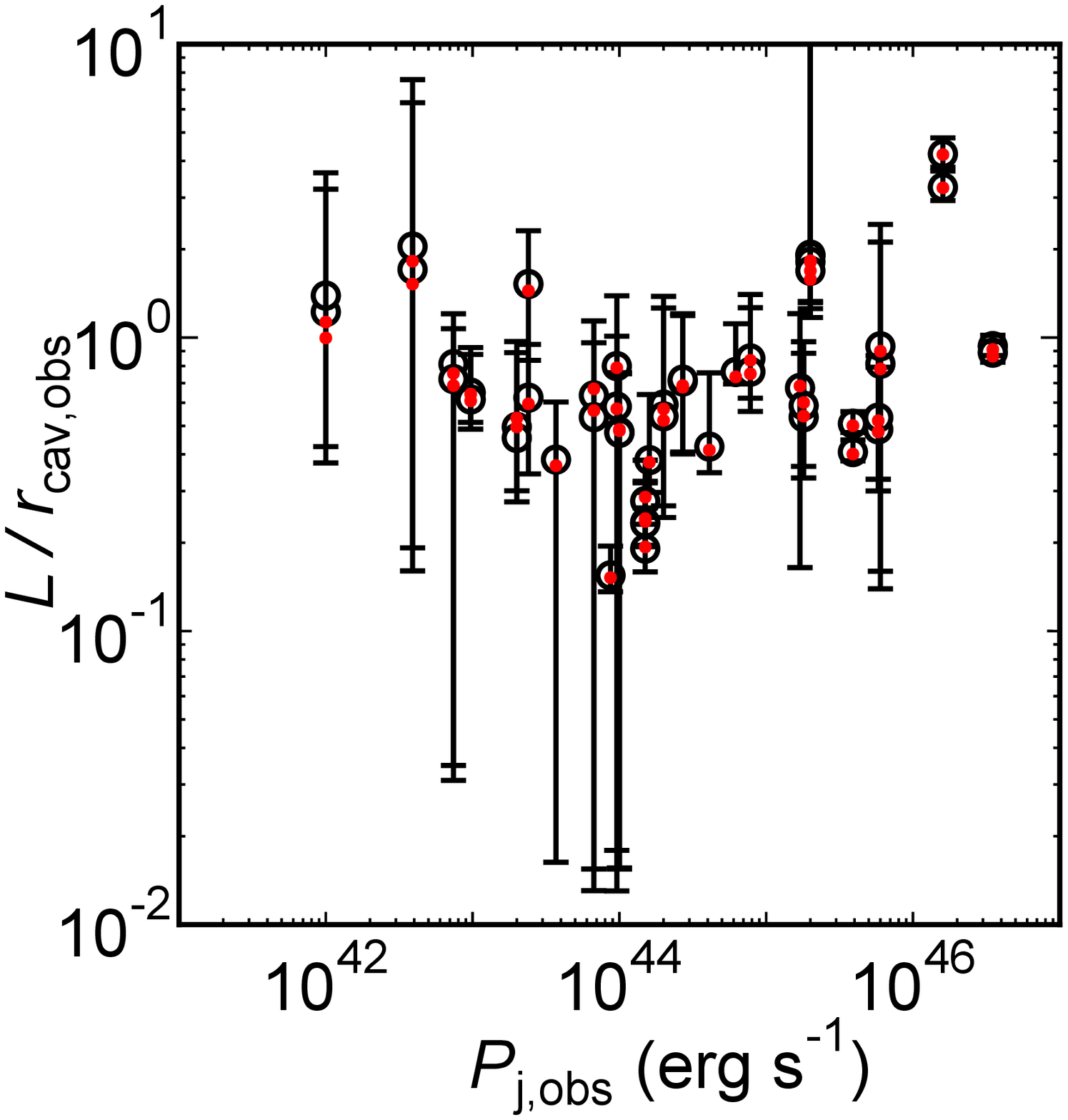}
  \end{center}
\caption{The ratio $L/r_{\rm cav,obs}$ for galaxies for which $r_{\rm
buo}$ can be determined. Open circles are for the FR\,I low-temperature
model and filled circles are for the FR\,I isoentropic model. Error bars
are drawn for the former; they are almost the same for the latter. Some
galaxies have multiple cavities, and thus they have multiple values of
$L/r_{\rm cav,obs}$. Galaxies for which only lower or upper limits
of $r_{\rm bub}$ have been obtained are not included.}
\label{fig:rat_rbub}
\end{figure*}

\begin{figure*}
  \begin{center}
    \begin{tabular}{c}
      \begin{minipage}{0.5\hsize}
        \begin{center}
          \includegraphics[width=84mm]{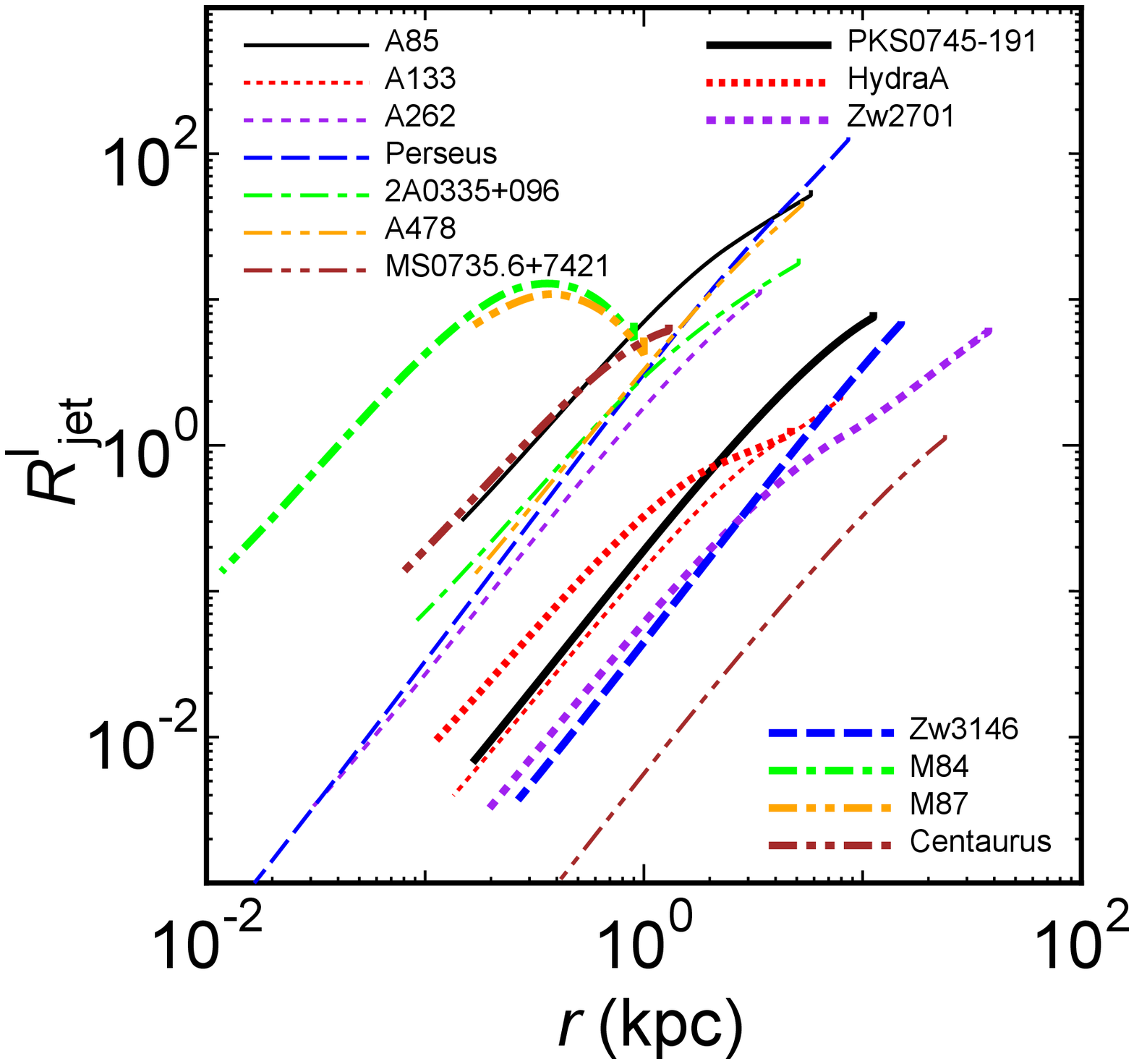}
        \end{center}
      \end{minipage}
      \begin{minipage}{0.5\hsize}
        \begin{center}
          \includegraphics[width=84mm]{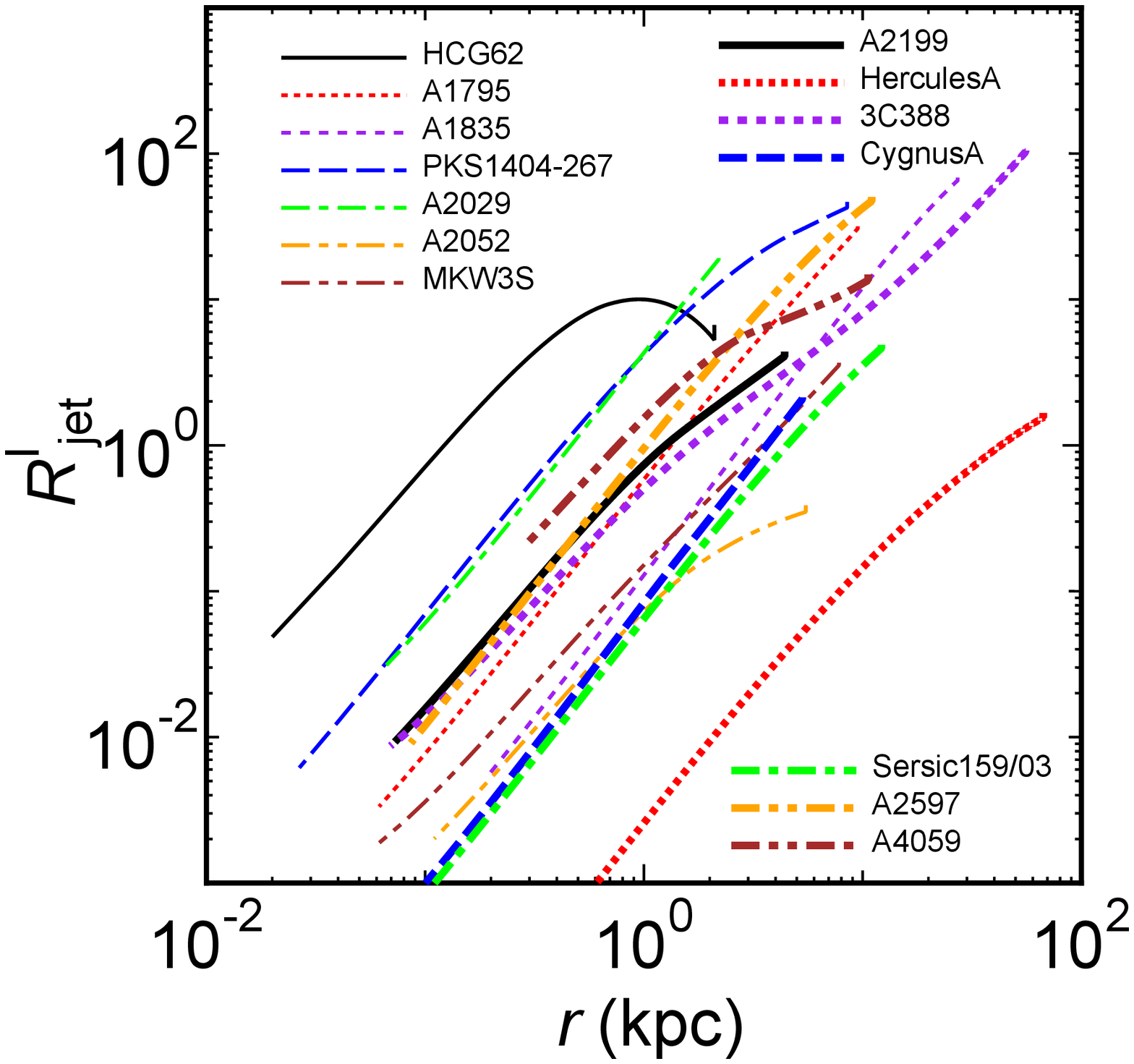}
        \end{center}
      \end{minipage}
    \end{tabular}
  \end{center}
 \caption{Most probable profiles of ${\cal R}^{\rm I}_{\rm jet}$ for
 the FR\,I isentropic model.} \label{fig:ReI}
\end{figure*}

\section{Data}
\label{sec:data}

Accounting for data uniformity and consistency, we study 28 bright
elliptical galaxies in clusters, for which the properties of the central
gas, the excavated cavity, etc., have been studied by
\authorcite{raf06a} (\yearcite{raf06a}; see their table~6). They are
BCGs except for M84. We show the parameters for the gravitational
potentials in table~\ref{tab:pot} (see also
appendix~\ref{sec:app_low}). The masses of the SMBHs were derived by
\citet{mcn11a}. They have been estimated using $R$-band absolute
magnitudes ($M_{\rm R}$) obtained by \citet{raf06a}.  Since
\citet{mcn11a} did not provide the error bars in $M_\bullet$, we take
them as 0.5~dex, based on the dispersion around the observed $M_{\rm
R}$--$M_\bullet$ relations (e.g. \cite{mcl02a}; see also \cite{mcc13a}
for massive galaxies). Using the $R$-band absolute magnitudes, the
galaxy masses, $M_{\rm gal}$, were estimated by \citet{raf06a} and are
consistent with $M_\bullet$. The effective radii of the galaxies,
$R_{\rm e}$, have been derived from the 2MASS All-Sky Extended Source
Catalogue \citep{skr06}\footnote{http:\slash\slash
irsa.ipac.caltech.edu\slash frontpage}. We assume the average of $R_{\rm
e}$ in the $J$, $H$ and $K$-bands, and take their scatter as the
error. The galaxy velocity dispersion, $\sigma$, has been obtained from
the HyperLeda database\citep{mak14}\footnote{http:\slash\slash\
leda.univ-lyon1.fr\slash}. However, no data have been found for 12
galaxies. For those galaxies, we take the error-weighted average of the
remaining 16 galaxies ($290\rm\: km\: s^{-1}$) as $\sigma$, and the
scatter of these galaxies as the error of $\sigma$ ($36\rm\: km\:
s^{-1}$).

In addition, the parameters for the clusters are shown in
table~\ref{tab:pot} (see also appendix~\ref{sec:app_low}). Most of them
are based on recent X-ray observations. If there are no appropriate
X-ray data, we adopt the data obtained through lensing observations or
kinematics of the member galaxies. We do not consider the contribution
of the cluster component to the total gravitational acceleration $g$ for
M84, HCG~62, and 3C~388, because M84 is not a BCG and there are no
appropriate data for the other two. For Hercules~A and Cygnus~A, we use
the cluster temperatures and the core radii derived by \citet{giz04a}
and \citet{smi02a}, respectively. Using the cluster mass--temperature
relation derived by \citet{sun09a}, we convert the temperatures into the
cluster masses, $M_{\rm vir}$. The core radii, $r_{\rm c}$, can be
converted into the characteristic radii, $r_{\rm s}=r_{\rm vir}/c_{\rm
vir}$, by using the relation $r_{\rm s}=r_{\rm c}/0.22$ \citep{mak98a}.

The boundary conditions $r_{\rm in}$, $n_{\rm e,in} (=n_{\rm e}(r_{\rm
in}))$, and $T_{\rm in}$ are shown in table~\ref{tab:obs}, and are the
same as those in table~6 of \citet{raf06a}. In their table, $r_{\rm
in}$, $n_{\rm e,in}$, and $T_{\rm in}$ are represented by $a$, $n_{\rm
e}$, and $kT$, respectively. Although \citet{raf06a} gave the average
densities and temperatures for $r<r_{\rm in}$ excluding the AGN, we
expect that most of the emission comes from $r\sim r_{\rm in}$, because
the density profiles near the galaxy centers are not very steep, i.e.,
$\alpha\lesssim 1$ for $\rho\propto r^{-\alpha}$, as is shown later
(figures~\ref{fig:nT} and~\ref{fig:nTe}). Note that the density and
temperature used are the deprojected ones. Thus, in general, the density
is higher and the temperature is lower than the projected ones because
the former increases and the latter decreases toward the galactic
center. The deprojected values should be identical to the actual ones as
long as the gas is spherically distributed and the density and
temperature change smoothly. However, if the gas is strongly disturbed
by AGN activities, the results may have some uncertainties. For example,
figure~10 of \citet{rus13b} shows that the density and temperature
profiles of some galaxies are somewhat irregular at their centers, which
may indicate errors of less than a factor of two.

The jet kinetic power, $P_{\rm j}$, can be obtained observationally, and
we denote it by $P_{\rm j,obs}$. It has been estimated as the ratio of
the enthalpy of cluster X-ray cavities to their buoyancy timescales
\citep{mcn11a}. The enthalpy is given by
\begin{equation}
\label{eq:Ecav}
 E_{\rm cav}=\frac{\gamma_{\rm c}}{\gamma_{\rm c}-1}p_{\rm s}V_{\rm c}\:,
\end{equation}
where $p_{\rm s}$ is the pressure of the gas surrounding the cavity, and
$V_{\rm c}$ is the cavity's volume. Note that while
equation~(\ref{eq:Ecav}) is appropriate for FR\,I objects (most of our
sample galaxies), it may underestimate the jet power for FR\,II objects
(Cygnus~A) by at most a factor of 10 \citep{ito08a}. Thus, $P_{\rm
j,obs}$ for Cygnus~A should be regarded as a lower limit. The jet power
can also be estimated from the Bondi accretion rate. The maximal power
released from the neighborhood of the SMBH through the Bondi accretion
is
\begin{equation}
 P_{\rm B} = \eta\dot{M}_{\rm B}c^2\:,
\end{equation}
where $\eta$ is the accretion efficiency assumed $\eta=0.1$.  We present
$P_{\rm j,obs}$ and $P_{\rm B}$ in tables~\ref{tab:obs}
and~\ref{tab:res}.  In general, $P_{\rm B}$ is much larger than $P_{\rm
j,obs}$.

The reduction factor (see equation~\ref{eq:condI}) is given by
$f_1=t_{c_{\rm s}}/t_{\rm buoy}$, where $t_{c_{\rm s}}$ is the migration
time of a cavity when the rising velocity is the sound velocity, and
$t_{\rm buoy}$ is the one when the rising velocity is the buoyant
velocity (e.g., \cite{bir04a}). Using X-ray observations, \citet{raf06a}
estimated both times for our sample galaxies (see their table~5), and we
adopt those values. If multiple values are given per galaxy because
there are multiple cavities, we take the average. In general, the
reduction factor is $0.5\lesssim f_1\lesssim 1$.

\section{Results}
\label{sec:results}

Using input parameters shown in tables~\ref{tab:pot} and~\ref{tab:obs},
we calculate the evolution of cocoons. Output parameters are summarized
in tables~\ref{tab:res}--\ref{tab:jetII}.

\subsection{Hot gas profiles}
\label{sec:hot}

In this subsection, we invoke the Bondi accretion model, for reference
purposes only.  For the low-temperature model, we calculate the Bondi
accretion radii, $r_{\rm B}$, the density, $n_{\rm e,B}=n_{\rm e}(r_{\rm
B})$, and the temperature, $T_{\rm B}=T(r_{\rm B})$, at these radii, and
show them in table~\ref{tab:res}. We also present the Bondi accretion
rates. The Bondi radii we obtain are substantially larger than those in
\citet{raf06a} because of smaller $T_{\rm B}$ we adopt. Using Monte
Carlo simulations, we estimate the uncertainties in the results. Each
input parameter is randomly perturbed with a Gaussian distribution of
the perturbations, with an amplitude determined by the error bar of the
parameter. We obtain $10^3$ different realizations.

Figure~\ref{fig:nT} shows the density and the temperature profiles
between $r_{\rm B}$ and $r_{\rm in}$ for the low-temperature
model. While the density profiles can be represented by a power-law for
most galaxies, some profiles show noticeable bends. In
figure~\ref{fig:nTe}, we show the density and the temperature profiles
for the isentropic model. Sharp bends in the temperature profiles
correspond to the radii $r_{\rm s}$ where $t_{\rm cool}/t_{\rm ff}=10$
inside which the entropy is constant. The galaxies with monotonically
increasing temperatures toward the centers have the ratios of $t_{\rm
cool}/t_{\rm ff}<10$ at $r=r_{\rm in}$ (e.g. PKS~0745--191). Those with
monotonically decreasing temperatures have the ratios of $t_{\rm
cool}/t_{\rm ff}>10$ for $r_{\rm B} < r < r_{\rm in}$ (e.g. MKS~3S);
their profiles are the same as those in figure~\ref{fig:nT}. Except for
the last ones ($t_{\rm cool}/t_{\rm ff}>10$ for $r_{\rm B} < r < r_{\rm
in}$), the central densities derived based on the isentropic model are
much smaller than those on the low-temperature model
(figures~\ref{fig:nT} and~\ref{fig:nTe}). From
conditions~(\ref{eq:condI}) and~(\ref{eq:condII}), we define
\begin{equation}
\label{eq:RjetI}
 {\cal R}^{\rm I}_{\rm jet}(r)
= \frac{1}{P_{\rm j,obs}}\left[\xi(r) + 
\frac{\gamma_{\rm c}}{\gamma_{\rm c}-1}
4\pi r^2 f_1 c_{\rm s}(r) p(r)\right]\:,
\end{equation}
\begin{equation}
\label{eq:RjetII}
 {\cal R}^{\rm II}_{\rm jet}(r)
= \frac{2\rho(r)f_1^2 c_{\rm s}(r)^2 c A_{\rm h}(r)}{P_{\rm j,obs}}\:,
\end{equation}
Therefore, if ${\cal R}^i_{\rm jet}<1$ ($i=$~I or II), a jet with
$P_{\rm j}=P_{\rm j,obs}$ can progress in the ambient medium with
velocities larger than the buoyant velocities (phase A in
figure~\ref{fig:cocoon}). Equations~(\ref{eq:RjetI})
and~(\ref{eq:RjetII}), and (\ref{eq:Ah}) show that smaller $f_1$ and
$f_2$ give smaller ${\cal R}^i_{\rm jet}$. Using the density and
temperature profiles obtained in this subsection, we calculate the
profiles of ${\cal R}^i_{\rm jet}$.

\subsection{FR\,I type evolution and cavity ascent}
\label{sec:FRI}

Next, we discuss the results obtained for the case when cocoons expand
by the pressure inside them (section~\ref{sec:FRImodel}).  We show the
profiles of ${\cal R}^{\rm I}_{\rm jet}$ for the low-temperature model
in figure~\ref{fig:RI}. In general, they correspond to an increasing
function of $r$, because the pressure of the hot gas $p$ decreases
slower than $r^{-2}$ (equation~\ref{eq:RjetI}). However, ${\cal R}^{\rm
I}_{\rm jet}$ for a number of galaxies decrease in the outer region
because $p$ decreases faster than $r^{-2}$. These objects include a
single galaxy (M84), or belong to a galaxy group (HCG~62) or a
low-temperature cluster (M87), for which the contribution of the cluster
component can be ignored. On the contrary, galaxies in massive clusters
have monotonically increasing ${\cal R}^{\rm I}_{\rm jet}$.

We estimate the size of cavities by deriving the buoyancy radius $r_{\rm
buo}$ within which ${\cal R}^{\rm I}_{\rm jet}<1$
(figure~\ref{fig:PrbuoI}a) for our sample galaxies (see also
table~\ref{tab:jetI}). In figure~\ref{fig:PrbuoI}a, the radius $r_{\rm
buo}$ is an increasing function of $P_{\rm j,obs}$, and $r_{\rm
buo}<10$~kpc for most of them. The buoyancy radius $r_{\rm buo}$ does
not necessary show the observed positions of the cavities, because the
cavities rise in the ambient medium via buoyancy. We expect that the
growth of a cocoon stops at $r\sim r_{\rm buo}$ and thus the initial
size of the cavity is $L_{\rm ci}\sim r_{\rm buo}$ (phase B in
figure~\ref{fig:cocoon}).  The sizes of a significant fraction of
cocoons is $r_{\rm buo}\lesssim 1$~kpc ($P_{\rm j,obs}\lesssim
10^{44}\rm\: erg\: s^{-1}$ in figure~\ref{fig:PrbuoI}a). The radius
$r_{\rm buo}$ cannot be compared directly with the observed size of the
cavities $r_{\rm buo,obs}$ because of the evolution in phases B and D
(figure~\ref{fig:cocoon}). The details of the corrections are described
in appendix~\ref{sec:app_cavity}. In general, the ratio of the final
size of the cavity $L$ to the initial size $L_{\rm ci}$
(figure~\ref{fig:cocoon}) is less than a factor of a few.  In
figure~\ref{fig:rat_rbub}, we show the ratio $L/r_{\rm cav,obs}$. We
assume that $r_{\rm cav,obs}=\sqrt{ab}/2$, where $a$ and $b$ are the
semimajor and semiminor axes obtained by \citet{raf06a}, respectively.
We did not include galaxies for which only lower or upper limits of
$r_{\rm bub}$ have been obtained in
figure~\ref{fig:PrbuoI}. Figure~\ref{fig:rat_rbub} shows that $L/r_{\rm
cav,obs}\sim 0.5$ on average and indicates that the predicted radius of
the cavities is consistent with the observed radius within a factor of a
few. Note that $\xi$ in equation~(\ref{eq:RjetI}) affects $L/r_{\rm
cav,obs}$ by only less than a few percent in most cases.

One may assume that the actual jet power, $P_{\rm j}$, is represented
by $P_{\rm B}$ rather than $P_{\rm j,obs}$, if the Bondi accretion is
realized and if most of the jet energy has escaped from the cavities.
In this case, we replace $P_{\rm j,obs}$ in equation~(\ref{eq:RjetI}) by
$P_{\rm B}$:
\begin{equation}
\label{eq:RBI}
 {\cal R}^{\rm I}_{\rm B}(r)
= \frac{1}{P_{\rm B}}\left[\xi(r) + 
\frac{\gamma_{\rm c}}{\gamma_{\rm c}-1}
4\pi r^2 f_1 c_{\rm s}(r) p(r)\right]\:,
\end{equation}
In general, ${\cal R}^{\rm I}_{\rm B}$ is smaller than ${\cal R}^{\rm
I}_{\rm jet}$ , because $P_{\rm B}>P_{\rm j,obs}$ (tables~\ref{tab:obs}
and~\ref{tab:res}). We calculate the buoyancy radius, $r_{\rm buo}$,
within which ${\cal R}^{\rm I}_{\rm B}<1$ for our sample galaxies
(table~\ref{tab:jetI}). The buoyancy radii, $r_{\rm buo}$, for $P_{\rm
j}=P_{\rm B}$ are much larger than those for $P_{\rm j}=P_{\rm j,obs}$.
They can be $r_{\rm buo}\gtrsim 5$~kpc, considering the errors. This
means that, if $P_{\rm j}\approx P_{\rm B}$ is realized in actual
galaxies, the jet power is large enough or even exceeds the required
amount to explain the sizes of the observed cavities in most of the
galaxies.

Next, we show the profiles of ${\cal R}^{\rm I}_{\rm jet}$ for the
isentropic model in figure~\ref{fig:ReI}. For a given radius, ${\cal
R}^{\rm I}_{\rm jet}$ in figure~\ref{fig:ReI} are smaller than those in
figure~\ref{fig:RI} (see also table~\ref{tab:jetI}), because the
pressure of the hot gas outside the cocoon ($p$) is generally smaller in
the isentropic models. This is especially true in the innermost regions
of the galaxies where the profiles of the hot gas are different between
the isentropic and low-temperature models (figures~\ref{fig:nT}
and~\ref{fig:nTe}). However, the overall tendency is not different
between figure~\ref{fig:RI} and figure~\ref{fig:ReI}; ${\cal R}^{\rm
I}_{\rm jet}$ is the increasing function of $r$ for most galaxies. This
means that the expansion of a cocoon stops at a buoyancy radius $r_{\rm
buo}$. Figure~\ref{fig:PrbuoI}b shows the relation between $P_{\rm
j,obs}$ and $r_{\rm buo}$ for the isentropic model. This figure is
nearly the same as figure~\ref{fig:PrbuoI}a, although $r_{\rm buo}$ of
the former is slightly smaller (by less than a factor of a few). This
happens because the difference in the pressure $p$ at $r\sim r_{\rm
buo}$ is not large between the isentropic and low-temperature models. In
figure~\ref{fig:rat_rbub}, we show the ratios $L/r_{\rm cav,obs}$ for
this model --- they are similar to those for the low-temperature model.

\subsection{FR\,II type evolution}
\label{sec:FRII}

\begin{figure*}
  \begin{center}
    \begin{tabular}{c}
      \begin{minipage}{0.5\hsize}
        \begin{center}
          \includegraphics[width=84mm]{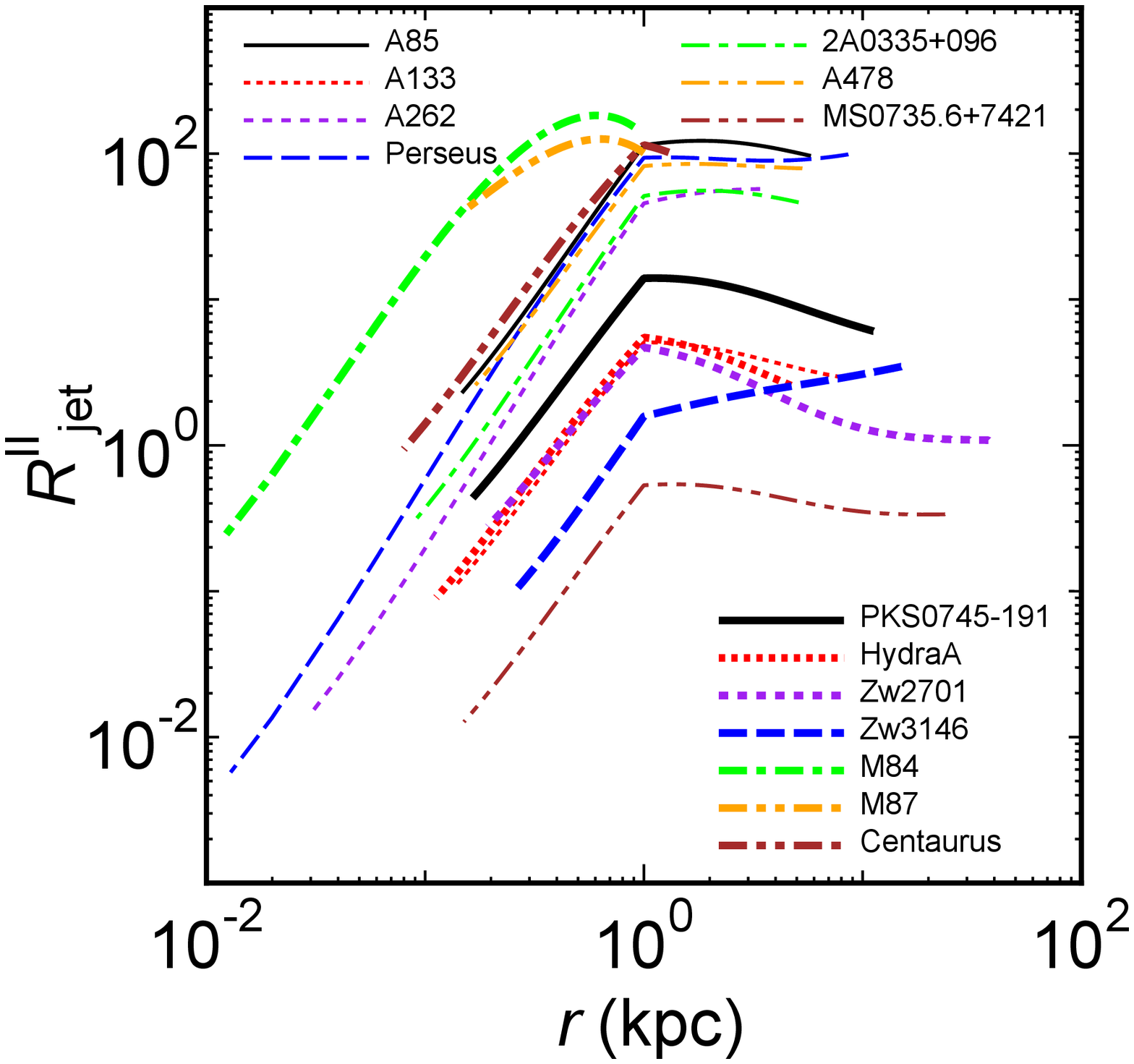}
        \end{center}
      \end{minipage}
      \begin{minipage}{0.5\hsize}
        \begin{center}
          \includegraphics[width=84mm]{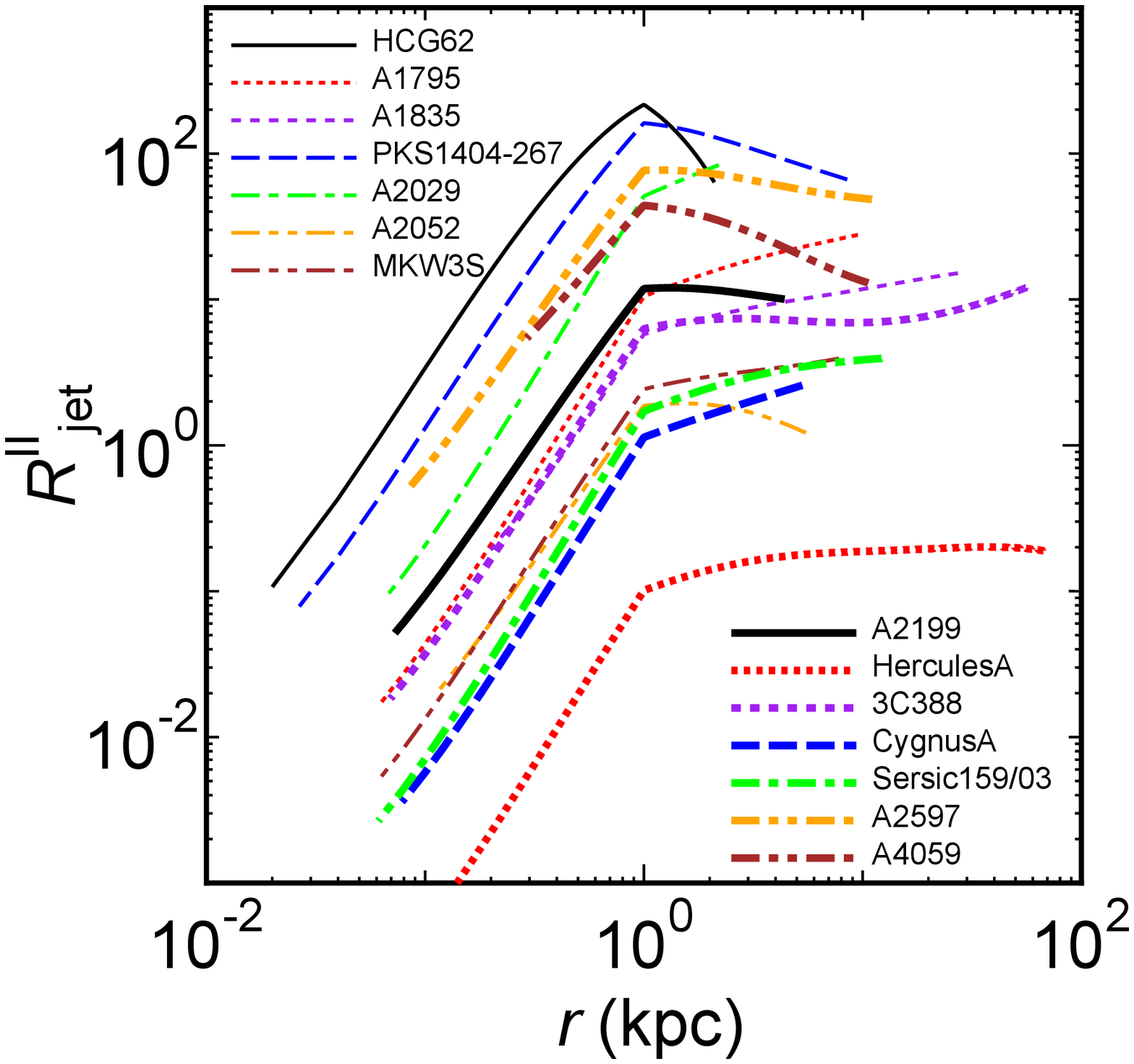}
        \end{center}
      \end{minipage}
    \end{tabular}
  \end{center}
    \caption{Most probable profiles of ${\cal R}^{\rm II}_{\rm jet}$ for
the FR\,II low-temperature model.}  \label{fig:RII}
\end{figure*}

In this subsection, we provide results for the jet momentum-driven
expansion of the cocoon (section~\ref{sec:FRIImodel}).  We show the
profiles of ${\cal R}^{\rm II}_{\rm jet}$ for the low-temperature model
in figure~\ref{fig:RII}. Most of them have a bend at $r\sim 1$~kpc,
where the index of $r_{\rm HS}$ changes (equation~\ref{eq:rHS}). For
larger radii, most galaxies have almost constant values of ${\cal
R}^{\rm II}_{\rm jet}$. This suggests that the condition whether the jet
can break through the central region or not depends on the values of
${\cal R}^{\rm II}_{\rm jet}$ at $r\gtrsim 1$~kpc as long as the ambient
medium is in hydrostatic equilibrium.

\begin{figure*}
  \begin{center}
    \begin{tabular}{c}
      \begin{minipage}{0.5\hsize}
        \begin{center}
          \includegraphics[width=84mm]{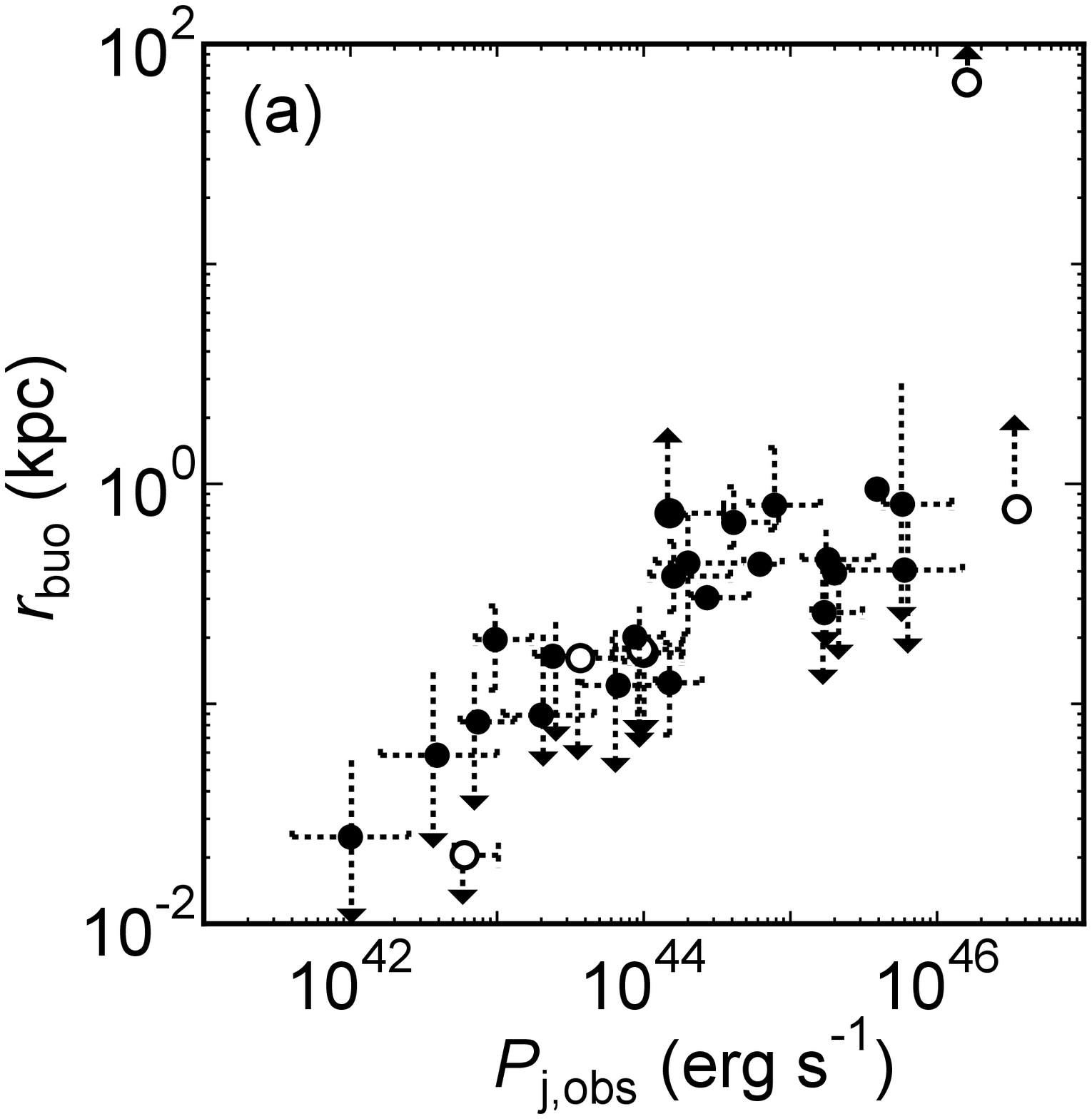}
        \end{center}
      \end{minipage}
      \begin{minipage}{0.5\hsize}
        \begin{center}
          \includegraphics[width=84mm]{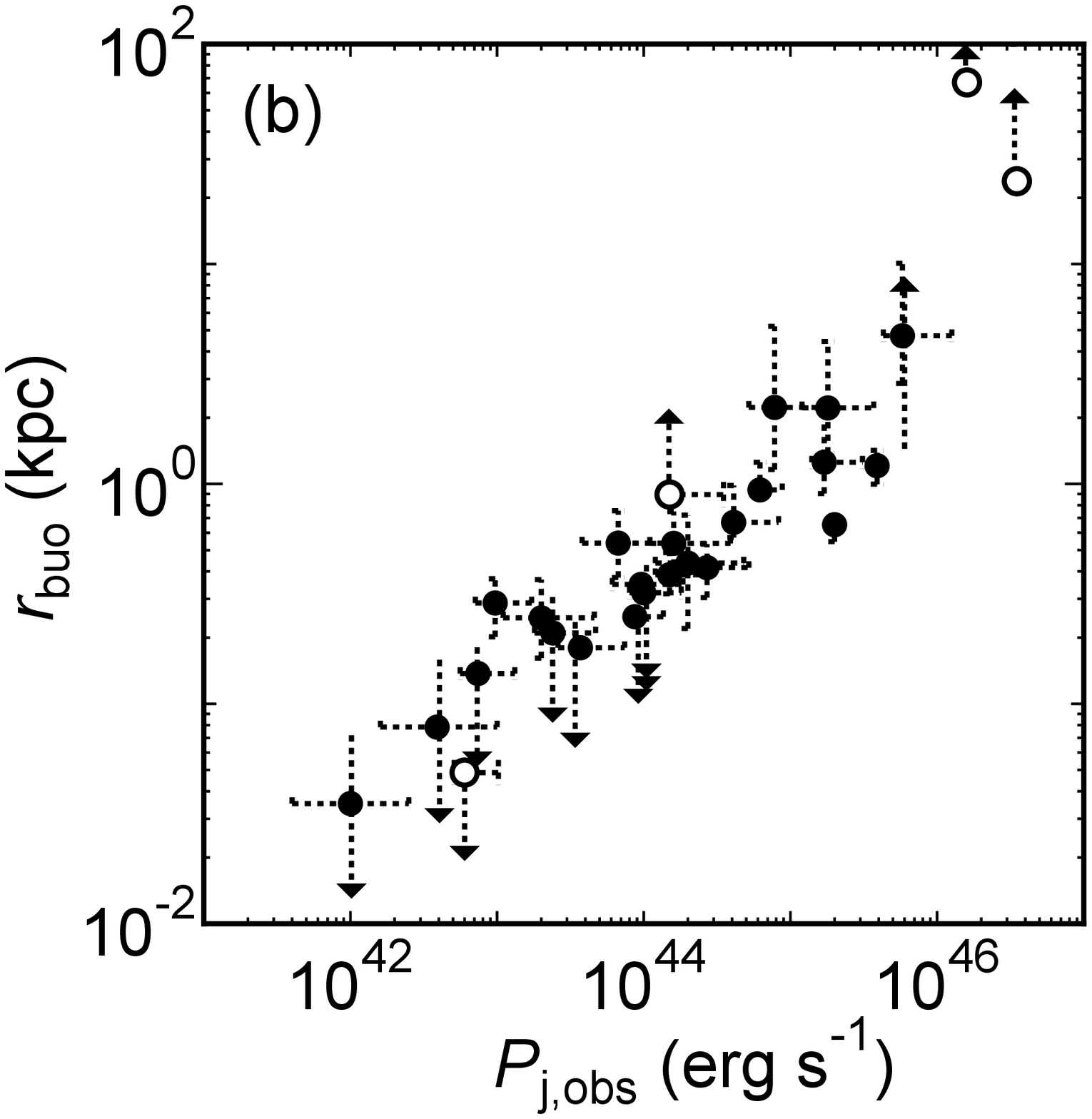}
        \end{center}
      \end{minipage}
    \end{tabular}
  \end{center}
\caption{(a) Relation between the jet power, $P_{\rm j,obs}$, and
$r_{\rm buo}$ for the FR\,II low-temperature model with $P_{\rm
j}=P_{\rm j,obs}$. Filled circles show the most probable values. Arrows
indicate that lower or upper limits are not determined. Open circles
show upper or lower limits. If $r_{\rm buo}>r_{\rm in}$ for all
realizations ($r_{\rm buo}=+\infty$ in table~\ref{tab:jetII}), we show
$r_{\rm in}$ as the lower limit. (b) Same as (a) but for the FR~II
isentropic model.}  \label{fig:PrbuoII}
\end{figure*}

\begin{figure*}
  \begin{center}
   \includegraphics[width=84mm]{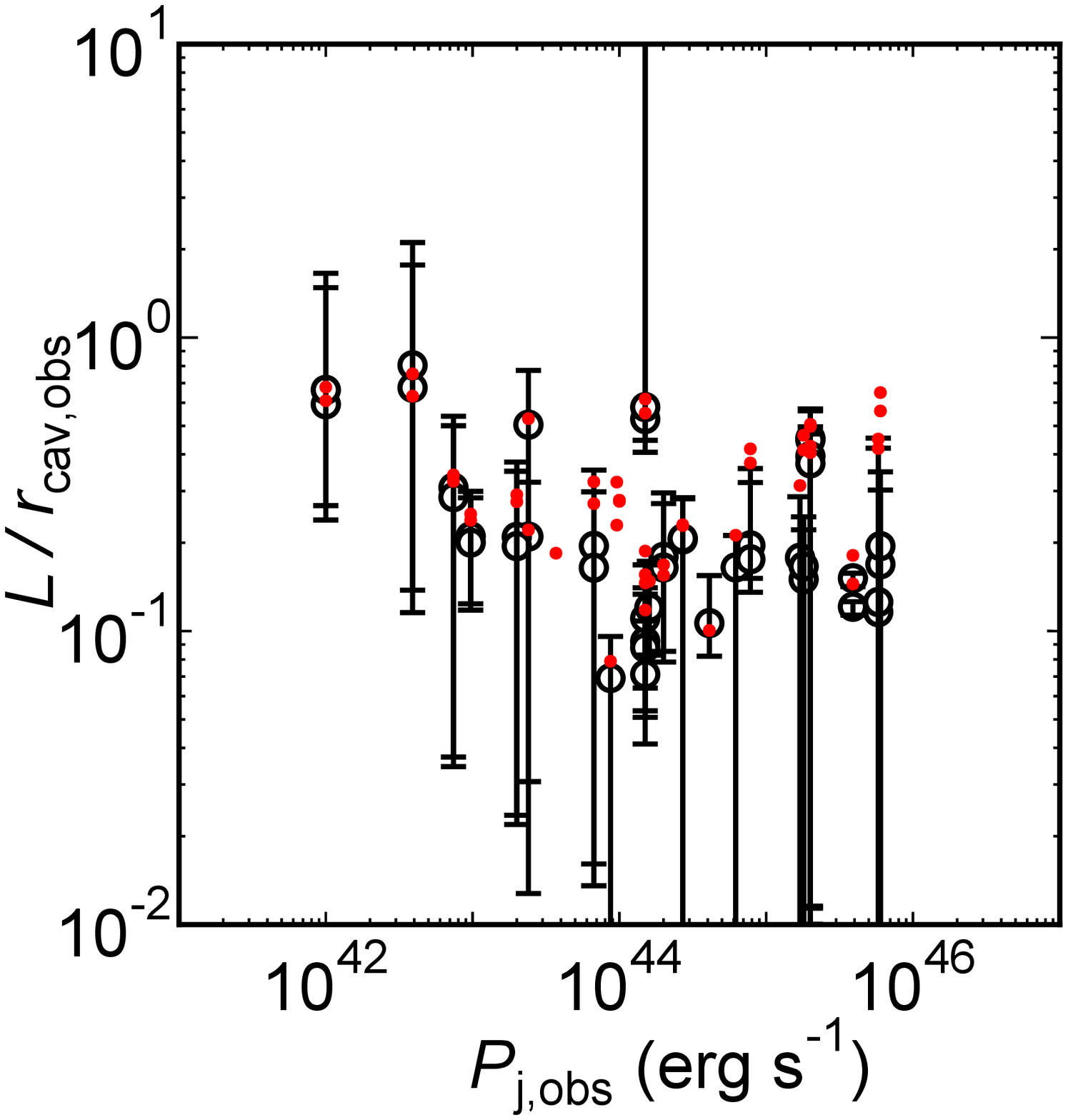}
  \end{center}
\caption{The ratio $L/r_{\rm cav,obs}$ for galaxies for which $r_{\rm
buo}$ can be determined. Open circles are for the FR\,II low-temperature
model and filled circles are for the FR\,II isentropic model. Error
bars are drawn for the former; they are almost the same for the
latter. Some galaxies have multiple cavities, and thus they have
multiple values of $L/r_{\rm cav,obs}$. The galaxies for which only
lower or upper limits of $r_{\rm bub}$ have been obtained are not
included.}  \label{fig:rat_rbub2}
\end{figure*}

We show the buoyancy radius $r_{\rm buo}$ within which ${\cal R}^{\rm
II}_{\rm jet}<1$ (figure~\ref{fig:PrbuoII}a) for our sample galaxies
(see also table~\ref{tab:jetII}). While $r_{\rm buo}$ is the increasing
function of $P_{\rm j,obs}$, the absolute values are smaller than those
in figure~\ref{fig:PrbuoI}a except for these with $r_{\rm buo}>r_{\rm
in}$. This is because ${\cal R}^{\rm II}_{\rm jet}$ tends to be larger
than ${\cal R}^{\rm I}_{\rm jet}$ at a given radius $r\lesssim 1$~kpc
(figures~\ref{fig:RI} and~\ref{fig:RII}, and tables~\ref{tab:jetI}
and~~\ref{tab:jetII}). Thus, if a cocoon expands following the FR\,II
type evolution, the breakout of the jets from the central region of the
galaxy is more difficult compared with the case when it follows the
FR\,I type evolution. This can mean that, if a cocoon starts to expand
following the FR\,II type evolution, its expansion switches to the FR\,I
type evolution around or within the buoyant radius calculated for the
FR\,II type evolution. We shall discuss this issue in
section~\ref{sec:FRI-II}. In figure~\ref{fig:rat_rbub2}, we show the
ratios of the predicted cavity size $L$ to the observed one $r_{\rm
cav,obs}$. The values of $L/r_{\rm cav,obs} (\sim 0.2)$ are generally
smaller than those in figure~\ref{fig:rat_rbub}, which indicates that
the predicted cavity sizes are less consistent with the
observations. This may also indicate that the FR\,I type evolution is
preferable to the FR\,II type evolution in clusters.

If the jet power is given by the Bondi power ($P_{\rm j}=P_{\rm
B}$), the condition of jet breakout is given by
\begin{equation}
\label{eq:RBII}
 {\cal R}^{\rm II}_{\rm B}(r)
= \frac{2\rho(r)f_1^2 c_{\rm s}(r)^2 c A_{\rm h}(r)}{P_{\rm B}}\:.
\end{equation}
Since $P_{\rm B}>P_{\rm j,obs}$, the buoyancy radii $r_{\rm buo}$ for
${\cal R}^{\rm II}_{\rm B}$ are larger than those for ${\cal R}^{\rm
II}_{\rm jet}$ (table~\ref{tab:jetII}), which makes the breakout
easier. However, $r_{\rm buo}$ for ${\cal R}^{\rm II}_{\rm B}$ are
smaller than those for ${\cal R}^{\rm I}_{\rm B}$ (tables~\ref{tab:jetI}
and~\ref{tab:jetII}).

\begin{figure*}
  \begin{center}
    \begin{tabular}{c}
      \begin{minipage}{0.5\hsize}
        \begin{center}
          \includegraphics[width=84mm]{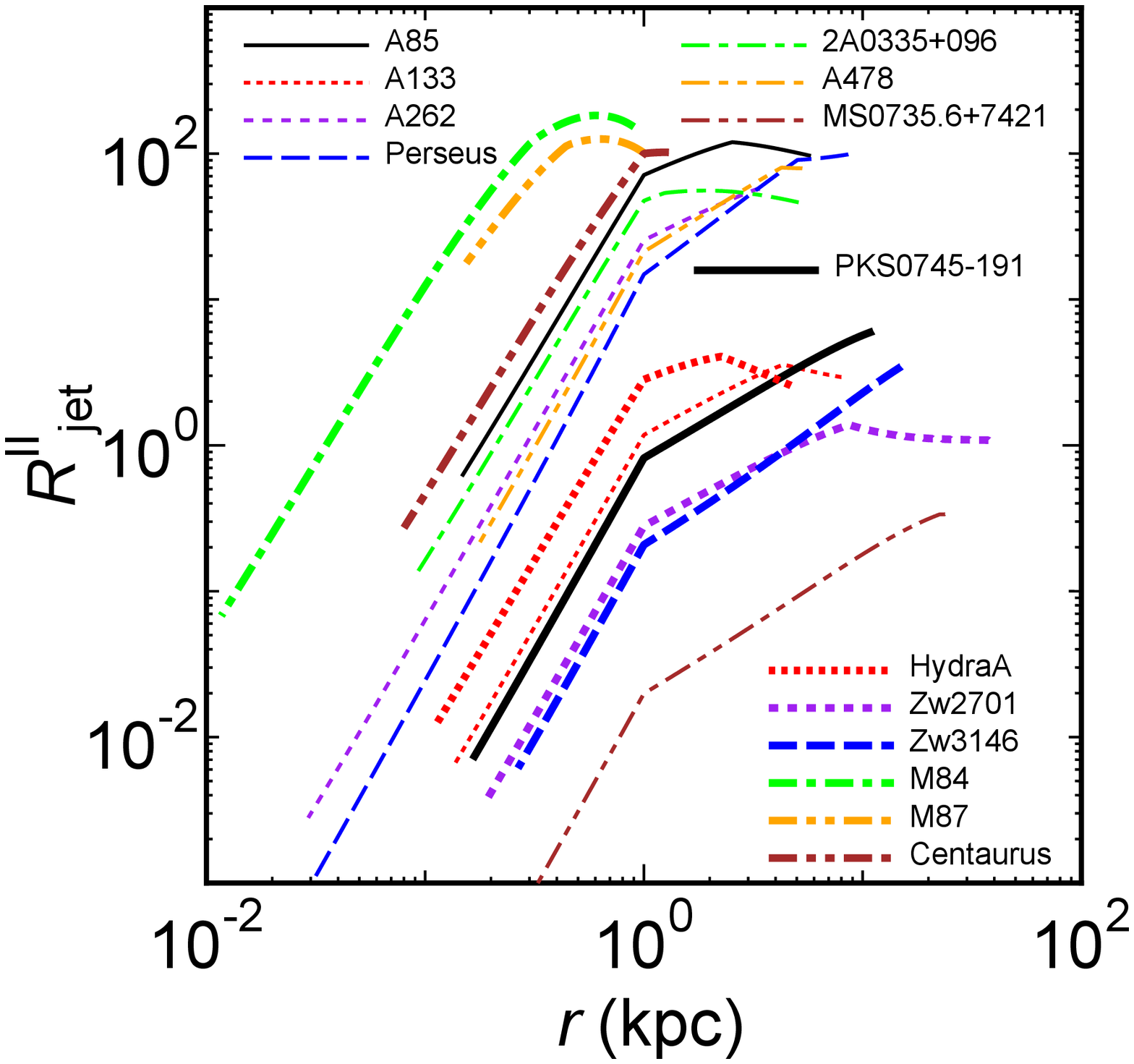}
        \end{center}
      \end{minipage}
      \begin{minipage}{0.5\hsize}
        \begin{center}
          \includegraphics[width=84mm]{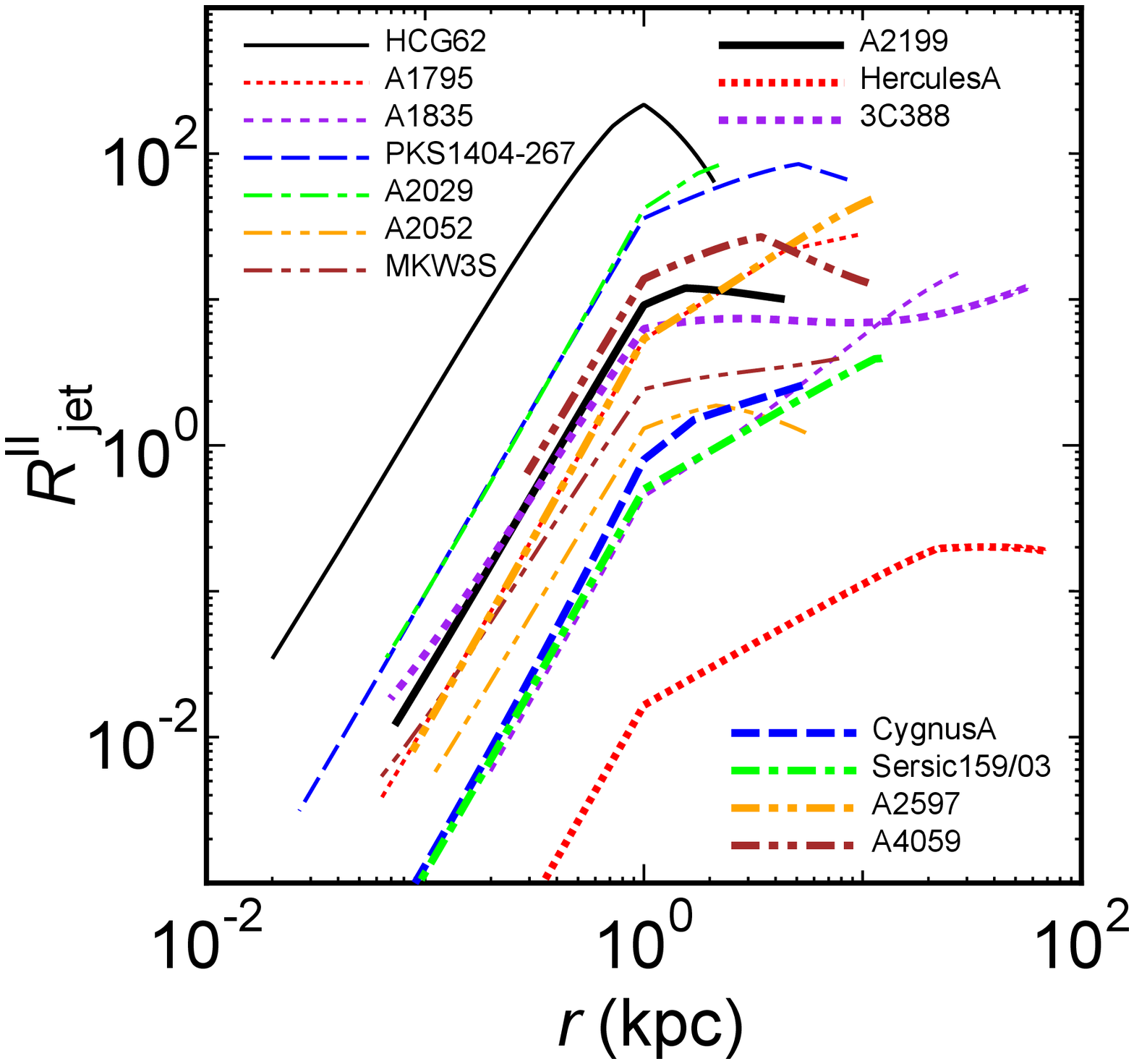}
        \end{center}
      \end{minipage}
    \end{tabular}
  \end{center}
    \caption{Most probable profiles of ${\cal R}^{\rm II}_{\rm jet}$ for
   the FR\,II isentropic model.}  \label{fig:ReII}
\end{figure*}

Figure~\ref{fig:ReII} shows the profiles of ${\cal R}^{\rm II}_{\rm
jet}$ for the isentropic model. Some galaxies exhibit a bend at
$r=1$~kpc, which corresponds to the index change of the size of the hot
spot (equation~\ref{eq:rHS}). Some galaxies (e.g. A85) show another bend
at the radius $r_{\rm s}$ where $t_{\rm cool}/t_{\rm ff}=10$ (see
figure~\ref{fig:nTe}). Outside this radius, the profile is the same as
that in the low-temperature model (figure~\ref{fig:RII}). Some galaxies
in figure~\ref{fig:ReII} (e.g. PKS~0745--191) have monotonically,
outwardly increasing profiles of ${\cal R}^{\rm II}_{\rm jet}$ for
$r<r_{\rm in}$ because $t_{\rm cool}/t_{\rm ff}(r_{\rm
in})<10$. However, it is unlikely that they continue to increase
endlessly for $r>r_{\rm in}$, because the hot gas is expected to be
isentropic only in the innermost region of galaxies, and because the
results of the low-temperature model (figure~\ref{fig:RII}) indicate
that ${\cal R}^{\rm II}_{\rm jet}$ does not change much at $r>1$~kpc.
Compared with figure~\ref{fig:RII}, the isentropic model gives smaller
${\cal R}^{\rm II}_{\rm jet}$ in the inner region
(figure~\ref{fig:ReII}) because of the lower density of the ambient gas
(figure~\ref{fig:nTe}). The difference of ${\cal R}^{\rm II}_{\rm jet}$
(equation~\ref{eq:RjetII}) between the low-temperature and isentropic
models is more significant than with ${\cal R}^{\rm I}_{\rm jet}$
(equation~\ref{eq:RjetI}), because the difference in the density
profiles of the hot gas between the low-temperature and isentropic
models is more significant than that in the pressure profiles.

Figure~\ref{fig:PrbuoII}b shows the relation between $P_{\rm j,obs}$ and
$r_{\rm buo}$ for the isentropic model.  Although the buoyant radii
$r_{\rm buo}$ are sightly larger than those shown in
figure~\ref{fig:PrbuoII}a for many of the galaxies, the difference is
small (see also table~\ref{tab:jetII}).  Therefore, the isentropic model
does not allow for the smooth penetration of the FR\,II type jets in the
central dense regions of those galaxies, either (see also
figure~\ref{fig:rat_rbub2}). Since real gas profiles are expected to lie
between the low-temperature and the isentropic models, we conclude that
the FR\,II type jets in most galaxies cannot break out of the central
region. This result may be closely related to the observations that the
morphology of $\sim 1$~kpc-scale low power compact radio sources tends
to be irregular (e.g. \cite{kun05a}).

\begin{figure*}
  \begin{center}
   \includegraphics[width=84mm]{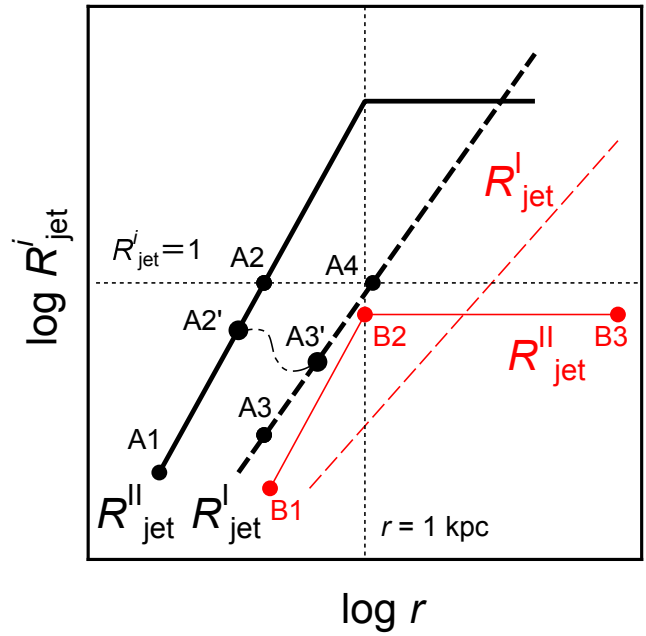}
  \end{center}
\caption{Schematic figure for the relation between ${\cal R}^{\rm
	 I}_{\rm jet}$ and $r$ (dashed lines) and ${\cal R}^{\rm
	 II}_{\rm jet}$ and $r$ (solid lines). Thick lines (smaller
	 $P_{\rm j,obs}$) and thin
	 lines (larger $P_{\rm j,obs}$) show the
	 cases where the cocoon finally becomes FR\,I and FR\,II,
	 respectively. The buoyant radii $r_{\rm buo}$ are the intersections of these
	 lines with ${\cal R}^i_{\rm jet}=1$.}  \label{fig:FR}
\end{figure*}

\begin{figure*}
  \begin{center}
    \begin{tabular}{c}
      \begin{minipage}{0.5\hsize}
        \begin{center}
          \includegraphics[width=84mm]{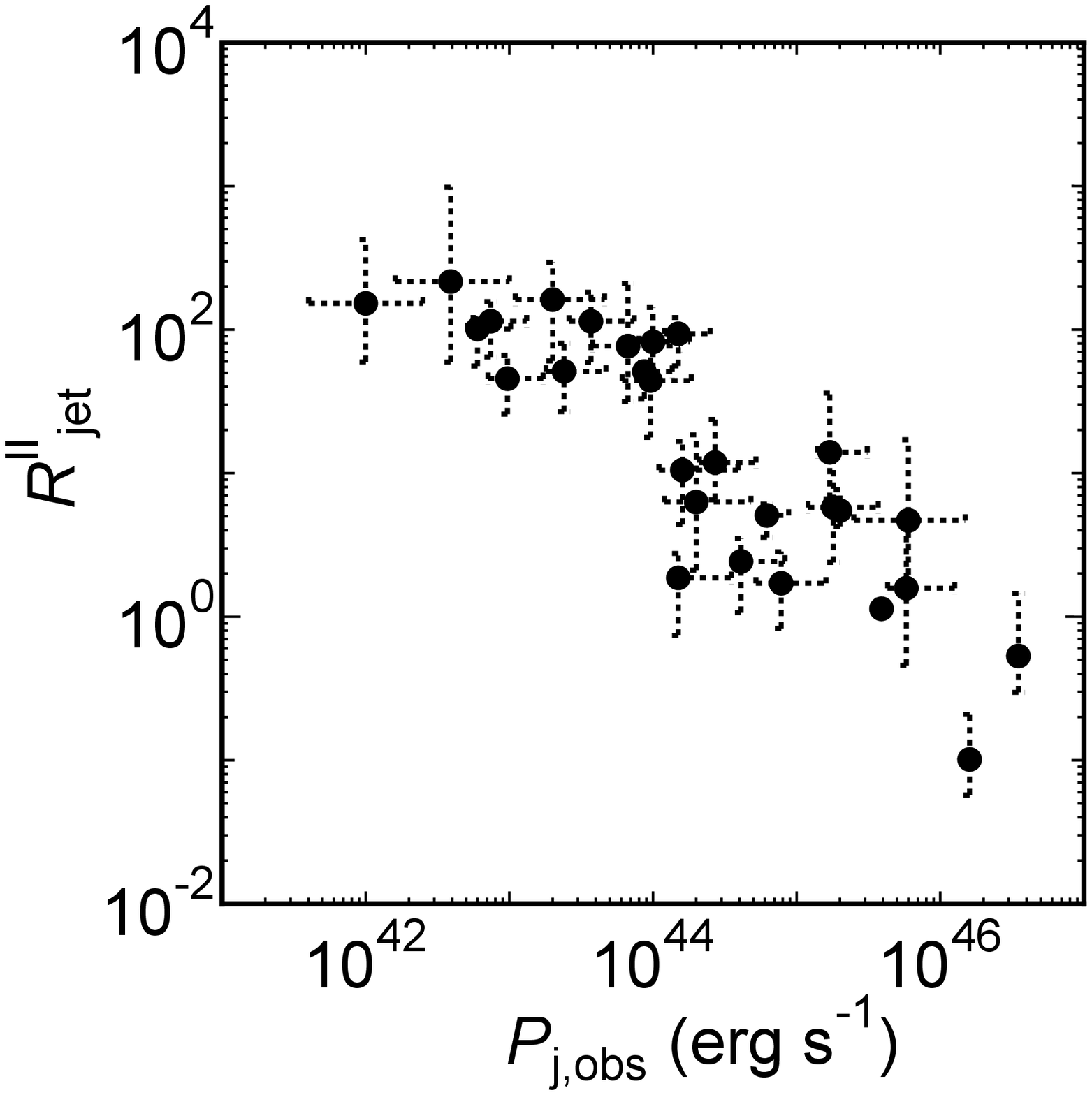}
\caption{Relation between jet power $P_{\rm j,obs}$ and ${\cal R}^{\rm
	 II}_{\rm jet}(r=\rm 1 kpc)$ for the FR\,II low-temperature model.}\label{fig:PRI}
        \end{center}
      \end{minipage}
      \begin{minipage}{0.5\hsize}
        \begin{center}
          \includegraphics[width=84mm]{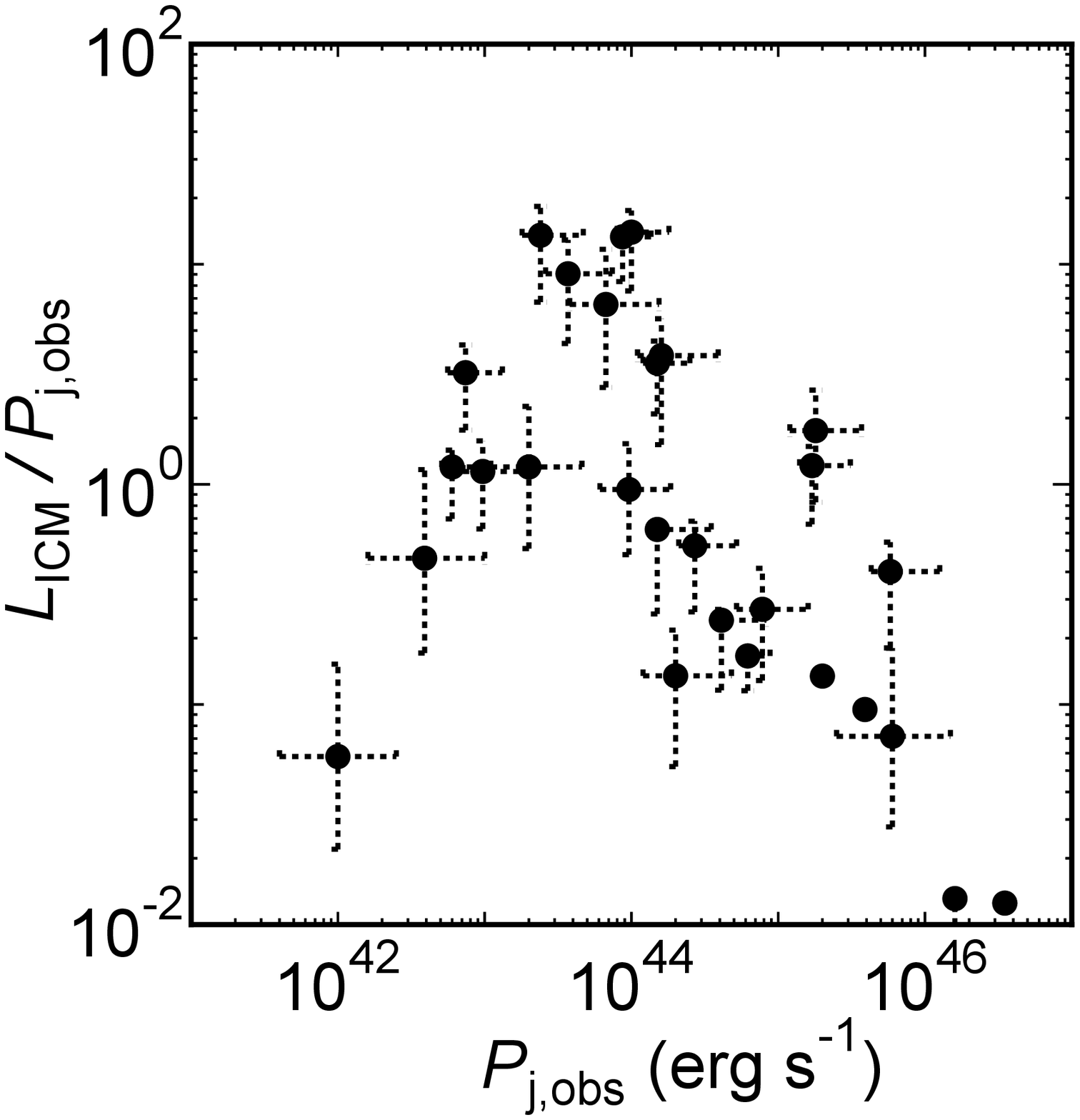}
\caption{Relation between $P_{\rm j,obs}$ and $L_{\rm ICM}$.}  \label{fig:PLICM}
        \end{center}
      \end{minipage}
    \end{tabular}
 \end{center}
\end{figure*}

\section{Discussion}
\label{sec:dis}

\subsection{Summary of the results}
\label{sec:sol}

We find that, with the jet power estimated from the observations of
X-ray cavities ($P_{\rm j,obs}$), jets have difficulty to progress
through the dense central region of the host galaxies faster than the
buoyant velocities of the cocoons. This appears to be the case for a
significant fraction of our sample galaxies. Such a statement holds even
if we assume the FR\,I type evolution, which makes jet propagation
easier than the FR\,II type evolution. Therefore, the cocoons should
have been plausibly broken by the buoyant force before they have grown
up. We estimate the sizes of the cavities originated from the broken
cocoons, and find that they are consistent with the observations within
a factor of two for the FR\,I type evolution. At the same time, the
discrepancy is larger for the FR\,II type evolution. The resulting
``easiness'' of jet propagation in the FR\,I type evolution may explain
why FR\,I sources are more frequent over FR\,II sources in clusters
\citep{pre88a,mil02a}.

\subsection{Dichotomy between FR\,I and II}
\label{sec:FRI-II}

The advance velocity of a newborn jet, $v_{\rm h}$, can be very large
($v_{\rm h}\sim c$; \cite{kaw08a}). Thus, the cocoon may initially
expand by the jet momentum (FR\,II type evolution) even for galaxies
with smaller $P_{\rm j,obs}$ (from point A1 to A2 in
figure~\ref{fig:FR}). However, the results in section~\ref{sec:FRII}
show that this type of evolution does not last for a long time for
galaxies with smaller $P_{\rm j,obs}$ because of small buoyant radii,
$r_{\rm buo}$ (point A2). Since the advance velocity of a jet decreases
to the sound velocity at $r\lesssim r_{\rm buo}$, the jet momentum can
be insufficient to inflate the cocoon further. However, the energy
injection by the jet is still substantial at this point because the
buoyant radii, $r_{\rm buo}$, for the FR\,I evolution are larger than
those for the FR\,II evolution, for most of our samples
(sections~\ref{sec:FRI} and~\ref{sec:FRII}, see also
tables~\ref{tab:jetI} and~\ref{tab:jetII}). That is, the location of the
cocoon jumps from A2 to A3 and then moves toward A4 along the line of
the FR\,I type evolution (thick dashed line in figure~\ref{fig:FR}). Of
course, the actual transition could be smooth. Since ${\cal R}^{\rm
I}_{\rm jet}\sim f_1 c_{\rm s}/v_{\rm I}$ (equations~\ref{eq:FRI}
and~\ref{eq:RjetI}) and ${\cal R}^{\rm II}_{\rm jet}=(f_1 c_{\rm
s}/v_{\rm h})^2$ (equations~\ref{eq:FRII} and~\ref{eq:RjetII}), the
cocoon may track from A2$'$ and A3$'$, where ${\cal R}^i_{\rm jet}\sim
f_1^2$ or $f_1$, although the detailed discussion on the track is beyond
the scope of the paper. The expansion of the cocoon finally stops at
point A4. This can explain why most of our objects are known as FR\,I
sources.

The profiles of ${\cal R}^{\rm II}_{\rm jet}$ bent at $r=1$~kpc
(low-temperature and isentropic models) and $r=r_{\rm s}$ (isentropic
model) show that some cocoons with relatively large $P_{\rm j,obs}$ can
continue to evolve as FR~II sources, if they start evolving as FR\,IIs
(thin solid line in figure~\ref{fig:FR}). This happens because the
profiles of ${\cal R}^{\rm II}_{\rm jet}$ for the low-temperature model
become almost flat at $r>1$~kpc (figure~\ref{fig:RII}) and those for the
isentropic model appear to be the same as those for the low-temperature
model at $r>r_{\rm s}$ (figure~\ref{fig:ReII}). Thus, if ${\cal R}^{\rm
II}_{\rm jet}$ of the low-temperature model is less than around unity at
$r\sim 1$~kpc, it also remains less than around unity at $r>1$~kpc, for
both the low-temperature and isentropic models. If this happens, the
cocoon can grow as an FR~II (point B1 $\rightarrow$ B2 $\rightarrow$ B3
in figure~\ref{fig:FR}). Even if a cocoon initially tracks the path of
${\cal R}^{\rm I}_{\rm jet}$ (thin dashed line in figure~\ref{fig:FR}),
it may transfer to the path of ${\cal R}^{\rm II}_{\rm jet}$ (thin solid
line in figure~\ref{fig:FR}) at large radii where ${\cal R}^{\rm
II}_{\rm jet} \ll {\cal R}^{\rm I}_{\rm jet}$.

Figure~\ref{fig:PRI} shows the relation between ${\cal R}^{\rm II}_{\rm
jet} (r=1\rm\: kpc)$ for the low-temperature model and $P_{\rm j,obs}$.
We observe that ${\cal R}^{\rm II}_{\rm jet}$ decreases
monotonically. This relation can be explained by
equation~(\ref{eq:RjetII}), and by the fact that $P_{\rm j,obs}$ varies
in a wider range (figure~\ref{fig:PrbuoII}) than the profiles of the hot
gas (figure~\ref{fig:nT}) among our sample galaxies. We are only
interested in the normalization of the correlation and find that ${\cal
R}^{\rm II}_{\rm jet}\lesssim 1$ at $P_{\rm j,obs}\gtrsim 10^{46}\rm\:
erg\: s^{-1}$. Although ${\cal R}^{\rm II}_{\rm jet}$ of some galaxies
with $10^{44}\lesssim P_{\rm j,obs}\lesssim 10^{46}\rm\: erg\: s^{-1}$
is slightly larger than unity (figure~\ref{fig:PRI}), it can be less
than unity if $P_{\rm jet}$ is underestimated or $\rho$ is overestimated
by only a factor of few. In this case, a significant fraction of
galaxies with $P_{\rm jet}\gtrsim 10^{44}\rm\: erg\: s^{-1}$ could have
evolved as FR\,II type and created cavities with $r_{\rm buo}>10$~kpc.

In summary, some of the jets with $P_{\rm j,obs}\gtrsim
10^{44}\rm\: erg\: s^{-1}$ could extend well beyond the galaxies without
being much affected by the friction with the surrounding gas and could
turn to be FR\,II sources. Cygnus~A, which is known as FR~II, is
included in this class of galaxies. Some FR~Is with $P_{\rm
j,obs}\gtrsim 10^{44}\rm\: erg\: s^{-1}$ might have been FR\,IIs until
fairly recently. On the other hand, ${\cal R}^{\rm II}_{\rm jet}$ is
generally much larger than unity at $P_{\rm j,obs}\lesssim 10^{44}\rm\:
erg\: s^{-1}$ (figure~\ref{fig:PRI}), and ${\cal R}^{\rm II}_{\rm
jet}<1$ may not be achieved even considering uncertainties. Thus, jets
with $P_{\rm j,obs}\lesssim 10^{44}\rm\: erg\: s^{-1}$ could be strongly
decelerated by the ambient medium and might become FR~Is soon after
their launch. Therefore, $P_{\rm j,obs}\gtrsim 10^{44}\rm\: erg\:
s^{-1}$ can be a necessary condition to create FR\,IIs in BCGs. FR\,II
sources appear rare in clusters perhaps because the AGN there are rarely
experiencing activity strong enough to break out the central
region. \citet{led96a} indicated that the FR~I/II division is a
function of an optical luminosity as well as of a radio luminosity. It
would be interesting to study the relation between $P_{\rm j,obs}$ and
optical and radio luminosities for a larger sample.

\subsection{ICM heating}

The threshold $P_{\rm j,obs}\sim 10^{44}\rm\: erg\: s^{-1}$ may have
another interesting implication. Figure~\ref{fig:PLICM} shows the
relation between $P_{\rm j,obs}$ and the ratio $P_{\rm j,obs}/L_{\rm
ICM}$, where $L_{\rm ICM}$ is the X-ray luminosity of the intracluster
medium (ICM), inside the cooling radius of the host cluster, which is
offset to be consistent with the spectra, $L_{\rm ICM}=L_{\rm Xc}-L_{\rm
cool}$. We have adopted the luminosities derived by \citet{raf06a};
$L_{\rm Xc}$ is the X-ray luminosity for which the gas cooling time is
less than the look-back time for $z=1$ ($7.7\times 10^9$~yr), and
$L_{\rm cool}$ is the associated luminosity of the gas cooling to low
temperatures, derived from the X-ray spectrum. We assume $L_{\rm
cool}=0$ for A1835 because $L_{\rm cool}$ could not be detected
\citep{raf06a}. This assumption will not affect the results strongly,
because $L_{\rm cool}\ll L_{\rm Xc}$ for most of the other clusters.

Luminosities $L_{\rm ICM}$ are shown in table~\ref{tab:obs}. In
figure~\ref{fig:PLICM}, $L_{\rm ICM}/P_{\rm j,obs}\gtrsim 1$ for most
galaxies with $P_{\rm j,obs}\lesssim 10^{44}\rm\: erg\: s^{-1}$, while
$L_{\rm ICM}/P_{\rm j,obs}\lesssim 1$ in general for $P_{\rm
j,obs}\gtrsim 10^{44}\rm\: erg\: s^{-1}$. Note that, although the
distribution appears to peak at $P_{\rm j,obs}\sim 3\times 10^{44}\rm\:
erg\: s^{-1}$ (figure~\ref{fig:PLICM}), we avoid discussing this,
because the peak is blurred if we omit the object with the lowest jet
power (M84; $P_{\rm j,obs} = 1\times 10^{42}\rm\: erg\: s^{-1}$). If the
jet heats up the cool core of the host clusters, the radiative cooling
of the hot gas is compensated by the jet power and $L_{\rm ICM}/P_{\rm
j,obs}\sim 1$ is expected. Since figure~\ref{fig:PRI} shows that
the jets with $P_{\rm j,obs}\gtrsim 10^{46}\rm\: erg\: s^{-1}$ can
evidently break out of the central region (${\cal R}^{\rm II}_{\rm
jet}<1$; see point B3 in figure~\ref{fig:FR}), results shown in
figure~\ref{fig:PLICM} indicate that the jets with $P_{\rm j,obs}\gtrsim
10^{46}\rm\: erg\: s^{-1}$ have broken out of the most dense region in
the center, where the radiative cooling is most efficient, and thus the
jet power is not used effectively to compensate for the radiative
cooling.

For jets with $P_{\rm j,obs}\lesssim 10^{44}\rm\: erg\: s^{-1}$ (mostly
$L_{\rm ICM}/P_{\rm j,obs}> 1$), the energy created around AGN can be
conveyed to the ICM in some hidden form, such as a thermal conduction
and cosmic rays (e.g., \cite{rus02a,guo08a,fuj12a,fuj13c}).  The jets
with $10^{44}\lesssim P_{\rm j,obs}\lesssim 10^{46}\rm\: erg\: s^{-1}$
can represent the mix of the two populations --- a kind of a gray
zone. Alternatively, figure~\ref{fig:PLICM} can indicate that the jet
power fluctuates and $L_{\rm ICM}/P_{\rm j,obs}\sim 1$ in a long-time
average.

\section{Conclusions}
\label{sec:con}

We have studied the evolution of AGN jets and associated cocoons in the
brightest elliptical galaxies (BCGs) in clusters. Using observational
data of the jet power estimated from the size of X-ray cavities and the
gas properties of the ambient hot gas, we have analyzed whether these
jets can propagate in the host galaxy with sufficiently large
velocities. For this purpose, we consider the balance between the
pressure inside the cocoon and the thermal pressure outside the cocoon
(FR\,I type evolution), and the balance between the momentum flux of the
jet and the ram pressure of the ambient gas (FR\,II type
evolution). Since the hot gas profiles in the innermost region of
galaxies are not known, we have extrapolated the observed profiles based
on two extreme models. In the low-temperature model, we assume that the
gas temperature reflects the virial temperature of the galaxy. In the
isentropic model, the entropy of the gas is constant due to the thermal
instabilities. The former and the latter give higher and lower central
densities, respectively. The actual density is expected to lie in
between.

We find that most jets with observed powers of $\lesssim 10^{44}\rm\:
erg\: s^{-1}$ have difficulty to penetrate the dense central regions of
their host galaxies with sufficiently high velocities to create large
cocoons.  If they start evolving as FR\,IIs, for which the evolution is
driven by the jet momentum, the velocity of the jet head quickly falls
below the sound speed of the ambient gas. Thus, the cocoons may change
into FR\,Is, for which the evolution is driven by the pressure inside
the cocoons. This may be the reason why FR\,I sources are common in
clusters. However, even for the FR\,I evolution, the expansion velocity
gradually decreases down to the buoyant velocity of the cocoons. This
indicates that the cocoons could be destroyed by the buoyancy
force. From the observed jet powers, we predict the size of cavities,
which are the relics of the cocoons, and found that it is consistent
with the observed one within a factor of few, if the FR\,I type
evolution is realized.

Our results also indicate that some of the jets
with powers of $\gtrsim 10^{44}\rm\: erg\: s^{-1}$ are less affected by
the ambient medium and can serve as FR\,II sources. Most of the power
can be released outside of the most dense region at the galactic center
and thus may not efficiently compensate the cooling. On the other hand,
jets with powers of $\lesssim 10^{44}\rm\: erg\: s^{-1}$ are strongly
disturbed by the ambient medium and become FR\,I sources.

\begin{ack}
This work was supported by the
International Joint Research Promotion Program and Challenge Support
Program by Osaka University, and by KAKENHI No. 15K05080 (Y.F). NK
acknowledges the financial support of Grant-in-Aid for Young Scientists
(B:25800099). I.S. acknowledges partial support from the NSF and
STScI. STScI is operated by AURA, Inc., under NASA contract NAS 5-26555.
\end{ack}

\appendix 
\section{Low-temperature model}
\label{sec:app_low}

In this model, we make two assumptions. First, we assume that the hot
gas outside the Bondi radius is in a nearly hydrostatic equilibrium:
\begin{equation}
\label{eq:hydroeq}
 -\frac{dp}{dr} = \rho g\:,
\end{equation}
where $p(r)$ is the thermal gas pressure, and $g(r)$ is the
gravitational acceleration. Second, we assume that the gas temperature
near the SMBH (i.e., at $r\sim r_{\rm B}$) reflects the velocity
dispersion $\sigma$ or the virial temperature $T_{\rm gal,vir}$ of the
host galaxy:
\begin{equation}
\label{eq:T0}
 T_0 = \zeta^{-1} \frac{\mu m_{\rm p} \sigma^2}{k}\sim T_{\rm gal,vir}\:,
\end{equation}
where $k$ is the Boltzmann constant, and $\zeta$ is the constant of
order of unity. Following \citet{mat01a}, we adopt $\zeta=0.5$ for
massive elliptical galaxies, including BCGs. The second assumption is
based on the first one, because the left-hand side of
equation~(\ref{eq:hydroeq}) can be approximated by $-dp/dr\sim p/r=n k
T/r$, where $n$ is the number density of the gas, and the right-hand
side can be approximated by
\begin{equation}
 \rho g = \rho\frac{G M(<r)}{r^2}\sim n\frac{kT_{\rm gal,vir}}{r}\:,
\end{equation}
where $M(<r)$ is the gravitational mass within the radius $r$. The
second assumption (equation~\ref{eq:T0}), is generally consistent
with {\it ROSAT} X-ray observations \citep{mat01a}. 

Assuming that the temperature profile depends on the galaxy size, we
interpolate it between $r=r_{\rm B}$ and $r_{\rm in}$ as
\begin{equation}
\label{eq:T}
 T(r) = T_0 + (T_{\rm in} - T_0)\frac{\tanh(r/R_{\rm e})}{\tanh(r_{\rm
  in}/R_{\rm e})}\:,
\end{equation}
where $T_{\rm in}=T(r_{\rm in})$ and $R_{\rm e}$ is the effective radius
(half-light radius) of the galaxy. The resulting temperature profile
between $T_{\rm in}$ and $T_0$ nicely mimics the observed profiles
(e.g. \cite{chu03a}). In general, the temperature decreases toward the
galaxy center. Once we fix $T(r)$, the Bondi accretion radius, $r_{\rm
B}$, can be obtained numerically by solving the equation
\begin{equation}
\label{eq:rB}
 r_{\rm B} = \frac{2 G M_\bullet}{c_{\rm s}(T(r_{\rm B}))^2}
\end{equation}
for a given SMBH mass $M_\bullet$ \citep{bon52a}. If the angular
momentum of the gas can be ignored and the gas is adiabatic, the
accretion onto the SMBH can follow the Bondi accretion. Since the direct
application of the Bondi accretion model provides a grossly
oversimplified picture (e.g., \cite{sok06a,piz10a,mcn11a}), we consider
the Bondi accretion just as a reference. The Bondi accretion rate is
given by
\begin{equation}
\label{eq:dotMB}
 \dot{M}_{\rm B} = 4\pi\lambda_{\rm c} 
(G M_\bullet)^2 c_{\rm s,B}^{-3}\rho_{\rm B}
= \pi\lambda_{\rm c} c_{\rm s,B} \rho r_{\rm B}^2\:,
\end{equation}
where $\rho_{\rm B}=\rho(r_{\rm B})$ and $c_{\rm s,B}=c_{\rm s}(r_{\rm
B})$ are the density and the sound speed at the Bondi radius
\citep{bon52a}. The coefficient $\lambda_{\rm c}$ depends on the
adiabatic index of the accreting gas ($\gamma$), and we assume
$\gamma=5/3$ and $\lambda_{\rm c}=0.25$. 

The equation of the hydrostatic equilibrium (equation~\ref{eq:hydroeq})
can be written as
\begin{equation}
\label{eq:drhodr}
 \frac{d\rho}{dr} = -\frac{\rho}{T}\left(\frac{\mu m_{\rm p}}{k}g 
+ \frac{dT}{dr}\right) \:.
\end{equation}
In general, the first term on the right hand side dominates over the
second term. Since $T(r)$ has been determined by equation~(\ref{eq:T}),
$\rho(r)$ can be obtained by numerically integrating the
equation~(\ref{eq:drhodr}) and setting $\rho_{\rm in}=\rho(r_{\rm
in})$ and $g(r)$.  The electron number density is defined as $n_{\rm
e}=\rho/(1.13\: m_{\rm p})$.

The gravitational acceleration $g(r)$ is given by three components, i.e.,
$g=g_\bullet + g_{\rm gal} + g_{\rm cl}$, where $g_\bullet$ is the SMBH
contribution, $g_{\rm gal}$ is the galaxy contribution, and $g_{\rm cl}$
is the cluster contribution \citep{mat06a,guo14a}. The acceleration from
an SMBH is
\begin{equation}
\label{eq:gBH}
 g_\bullet(r) = \frac{G M_\bullet}{r^2}\:.
\end{equation}
The acceleration from a galaxy with the Hernquist profile \citep{her90a}
is
\begin{equation}
 g_{\rm gal}(r) = \frac{G M_{\rm gal}}{(r + r_{\rm H})^2}\:,
\end{equation}
where $M_{\rm gal}$ is the stellar mass of the galaxy, and $r_{\rm
H}=R_{\rm e}/1.815$. Although the Hernquist profile may not be a good
approximation for the outer part of BCGs (e.g. \cite{gra96a}), it does
not affect our results because we are mostly interested in the inner
part. The cluster acceleration for the NFW profile \citep{nav96a} is
\begin{equation}
 g_{\rm cl}(r) = \frac{G M_{\rm vir}}{r^2}
\frac{\log(1+y)-y/(1+y)}
{\log(1+c_{\rm vir})-c_{\rm vir}/(1+c_{\rm vir})}\:,
\end{equation}
where $y=c_{\rm vir} r/r_{\rm vir}$, and $c_{\rm vir}$ is the
concentration parameter. The cluster virial radius, $r_{\rm vir}$,
is defined as the radius at which the average cluster density is
$\Delta(z)$ times the critical density $\rho_{\rm crit}(z)$ at the
cluster redshift $z$:
\begin{equation}
\label{eq:rvir}
 r_{\rm vir}=\left(\frac{3 M_{\rm vir}}{4\pi 
\Delta(z)\rho_{\rm crit}(z)}\right)^{1/3}\:.
\end{equation}
For $\Delta(z)$, we use the fitting formula of \citet{bry98a}:
$\Delta=18\pi^2 + 82x -39x^2$, where $x=\Omega_{\rm m}(z)-1$.

To summarize, the required parameters are $z$, $M_\bullet$, $M_{\rm
gal}$, $R_{\rm e}$, $\sigma$, $c_{\rm vir}$, and $M_{\rm vir}$, and the
boundary conditions $r_{\rm in}$, $\rho_{\rm in}$, and $T_{\rm in}$
in order to derive $r_{\rm B}$ and $\dot{M}_{\rm B}$. First, $T(r)$ is
determined by equations~(\ref{eq:T0}) and~(\ref{eq:T}) for given
$\sigma$, $T_{\rm in}$, $R_{\rm e}$, and $r_{\rm in}$. The Bondi
radius $r_{\rm B}$ is obtained by solving equation~(\ref{eq:rB}) for
given $T(r)$ and $M_\bullet$. Then, $\rho_{\rm B}=\rho(r_{\rm B})$ is
estimated by integrating equation~(\ref{eq:drhodr}) from $r=r_{\rm in}$
to $r_{\rm B}$ using equations~(\ref{eq:gBH})--(\ref{eq:rvir}) for given
$T(r)$, $M_\bullet$, $M_{\rm gal}$, $R_{\rm e}$, $M_{\rm vir}$, $c_{\rm
vir}$ and $z$. Finally, the Bondi accretion rate is given by
equation~(\ref{eq:dotMB}).

\section{Correction of the cavity sizes}
\label{sec:app_cavity}

We need to consider two corrections associated with the evolution in
phases B and D (figure~\ref{fig:cocoon}), when we compare the predicted
cavity size $L$ with the observation $r_{\rm cav,obs}$. First, the size
of a cavity increases as it rises via buoyancy keeping pressure balance
with the ambient gas (phase C$\rightarrow$D or C'$\rightarrow$D' in
figure~\ref{fig:cocoon}). If the cavity is adiabatic, the size changes
with the distance from the galactic center as $L(r)\sim L_{\rm ci}
[p(r)/p(r_{\rm buo})]^{-1/(3\gamma_{\rm c})}=L_{\rm ci} [p(r_{\rm
buo})/p(r)]^{1/4}$, where $p(r)$ is the pressure of the ambient medium.

Second, $P_{\rm j,obs}$ in equations~(\ref{eq:RjetI})
and~(\ref{eq:RjetII}) does not involve the effect of the buoyant rise in
phases B and D. This leads to an underestimate of jet power. Assuming
that the ambient pressure and the volume of a given cavity at phases B
and D are $p_{\rm B}$ ($=p(r_{\rm buo})$), $p_{\rm D}$, $V_{\rm B}$
($=4\pi r_{\rm buo}^3/3$), and $V_{\rm D}$, respectively, we obtain
$V_{\rm B} = V_{\rm D} (p_{\rm D}/p_{\rm B})^{1/\gamma_{\rm c}}$ or
$p_{\rm B} V_{\rm B} = p_{\rm D} V_{\rm D} (p_{\rm B}/p_{\rm
D})^{1-1/\gamma_{\rm c}} = p_{\rm D} V_{\rm D} (p_{\rm B}/p_{\rm
D})^{1/4}$. Thus, the enthalpy (equation~\ref{eq:Ecav}) estimated in
phase D is underestimated by a factor of $(p_{\rm B}/p_{\rm D})^{1/4}$,
compared with phase B. Since the observed jet power $P_{\rm j,obs}$ is
proportional to the enthalpy, it is underestimated by the same
factor. If the actual $P_{\rm j,obs}$ is larger than that we adopted,
${\cal R}^{\rm I}_{\rm jet}$ should be smaller by a factor of $(p_{\rm
B}/p_{\rm D})^{1/4}$ for a given radius $r$
(equation~\ref{eq:RjetI}). Figure~\ref{fig:RI} shows that ${\cal R}^{\rm
I}_{\rm jet}\propto r^\beta$, where $\beta\approx 1.4$ ($\beta\approx
1.5$, 2, and 2.4 in figures~\ref{fig:ReI}, \ref{fig:RII}, and
\ref{fig:ReII}, respectively). Thus, $r_{\rm buo}$ should be larger by
$(p_{\rm B}/p_{\rm D})^{1/(4\beta)}$ with this second correction. Note
that the pressure difference at $r=r_{\rm buo}(p_{\rm B}/p_{\rm
D})^{1/(4\beta)}$ with that at $r=r_{\rm buo}$ does not affect the
following results and can be ignored. Thus, combined with the first
correction, the cavity radius at the phase D should be $L \sim r_{\rm
buo}(p_{\rm B}/p_{\rm D})^{1/4(1+1/\beta)}$.

We have used the pressure profiles constructed from the density and
temperature profiles in figures~\ref{fig:nT} and~\ref{fig:nTe}, and
assumed that $p_{\rm B}$ is the pressure at $r=r_{\rm buo}$, and $p_{\rm
D}$ is that at the observed position of a cavity, $R$, obtained by
\citet{raf06a}. If $R>r_{\rm in}$, we extrapolate the pressure profile
at $r\sim r_{\rm in}$ assuming that it is given by a power-law. We find
that the correction factor is $1\lesssim (p_{\rm B}/p_{\rm
D})^{1/4(1+1/\beta)}\lesssim 10$, and is $\lesssim 3$ for most galaxies
and cavities for both the FR\,I and II evolutions.

\newpage

\begin{table*}
  \tbl{Parameters for Gravitational Potentials.}{%
  \begin{tabular}{lcccccccc}
      \hline
   System & $z$ & $M_\bullet$ & $M_{\rm gal}$ & $R_{\rm e}$
     & $\sigma$ & $c_{\rm vir}$ & $M_{\rm vir}$ & References$^a$
 \\
          &     & $(10^9\: M_\odot)$ & $(10^{11}\: M_\odot)$ 
     & (kpc)
     & $(\rm km\: s^{-1})$ &  & $(10^{14}\: M_\odot)$ &  \\
      \hline
A85           & 0.055&
 7.0&
$ 31.0\pm 1.0$&
$ 16.3\pm
 0.03$&
$ 348\pm 19$&
$ 4.25
^{+ 0.76}_{- 0.96}$&
$12.33
^{+ 1.78}
_{- 1.34}$&
    1
\\
A133          & 0.060&
 3.0&
$ 17.9\pm 0.4$&
$ 14.6\pm
 0.43$&
$ 236\pm 11$&
$ 6.35
^{+ 0.53}_{- 0.53}$&
$ 5.64
^{+ 0.88}
_{- 0.77}$&
    2
\\
A262          & 0.016&
 0.6&
$  4.9\pm 0.1$&
$ 10.4\pm
 0.58$&
$ 230\pm  10$&
$ 8.84
^{+ 0.69}_{- 0.69}$&
$ 1.15
^{+ 0.092}
_{- 0.15}$&
    3
\\
Perseus       & 0.018&
 0.34&
$ 19.2\pm 0.1$&
$ 11.3\pm
 0.43$&
$ 259\pm 13$&
$ 8.08
^{+ 0.35}_{- 0.35}$&
$ 6.81
^{+ 0.63}
_{- 0.72}$&
    4
\\
2A~0335+096   & 0.035&
 3.0&
$ 18.0\pm 1.0$&
$ 15.0\pm
 1.4$&
$ 290\pm 36$&
$ 7.44
^{+ 0.42}_{- 0.42}$&
$ 2.11
^{+ 0.24}
_{- 0.29}$&
    3
\\
A478          & 0.081&
 5.8&
$ 28.0\pm 1.0$&
$ 15.8\pm
 3.1$&
$ 290\pm 36$&
$ 5.15
^{+ 0.45}_{- 0.49}$&
$ 16.6
^{+  2.0}
_{-  2.6}$&
    5
\\
MS~0735.6+7421& 0.216&
 5.0&
$ 24.0\pm 1.0$&
$ 15.1\pm
 3.8$&
$ 290\pm 36$&
$ 4.37
^{+ 0.22}_{- 0.23}$&
$  9
^{+ 0.40}
_{- 0.75}$&
    6
\\
PKS~0745-191  & 0.103&
 5.5&
$ 27.0\pm 1.0$&
$ 16.1\pm
 3.3$&
$ 290\pm 36$&
$ 7.75
^{+ 2.15}_{- 1.41}$&
$ 14.9
^{+  6.7}
_{-  3.7}$&
    5
\\
Hydra~A       & 0.055&
 5.8&
$ 28.2\pm 0.7$&
$ 10.5\pm
 0.90$&
$ 362\pm 19$&
$15.90
^{+ 0.23}_{- 0.23}$&
$ 1.15
^{+ 0.44}
_{- 0.36}$&
    7
\\
Zw~2701       & 0.214&
 6.5&
$ 30.0\pm 1.0$&
$ 13.4\pm
 1.3$&
$ 290\pm 36$&
$ 3.30
^{+  1.2}_{-  1.2}$&
$10.86
^{+ 2.57}
_{- 5.86}$&
    8
\\
Zw~3146       & 0.291&
 9.0&
$ 13.5\pm 6.9$&
$ 17.4\pm
 7.6$&
$ 290\pm 36$&
$ 4.19
^{+ 0.18}_{- 0.31}$&
$ 9.29
^{+ 1.04}
_{- 0.55}$&
    9
\\
M84           & 0.0035&
 0.36&
$  4.3\pm 1.3$&
$ 2.45\pm
 0.06$&
$ 282\pm  3$&
 $\cdots$&
 $\cdots$&
 $\cdots$
\\
M87           & 0.0042&
 6.4&
$ 11.0\pm 3.3$&
$ 3.67\pm
 0.13$&
$ 336\pm  5$&
$ 3.84
^{+ 0.91}_{- 0.92}$&
$ 5.78
^{+ 0.59}
_{-  1.5}$&
   10
\\
Centaurus     & 0.011&
 2.0&
$ 11.6\pm 0.1$&
$ 9.44\pm
 0.24$&
$ 254\pm  7$&
$ 7.75
^{+ 0.77}_{- 0.78}$&
$ 4.09
^{+ 0.32}
_{- 0.62}$&
    4
\\
HCG~62        & 0.014&
 0.65&
$ 13.5\pm 6.9$&
$ 6.87\pm
 0.04$&
$ 290\pm 36$&
 $\cdots$&
 $\cdots$&
 $\cdots$
\\
A1795         & 0.063&
 2.2&
$ 13.4\pm 0.6$&
$ 20.8\pm
 0.23$&
$ 302\pm  9$&
$ 6.16
^{+ 1.14}_{- 1.14}$&
$ 10.8
^{+  2.7}
_{-  2.4}$&
    5
\\
A1835         & 0.253&
 6.7&
$ 13.5\pm 6.9$&
$ 18.4\pm
 0.35$&
$ 290\pm 36$&
$ 4.18
^{+ 0.63}_{- 0.41}$&
$ 24.3
^{+  4.4}
_{-  4.9}$&
    5
\\
PKS~1404-267  & 0.022&
 0.7&
$  5.7\pm 0.5$&
$ 6.03\pm
 0.12$&
$ 260\pm  7$&
$12.25
^{+ 1.09}_{- 6.07}$&
$ 1.77
^{+ 0.43}
_{- 0.31}$&
    1
\\
A2029         & 0.077&
 4.0&
$ 21.9\pm 0.2$&
$ 24.2\pm
 1.6$&
$ 391\pm 10$&
$ 8.86
^{+ 0.44}_{- 0.50}$&
$ 10.1
^{+ 0.99}
_{- 0.77}$&
    5
\\
A2052         & 0.035&
 2.0&
$ 11.0\pm 3.3$&
$ 15.7\pm
 0.27$&
$ 216\pm 12$&
$ 6.50
^{+ 0.71}_{- 0.71}$&
$ 2.96
^{+ 0.52}
_{- 0.77}$&
    3
\\
MKW~3S        & 0.045&
 2.0&
$ 11.2\pm 0.3$&
$ 11.6\pm
 2.3$&
$ 290\pm 36$&
$ 7.83
^{+ 0.55}_{- 0.55}$&
$ 2.90
^{+ 0.27}
_{- 0.38}$&
    3
\\
A2199         & 0.030&
 2.7&
$ 15.7\pm 0.2$&
$ 10.6\pm
 0.20$&
$ 307\pm  7$&
$10.40
^{+ 14.6}_{-  7.9}$&
$  7.1
^{+  3.4}
_{-  2.4}$&
   11
\\
Hercules~A    & 0.154&
 2.5&
$ 15.0\pm 4.0$&
$ 20.1\pm
 2.0$&
$ 290\pm 36$&
$ 3.51
^{+ 0.23}_{- 0.23}$&
$ 4.33
^{+ 0.54}
_{- 0.54}$&
12,13
\\
3C~388        & 0.092&
 4.5&
$ 23.0\pm 6.0$&
$ 11.9\pm
 1.2$&
$ 408\pm 26$&
 $\cdots$&
 $\cdots$&
 $\cdots$
\\
Cygnus~A      & 0.056&
 2.7&
$  9.0\pm 2.0$&
$ 15.6\pm
 0.77$&
$ 290\pm 36$&
$16.40
^{+ 0.25}_{- 0.25}$&
$ 8.33
^{+ 0.38}
_{- 0.38}$&
13,14
\\
Sersic~159/03 & 0.058&
 2.0&
$ 11.0\pm 2.0$&
$ 20.2\pm
 0.95$&
$ 290\pm 36$&
$ 8.57
^{+ 0.69}_{- 0.69}$&
$ 1.61
^{+ 0.12}
_{- 0.20}$&
    3
\\
A2597         & 0.085&
 1.5&
$  9.0\pm 1.0$&
$ 11.7\pm
 1.3$&
$ 210\pm 57$&
$ 7.60
^{+ 0.63}_{- 0.63}$&
$ 3.55
^{+ 0.43}
_{- 0.40}$&
   15
\\
A4059         & 0.048&
 8.7&
$ 38.2\pm 0.4$&
$ 18.7\pm
 0.10$&
$ 272\pm 13$&
$ 3.57
^{+ 0.68}_{- 0.96}$&
$ 4.45
^{+ 0.63}
_{- 0.62}$&
    1
\\

     \hline
    \end{tabular}}\label{tab:pot}
\begin{tabnote}
$^a$ References for cluster parameters. 
(1) \citet{waj10a}; (2) \citet{vik06a}; 
(3) \citet{pif05a}; (4) \citet{ett02b}; (5) \citet{sch07a}; 
(6) \citet{git07a}; (7) \citet{dav01a}; (8) \citet{ric10a}; 
(9) \citet{ett10a}; (10) \citet{mcl99a}; (11) \citet{lok06a};
(12) \citet{giz04a}, \citet{sun09a}; (14) \citet{smi02a};
(15) \citet{poi05a}
\end{tabnote}
\end{table*}

\newpage
\begin{table*}
  \tbl{Observational Data.}{%
  \begin{tabular}{lccccc}
      \hline
   System & $r_{\rm in}$ & $n_{\rm e,in}$ & $T_{\rm in}$ 
     & $P_{\rm j,obs}$  &  $L_{\rm ICM}$ \\
          &   (kpc)       & $(\rm cm^{-3})$ &    (keV) 
     & $(10^{42}\rm\: erg\: s^{-1})$ & $(10^{42}\rm\: erg\: s^{-1})$ \\
      \hline
A85           &  5.8&
$  0.107^{+  0.009}_{-  0.008}$&
$  2.1^{+  0.1}_{-  0.2}$&
$     37^{+     37}_{    -11}$&
$    335^{+     21}_{    -29 }$
\\
A133          &  8.0&
$  0.048^{+  0.004}_{-  0.005}$&
$  1.8^{+  0.1}_{-  0.1}$&
$    620^{+    260}_{    -20}$&
$    103^{+      3}_{     -3 }$
\\
A262          &  3.4&
$  0.065^{+  0.008}_{-  0.007}$&
$ 0.86^{+ 0.01}_{- 0.01}$&
$  9.7^{+  7.5}_{ -2.6}$&
$ 11.10^{+ 0.31}_{-0.46 }$
\\
Perseus       &  8.6&
$  0.150^{+  0.005}_{-  0.005}$&
$  4.4^{+  0.5}_{-  0.4}$&
$    150^{+    100}_{    -30}$&
$    533^{+      7}_{     -8 }$
\\
2A~0335+096   &  5.1&
$  0.056^{+  0.003}_{-  0.002}$&
$  1.4^{+  0.1}_{-  0.1}$&
$     24^{+     23}_{     -6}$&
$    325^{+      4}_{     -4 }$
\\
A478          &  5.3&
$   0.20^{+   0.01}_{-   0.02}$&
$  2.7^{+  0.3}_{-  0.3}$&
$    100^{+     80}_{    -20}$&
$   1400^{+     22}_{    -51 }$
\\
MS~0735.6+7421& 23.8&
$  0.067^{+  0.002}_{-  0.003}$&
$  3.2^{+  0.2}_{-  0.2}$&
  35000&
$    438^{+     11}_{    -17 }$
\\
PKS~0745-191  & 11.2&
$   0.14^{+   0.01}_{-   0.01}$&
$  2.6^{+  0.4}_{-  0.4}$&
$   1700^{+   1400}_{   -300}$&
$   2070^{+    120}_{   -125 }$
\\
Hydra~A       &  4.7&
$   0.15^{+   0.01}_{-   0.02}$&
$  2.6^{+  0.8}_{-  0.5}$&
$   2000^{+     50}_{    -50}$&
$    269^{+      4}_{     -4 }$
\\
Zw~2701       & 37.6&
$  0.024^{+  0.002}_{-  0.002}$&
$  3.3^{+  0.3}_{-  0.3}$&
$   6000^{+   8900}_{  -3500}$&
$    430^{+     18}_{    -32 }$
\\
Zw~3146       & 15.0&
$  0.177^{+  0.007}_{-  0.007}$&
$  3.1^{+  0.3}_{-  0.2}$&
$   5800^{+   6800}_{  -1500}$&
$   2330^{+    161}_{   -196 }$
\\
M84           &  0.9&
$  0.105^{+  0.007}_{-  0.007}$&
$ 0.57^{+ 0.01}_{- 0.01}$&
$  1.0^{+  1.5}_{ -0.6}$&
$  0.06^{+ 0.01}_{-0.01 }$
\\
M87           &  1.0&
$  0.191^{+  0.009}_{-  0.009}$&
$ 0.94^{+ 0.02}_{- 0.02}$&
$  6.0^{+  4.2}_{ -0.9}$&
$  7.20^{+ 0.20}_{-0.11 }$
\\
Centaurus     &  1.3&
$   0.23^{+   0.01}_{-   0.01}$&
$ 0.77^{+ 0.01}_{- 0.01}$&
$  7.4^{+  5.8}_{ -1.8}$&
$ 23.80^{+ 0.35}_{-0.35 }$
\\
HCG~62        &  2.1&
$  0.057^{+  0.007}_{-  0.005}$&
$ 0.67^{+ 0.01}_{- 0.01}$&
$  3.9^{+  6.1}_{ -2.3}$&
$  1.80^{+ 0.17}_{-0.24 }$
\\
A1795         &  9.5&
$  0.067^{+  0.005}_{-  0.005}$&
$  2.7^{+  0.6}_{-  0.4}$&
$    160^{+    230}_{    -50}$&
$    615^{+     10}_{    -19 }$
\\
A1835         & 27.2&
$  0.110^{+  0.003}_{-  0.003}$&
$  4.0^{+  0.3}_{-  0.3}$&
$   1800^{+   1900}_{   -600}$&
$   3160^{+     61}_{    -90 }$
\\
PKS~1404-267  &  8.5&
$  0.046^{+  0.002}_{-  0.002}$&
$  1.3^{+  0.1}_{-  0.1}$&
$     20^{+     26}_{     -9}$&
$     24^{+      1}_{     -1 }$
\\
A2029         &  2.2&
$   0.37^{+   0.04}_{-   0.03}$&
$  2.9^{+  0.3}_{-  0.2}$&
$     87^{+     49}_{     -4}$&
$   1160^{+      9}_{    -11 }$
\\
A2052         &  5.5&
$  0.017^{+  0.002}_{-  0.002}$&
$ 0.71^{+ 0.04}_{- 0.08}$&
$    150^{+    200}_{     -7}$&
$     94^{+      1}_{     -1 }$
\\
MKW~3S        &  7.8&
$  0.028^{+  0.006}_{-  0.009}$&
$  2.8^{+  0.8}_{-  0.5}$&
$    410^{+    420}_{    -44}$&
$     99^{+      3}_{     -4 }$
\\
A2199         &  4.4&
$  0.099^{+  0.005}_{-  0.005}$&
$  2.2^{+  0.2}_{-  0.1}$&
$    270^{+    250}_{    -60}$&
$    142^{+      1}_{     -3 }$
\\
Hercules~A    & 67.0&
$ 0.0111^{+ 0.0006}_{- 0.0005}$&
$  2.0^{+  0.2}_{-  0.2}$&
  16000&
$    210^{+      6}_{    -54 }$
\\
3C~388        & 55.6&
$ 0.0069^{+ 0.0004}_{- 0.0004}$&
$  3.0^{+  0.2}_{-  0.2}$&
$    200^{+    280}_{    -80}$&
$     27^{+      1}_{     -4 }$
\\
Cygnus~A      &  5.3&
$  0.132^{+  0.009}_{-  0.008}$&
$  5.2^{+  0.5}_{-  0.6}$&
   3900&
$    370^{+     11}_{    -11 }$
\\
Sersic~159/03 & 12.2&
$  0.056^{+  0.004}_{-  0.004}$&
$  1.8^{+  0.2}_{-  0.1}$&
$    780^{+    820}_{   -260}$&
$    211^{+      7}_{     -8 }$
\\
A2597         & 11.0&
$  0.073^{+  0.005}_{-  0.005}$&
$  1.6^{+  0.2}_{-  0.2}$&
$     67^{+     87}_{    -29}$&
$    440^{+     19}_{    -37 }$
\\
A4059         & 10.6&
$  0.022^{+  0.001}_{-  0.001}$&
$  2.1^{+  0.1}_{-  0.1}$&
$     96^{+     89}_{    -35}$&
$     91^{+      1}_{     -1 }$
\\

     \hline
    \end{tabular}}\label{tab:obs}
\end{table*}

\newpage
\begin{table*}
  \tbl{Parameters for the Bondi Accretion in the Low-Temperature
  Model.}{%
  \begin{tabular}{lccccc}
      \hline
   System & $r_{\rm B}$ & $n_{\rm e,B}$ & $T_{\rm B}$ 
   & $\dot{M}_{\rm B}$  & $P_{\rm B}$ \\
          &   (kpc)       & $(\rm cm^{-3})$ &    (keV) 
     & $(M_\odot\rm\: yr^{-1})$ & $(10^{44}\rm\: erg\: s^{-1})$  \\
      \hline
A85           &
$  0.15^{+  0.34}_{ -0.09}$&
$  2.73^{+  0.68}_{ -0.89}$&
$ 1.56^{+ 0.17}_{-0.14}$&
$  0.85^{+  6.37}_{ -0.72}$&
$     48^{+    361}_{    -41 }$
\\
A133          &
$  0.13^{+  0.29}_{ -0.08}$&
$  5.68^{+  1.87}_{ -2.27}$&
$ 0.73^{+ 0.09}_{-0.06}$&
$  1.01^{+  6.16}_{ -0.85}$&
$     58^{+    349}_{    -48 }$
\\
A262          &
$ 0.029^{+ 0.069}_{-0.018}$&
$  0.76^{+  0.20}_{ -0.15}$&
$ 0.67^{+ 0.06}_{-0.05}$&
$ 0.006^{+ 0.062}_{-0.005}$&
$ 0.35^{+ 3.52}_{-0.30 }$
\\
Perseus       &
$ 0.013^{+ 0.030}_{-0.008}$&
$    34^{+    12}_{    -9}$&
$ 0.86^{+ 0.10}_{-0.08}$&
$ 0.062^{+ 0.602}_{-0.053}$&
$  3.5^{+ 34.1}_{ -3.0 }$
\\
2A~0335+096   &
$ 0.092^{+ 0.222}_{-0.058}$&
$  1.27^{+  0.94}_{ -0.53}$&
$ 1.07^{+ 0.28}_{-0.23}$&
$  0.13^{+  1.29}_{ -0.11}$&
$  7.2^{+ 73.2}_{ -6.1 }$
\\
A478          &
$  0.17^{+  0.36}_{ -0.10}$&
$  7.93^{+ 10.53}_{ -4.15}$&
$ 1.12^{+ 0.31}_{-0.20}$&
$  2.78^{+ 20.03}_{ -2.33}$&
$    158^{+   1140}_{   -132 }$
\\
MS~0735.6+7421&
$  0.15^{+  0.34}_{ -0.09}$&
$    19^{+    60}_{   -12}$&
$ 1.09^{+ 0.30}_{-0.22}$&
$  5.26^{+ 64.70}_{ -4.61}$&
$    298^{+   3670}_{   -261 }$
\\
PKS~0745-191  &
$  0.17^{+  0.37}_{ -0.10}$&
$    25^{+    70}_{   -15}$&
$ 1.09^{+ 0.29}_{-0.21}$&
$  8.23^{+ 99.26}_{ -7.13}$&
$    466^{+   5630}_{   -404 }$
\\
Hydra~A       &
$  0.11^{+  0.25}_{ -0.07}$&
$    11^{+     7}_{    -5}$&
$ 1.69^{+ 0.20}_{-0.14}$&
$  2.10^{+ 13.76}_{ -1.76}$&
$    119^{+    780}_{   -100 }$
\\
Zw~2701       &
$  0.19^{+  0.43}_{ -0.12}$&
$    36^{+    54}_{   -24}$&
$ 1.10^{+ 0.30}_{-0.21}$&
$ 16.53^{+113.87}_{-14.52}$&
$    937^{+   6450}_{   -823 }$
\\
Zw~3146       &
$  0.27^{+  0.55}_{ -0.17}$&
$  5.72^{+ 20.06}_{ -3.37}$&
$ 1.11^{+ 0.37}_{-0.20}$&
$  4.89^{+ 68.78}_{ -4.25}$&
$    277^{+   3900}_{   -241 }$
\\
M84           &
$ 0.012^{+ 0.027}_{-0.007}$&
$  9.86^{+ 24.35}_{ -7.26}$&
$ 1.01^{+ 0.02}_{-0.02}$&
$ 0.016^{+ 0.215}_{-0.014}$&
$ 0.88^{+12.20}_{-0.81 }$
\\
M87           &
$  0.15^{+  0.46}_{ -0.10}$&
$  8.16^{+ 13.87}_{ -7.45}$&
$ 1.36^{+ 0.06}_{-0.24}$&
$  2.60^{+  3.45}_{ -2.26}$&
$    148^{+    196}_{   -128 }$
\\
Centaurus     &
$ 0.080^{+ 0.188}_{-0.050}$&
$  2.19^{+  0.34}_{ -0.70}$&
$ 0.82^{+ 0.04}_{-0.04}$&
$  0.15^{+  0.99}_{ -0.12}$&
$  8.3^{+ 56.2}_{ -7.0 }$
\\
HCG~62        &
$ 0.020^{+ 0.050}_{-0.013}$&
$  4.16^{+ 36.21}_{ -3.64}$&
$ 1.06^{+ 0.28}_{-0.24}$&
$ 0.020^{+ 0.643}_{-0.019}$&
$  1.1^{+ 36.4}_{ -1.1 }$
\\
A1795         &
$ 0.062^{+ 0.139}_{-0.039}$&
$  1.17^{+  0.21}_{ -0.20}$&
$ 1.17^{+ 0.08}_{-0.06}$&
$ 0.055^{+ 0.504}_{-0.047}$&
$  3.1^{+ 28.6}_{ -2.7 }$
\\
A1835         &
$  0.20^{+  0.44}_{ -0.12}$&
$  7.00^{+  7.83}_{ -3.96}$&
$ 1.10^{+ 0.30}_{-0.21}$&
$  3.36^{+ 28.44}_{ -2.97}$&
$    191^{+   1610}_{   -168 }$
\\
PKS~1404-267  &
$ 0.027^{+ 0.061}_{-0.017}$&
$  9.02^{+  3.17}_{ -4.10}$&
$ 0.86^{+ 0.05}_{-0.04}$&
$ 0.068^{+ 0.554}_{-0.060}$&
$  3.9^{+ 31.4}_{ -3.4 }$
\\
A2029         &
$ 0.067^{+ 0.146}_{-0.042}$&
$  1.72^{+  0.27}_{ -0.28}$&
$ 1.97^{+ 0.13}_{-0.08}$&
$  0.12^{+  1.02}_{ -0.10}$&
$  7.0^{+ 58.1}_{ -5.9 }$
\\
A2052         &
$  0.11^{+  0.26}_{ -0.07}$&
$  0.53^{+  0.46}_{ -0.27}$&
$ 0.60^{+ 0.06}_{-0.06}$&
$ 0.057^{+ 0.525}_{-0.050}$&
$  3.2^{+ 29.8}_{ -2.8 }$
\\
MKW~3S        &
$ 0.061^{+ 0.143}_{-0.038}$&
$  1.02^{+  1.13}_{ -0.52}$&
$ 1.08^{+ 0.29}_{-0.22}$&
$ 0.045^{+ 0.470}_{-0.039}$&
$  2.5^{+ 26.6}_{ -2.2 }$
\\
A2199         &
$ 0.073^{+ 0.160}_{-0.046}$&
$  3.81^{+  4.85}_{ -1.18}$&
$ 1.22^{+ 0.07}_{-0.04}$&
$  0.26^{+  2.67}_{ -0.21}$&
$     15^{+    151}_{    -12 }$
\\
Hercules~A    &
$ 0.077^{+ 0.187}_{-0.048}$&
$  1.28^{+  2.00}_{ -0.70}$&
$ 1.07^{+ 0.28}_{-0.24}$&
$ 0.089^{+ 1.125}_{-0.080}$&
$  5.1^{+ 63.7}_{ -4.6 }$
\\
3C~388        &
$ 0.069^{+ 0.162}_{-0.043}$&
$  0.65^{+  1.27}_{ -0.40}$&
$ 2.13^{+ 0.28}_{-0.25}$&
$ 0.053^{+ 0.602}_{-0.048}$&
$  3.0^{+ 34.1}_{ -2.7 }$
\\
Cygnus~A      &
$ 0.078^{+ 0.165}_{-0.048}$&
$  3.39^{+  1.07}_{ -1.15}$&
$ 1.13^{+ 0.32}_{-0.20}$&
$  0.25^{+  1.85}_{ -0.21}$&
$     14^{+    105}_{    -12 }$
\\
Sersic~159/03 &
$ 0.061^{+ 0.149}_{-0.039}$&
$  1.02^{+  0.68}_{ -0.37}$&
$ 1.07^{+ 0.28}_{-0.24}$&
$ 0.046^{+ 0.514}_{-0.040}$&
$  2.6^{+ 29.1}_{ -2.2 }$
\\
A2597         &
$ 0.086^{+ 0.226}_{-0.056}$&
$    13^{+    41}_{    -9}$&
$ 0.57^{+ 0.35}_{-0.24}$&
$  0.84^{+ 17.09}_{ -0.78}$&
$     48^{+    969}_{    -44 }$
\\
A4059         &
$  0.29^{+  0.62}_{ -0.18}$&
$  4.28^{+  1.64}_{ -2.32}$&
$ 0.98^{+ 0.12}_{-0.07}$&
$  4.14^{+ 16.90}_{ -3.43}$&
$    235^{+    958}_{   -194 }$
\\

     \hline
    \end{tabular}}\label{tab:res}
\end{table*}

\newpage
\begin{table*}
  \tbl{Jet Propagation (FR~I type).}{%
  \begin{tabular}{lccp{0pt}ccp{0pt}cc}
      \hline
 &\multicolumn{2}{c}{Low-Temperature ($P_{\rm j}=P_{\rm j,obs}$)}
&&\multicolumn{2}{c}{Low-Temperature ($P_{\rm j}=P_{\rm B}$)}
&&\multicolumn{2}{c}{Isentropic ($P_{\rm j}=P_{\rm j,obs}$)}\\
\cline{2-3}
\cline{5-6}
\cline{8-9}\\[-5pt]
   System & ${\cal R}^{\rm I}_{\rm jet}(r=1\rm\; kpc)$ & $r_{\rm buo}$ &
   & ${\cal R}^{\rm I}_{\rm B}(r=1\rm\; kpc)$ 
& $r_{\rm buo}$ & 
   & ${\cal R}^{\rm I}_{\rm jet}(r=1\rm\; kpc)$ 
& $r_{\rm buo}$ \\
          & & (kpc)   &&  & (kpc) & 
 & & (kpc)  \\
      \hline
A85           &
$  7.87^{+  4.39}_{ -3.74}$&
$  0.24^{+  0.13}_{ -0.23}$&&
$ 0.060^{+ 0.313}_{-0.052}$&
$>  2.18 $&&
$  7.02^{+  3.47}_{ -3.55}$&
$  0.31^{+  0.14}_{ -0.30 }$
\\
A133          &
$  0.25^{+  0.05}_{ -0.07}$&
$  3.90^{+  1.79}_{ -0.35}$&&
$ 0.027^{+ 0.132}_{-0.022}$&
$>  5.75 $&&
$  0.14^{+  0.02}_{ -0.04}$&
$  3.90^{+  1.80}_{ -0.31 }$
\\
A262          &
$  2.11^{+  0.95}_{ -0.91}$&
$  0.62^{+  0.26}_{ -0.13}$&&
$  0.59^{+  3.54}_{ -0.53}$&
$  1.43 ^{+\infty}_{ -1.01}$&&
$  1.82^{+  0.83}_{ -0.77}$&
$  0.71^{+  0.27}_{ -0.14 }$
\\
Perseus       &
$  5.76^{+  1.60}_{ -2.30}$&
$  0.31^{+  0.12}_{ -0.05}$&&
$  2.48^{+ 15.55}_{ -2.24}$&
$  0.53^{+  2.51}_{ -0.36}$&&
$  3.08^{+  0.93}_{ -1.25}$&
$  0.56^{+  0.17}_{ -0.07 }$
\\
2A~0335+096   &
$  2.94^{+  1.58}_{ -1.39}$&
$  0.48^{+  0.25}_{ -0.22}$&&
$ 0.098^{+ 0.572}_{-0.088}$&
$>  1.35 $&&
$  2.93^{+  1.27}_{ -1.52}$&
$  0.50^{+  0.29}_{ -0.14 }$
\\
A478          &
$  5.16^{+  4.04}_{ -2.38}$&
$  0.31^{+  0.18}_{ -0.30}$&&
$ 0.033^{+ 0.149}_{-0.027}$&
$>  3.52 $&&
$  3.30^{+  1.78}_{ -1.53}$&
$  0.52^{+  0.20}_{ -0.51 }$
\\
MS~0735.6+7421&
$ 0.031^{+ 0.051}_{-0.013}$&
$ 21.86^{+  1.94}_{ -1.58}$&&
$ 0.037^{+ 0.178}_{-0.032}$&
$ 19.23^{+ 80.77}_{-15.58}$&&
$ 0.006^{+ 0.004}_{-0.001}$&
$ 21.84^{+  1.96}_{ -1.45 }$
\\
PKS~0745-191  &
$  0.82^{+  1.22}_{ -0.41}$&
$  1.18^{+  0.94}_{ -0.89}$&&
$ 0.030^{+ 0.151}_{-0.025}$&
$>  5.08 $&&
$  0.19^{+  0.12}_{ -0.08}$&
$  2.56^{+  1.09}_{ -0.57 }$
\\
Hydra~A       &
$  0.39^{+  0.14}_{ -0.08}$&
$  3.56 ^{+\infty}_{ -1.10}$&&
$ 0.066^{+ 0.355}_{-0.056}$&
$>  2.72 $&&
$  0.33^{+  0.08}_{ -0.06}$&
$  3.56 ^{+\infty}_{ -1.07 }$
\\
Zw~2701       &
$  0.27^{+  0.62}_{ -0.18}$&
$  6.85^{+ 10.97}_{ -5.68}$&&
$ 0.017^{+ 0.085}_{-0.014}$&
$> 13.48 $&&
$ 0.061^{+ 0.097}_{-0.038}$&
$  6.96^{+ 10.87}_{ -3.72 }$
\\
Zw~3146       &
$ 0.095^{+ 0.174}_{-0.071}$&
$  4.56^{+  2.86}_{ -1.75}$&&
$ 0.020^{+ 0.081}_{-0.017}$&
$ 12.07^{+ 87.93}_{ -8.16}$&&
$ 0.045^{+ 0.038}_{-0.030}$&
$  5.08^{+  2.58}_{ -1.26 }$
\\
M84           &
$  5.33^{+  7.76}_{ -3.19}$&
$ 0.033^{+ 0.053}_{-0.023}$&&
$ 0.060^{+ 0.718}_{-0.056}$&
$>  0.23 $&&
$  5.33^{+  7.76}_{ -3.19}$&
$ 0.040^{+ 0.047}_{-0.030 }$
\\
M87           &
$  4.21^{+  0.78}_{ -2.06}$&
$<  0.08 $&&
$ 0.002^{+ 0.012}_{-0.001}$&
$+\infty$&&
$  4.21^{+  0.78}_{ -2.06}$&
$<  0.08  $
\\
Centaurus     &
$  5.24^{+  1.76}_{ -2.29}$&
$  0.23^{+  0.11}_{ -0.22}$&&
$ 0.047^{+ 0.264}_{-0.041}$&
$+\infty$&&
$  5.18^{+  1.74}_{ -2.27}$&
$  0.28^{+  0.12}_{ -0.27 }$
\\
HCG~62        &
$ 10.01^{+ 35.21}_{ -7.22}$&
$  0.11^{+  0.29}_{ -0.10}$&&
$  0.35^{+  4.22}_{ -0.33}$&
$>  0.26 $&&
$ 10.01^{+ 28.39}_{ -7.22}$&
$  0.12^{+  0.28}_{ -0.08 }$
\\
A1795         &
$  0.67^{+  0.38}_{ -0.39}$&
$  1.26^{+  0.85}_{ -0.28}$&&
$  0.34^{+  1.99}_{ -0.30}$&
$  1.85^{+  5.46}_{ -1.22}$&&
$  0.58^{+  0.33}_{ -0.35}$&
$  1.33^{+  0.85}_{ -0.28 }$
\\
A1835         &
$  0.35^{+  0.40}_{ -0.22}$&
$  1.92^{+  1.25}_{ -0.73}$&&
$ 0.033^{+ 0.173}_{-0.028}$&
$  8.54 ^{+\infty}_{ -6.03}$&&
$  0.13^{+  0.08}_{ -0.08}$&
$  2.81^{+  1.34}_{ -0.61 }$
\\
PKS~1404-267  &
$  8.23^{+  7.04}_{ -5.22}$&
$  0.20^{+  0.19}_{ -0.08}$&&
$  0.42^{+  2.76}_{ -0.38}$&
$  2.86 ^{+\infty}_{ -2.45}$&&
$  4.20^{+  3.31}_{ -2.59}$&
$  0.42^{+  0.32}_{ -0.12 }$
\\
A2029         &
$  4.35^{+  0.71}_{ -1.53}$&
$  0.45^{+  0.12}_{ -0.05}$&&
$  0.54^{+  3.16}_{ -0.48}$&
$  1.39 ^{+\infty}_{ -0.88}$&&
$  4.31^{+  0.72}_{ -1.51}$&
$  0.46^{+  0.11}_{ -0.06 }$
\\
A2052         &
$ 0.078^{+ 0.033}_{-0.049}$&
$+\infty$&&
$ 0.036^{+ 0.213}_{-0.032}$&
$>  3.77 $&&
$ 0.073^{+ 0.020}_{-0.045}$&
  $+\infty$
\\
MKW~3S        &
$  0.15^{+  0.07}_{ -0.09}$&
$  3.50^{+  2.74}_{ -0.65}$&&
$  0.24^{+  1.32}_{ -0.22}$&
$  2.55 ^{+\infty}_{ -1.78}$&&
$  0.15^{+  0.07}_{ -0.09}$&
$  3.50^{+  2.74}_{ -0.65 }$
\\
A2199         &
$  0.76^{+  0.71}_{ -0.34}$&
$  1.25^{+  0.85}_{ -0.54}$&&
$  0.14^{+  0.72}_{ -0.12}$&
$>  1.11 $&&
$  0.74^{+  0.55}_{ -0.33}$&
$  1.25^{+  0.84}_{ -0.42 }$
\\
Hercules~A    &
$ 0.006^{+ 0.006}_{-0.002}$&
$ 42.80^{+  5.78}_{ -4.25}$&&
$  0.18^{+  1.11}_{ -0.17}$&
$  2.96^{+ 17.85}_{ -2.07}$&&
$ 0.003^{+ 0.001}_{-0.001}$&
$ 42.80^{+  5.89}_{ -4.25 }$
\\
3C~388        &
$  0.50^{+  0.97}_{ -0.34}$&
$  1.65^{+  2.21}_{ -0.91}$&&
$  0.34^{+  1.99}_{ -0.30}$&
$  2.28^{+ 16.12}_{ -1.69}$&&
$  0.50^{+  0.97}_{ -0.34}$&
$  1.65^{+  2.21}_{ -0.91 }$
\\
Cygnus~A      &
$ 0.085^{+ 0.013}_{-0.009}$&
$  3.64^{+  0.36}_{ -0.24}$&&
$  0.23^{+  1.26}_{ -0.20}$&
$  2.17 ^{+\infty}_{ -1.35}$&&
$ 0.083^{+ 0.012}_{-0.010}$&
$  3.64^{+  0.36}_{ -0.24 }$
\\
Sersic~159/03 &
$  0.10^{+  0.07}_{ -0.05}$&
$  4.15^{+  2.71}_{ -1.11}$&&
$  0.30^{+  1.90}_{ -0.27}$&
$  2.04^{+  8.99}_{ -1.38}$&&
$ 0.065^{+ 0.044}_{-0.033}$&
$  4.51^{+  2.50}_{ -1.10 }$
\\
A2597         &
$  3.37^{+  4.79}_{ -1.90}$&
$  0.41^{+  0.33}_{ -0.40}$&&
$ 0.047^{+ 0.471}_{-0.044}$&
$>  1.66 $&&
$  0.96^{+  0.88}_{ -0.56}$&
$  1.02^{+  0.62}_{ -0.30 }$
\\
A4059         &
$  2.39^{+  1.49}_{ -1.54}$&
$  0.45^{+  0.33}_{ -0.44}$&&
$ 0.010^{+ 0.038}_{-0.007}$&
$+\infty$&&
$  1.53^{+  0.84}_{ -0.96}$&
$  0.76^{+  0.38}_{ -0.75 }$
\\

     \hline
    \end{tabular}}\label{tab:jetI}
\end{table*}

\newpage
\begin{table*}
  \tbl{Jet Propagation (FR~II type).}{%
  \begin{tabular}{lccp{0pt}ccp{0pt}cc}
      \hline
 &\multicolumn{2}{c}{Low-Temperature ($P_{\rm j}=P_{\rm j,obs}$)}
&&\multicolumn{2}{c}{Low-Temperature ($P_{\rm j}=P_{\rm B}$)}
&&\multicolumn{2}{c}{Isentropic ($P_{\rm j}=P_{\rm j,obs}$)}\\
\cline{2-3}
\cline{5-6}
\cline{8-9}\\[-5pt]
   System & ${\cal R}^{\rm II}_{\rm jet}(r=1\rm\; kpc)$ & $r_{\rm buo}$ &
   & ${\cal R}^{\rm II}_{\rm B}(r=1\rm\; kpc)$ 
& $r_{\rm buo}$ & 
   & ${\cal R}^{\rm II}_{\rm jet}(r=1\rm\; kpc)$ 
& $r_{\rm buo}$ \\
          & & (kpc)   &&  & (kpc) & 
 & & (kpc)  \\
      \hline
A85           &
$114.29^{+ 68.14}_{-55.09}$&
$<  0.17 $&&
$  0.88^{+  7.75}_{ -0.73}$&
$>  0.38 $&&
$ 71.30^{+ 38.25}_{-34.89}$&
$  0.18^{+  0.07}_{ -0.17 }$
\\
A133          &
$  5.08^{+  1.02}_{ -1.51}$&
$  0.43^{+  0.12}_{ -0.42}$&&
$  0.55^{+  4.67}_{ -0.44}$&
$>  0.46 $&&
$  1.18^{+  0.11}_{ -0.34}$&
$  0.94^{+  0.32}_{ -0.04 }$
\\
A262          &
$ 45.59^{+ 20.74}_{-19.77}$&
$  0.20^{+  0.08}_{ -0.08}$&&
$ 12.77^{+133.17}_{-11.27}$&
$  0.33^{+  0.50}_{ -0.22}$&&
$ 25.43^{+ 12.90}_{-10.26}$&
$  0.29^{+  0.08}_{ -0.09 }$
\\
Perseus       &
$ 93.68^{+ 27.89}_{-37.71}$&
$  0.12^{+  0.07}_{ -0.05}$&&
$ 40.28^{+449.81}_{-35.55}$&
$  0.18^{+  0.28}_{ -0.12}$&&
$ 14.87^{+  4.49}_{ -6.02}$&
$  0.39^{+  0.10}_{ -0.07 }$
\\
2A~0335+096   &
$ 51.24^{+ 29.27}_{-24.42}$&
$  0.16^{+  0.09}_{ -0.15}$&&
$  1.71^{+ 17.91}_{ -1.50}$&
$  0.77 ^{+\infty}_{ -0.50}$&&
$ 47.51^{+ 12.97}_{-27.34}$&
$  0.21^{+  0.10}_{ -0.20 }$
\\
A478          &
$ 82.17^{+ 60.54}_{-38.67}$&
$<  0.18 $&&
$  0.52^{+  4.40}_{ -0.42}$&
$>  0.48 $&&
$ 21.25^{+ 11.61}_{ -9.67}$&
$  0.32^{+  0.10}_{ -0.31 }$
\\
MS~0735.6+7421&
$  0.53^{+  0.92}_{ -0.24}$&
$>  0.77 $&&
$  0.63^{+  5.47}_{ -0.53}$&
$>  0.43 $&&
$ 0.020^{+ 0.014}_{-0.004}$&
  $+\infty$
\\
PKS~0745-191  &
$ 13.99^{+ 22.14}_{ -7.35}$&
$  0.26^{+  0.16}_{ -0.25}$&&
$  0.51^{+  4.44}_{ -0.42}$&
$>  0.47 $&&
$  0.82^{+  0.48}_{ -0.34}$&
$  1.25^{+  1.06}_{ -0.35 }$
\\
Hydra~A       &
$  5.49^{+  2.18}_{ -1.22}$&
$  0.39^{+  0.10}_{ -0.38}$&&
$  0.92^{+  8.97}_{ -0.76}$&
$>  0.33 $&&
$  2.83^{+  0.86}_{ -0.45}$&
$  0.65^{+  0.06}_{ -0.11 }$
\\
Zw~2701       &
$  4.69^{+ 12.42}_{ -3.08}$&
$  0.41^{+  0.32}_{ -0.40}$&&
$  0.30^{+  2.69}_{ -0.23}$&
$>  0.56 $&&
$  0.28^{+  0.47}_{ -0.17}$&
$  5.07 ^{+\infty}_{ -3.61 }$
\\
Zw~3146       &
$  1.58^{+  3.08}_{ -1.12}$&
$  0.81^{+  2.12}_{ -0.80}$&&
$  0.33^{+  2.53}_{ -0.27}$&
$>  0.68 $&&
$  0.21^{+  0.14}_{ -0.13}$&
$  4.72^{+  5.36}_{ -1.85 }$
\\
M84           &
$152.58^{+271.85}_{-92.89}$&
$ 0.025^{+ 0.037}_{-0.015}$&&
$  1.73^{+ 29.85}_{ -1.58}$&
$  0.24 ^{+\infty}_{ -0.19}$&&
$152.58^{+271.85}_{-92.89}$&
$ 0.035^{+ 0.037}_{-0.025 }$
\\
M87           &
$101.17^{+ 19.08}_{-45.33}$&
$<  0.03 $&&
$ 0.041^{+ 0.418}_{-0.027}$&
$+\infty$&&
$101.17^{+ 19.08}_{-45.33}$&
$<  0.06  $
\\
Centaurus     &
$114.57^{+ 42.52}_{-50.10}$&
$ 0.083^{+ 0.062}_{-0.073}$&&
$  1.02^{+  9.32}_{ -0.87}$&
$  0.98 ^{+\infty}_{ -0.69}$&&
$100.06^{+ 36.01}_{-44.12}$&
$  0.14^{+  0.06}_{ -0.13 }$
\\
HCG~62        &
$216.53^{+764.95}_{-157.07}$&
$ 0.058^{+ 0.095}_{-0.048}$&&
$  7.51^{+132.18}_{ -6.96}$&
$  0.27 ^{+\infty}_{ -0.19}$&&
$216.53^{+419.16}_{-156.81}$&
$ 0.078^{+ 0.093}_{-0.068 }$
\\
A1795         &
$ 10.50^{+  6.12}_{ -6.11}$&
$  0.38^{+  0.17}_{ -0.12}$&&
$  5.35^{+ 50.17}_{ -4.66}$&
$  0.50^{+  1.78}_{ -0.31}$&&
$  5.35^{+  3.14}_{ -3.13}$&
$  0.54^{+  0.21}_{ -0.11 }$
\\
A1835         &
$  5.78^{+  7.08}_{ -3.39}$&
$  0.45^{+  0.21}_{ -0.44}$&&
$  0.55^{+  4.57}_{ -0.45}$&
$  6.47 ^{+\infty}_{ -5.97}$&&
$  0.45^{+  0.28}_{ -0.24}$&
$  2.22^{+  2.23}_{ -0.84 }$
\\
PKS~1404-267  &
$161.94^{+132.74}_{-101.28}$&
$ 0.089^{+ 0.071}_{-0.079}$&&
$  8.34^{+ 86.51}_{ -7.29}$&
$  0.33^{+  0.65}_{ -0.23}$&&
$ 36.04^{+ 30.18}_{-21.24}$&
$  0.25^{+  0.12}_{ -0.08 }$
\\
A2029         &
$ 51.09^{+  8.01}_{-17.98}$&
$  0.20^{+  0.08}_{ -0.19}$&&
$  6.40^{+ 59.01}_{ -5.55}$&
$  0.47^{+  0.85}_{ -0.29}$&&
$ 42.12^{+ 10.85}_{-12.66}$&
$  0.25^{+  0.07}_{ -0.24 }$
\\
A2052         &
$  1.87^{+  0.89}_{ -1.13}$&
$  0.74 ^{+\infty}_{ -0.17}$&&
$  0.87^{+  8.27}_{ -0.73}$&
$>  0.37 $&&
$  1.31^{+  0.29}_{ -0.77}$&
$  0.89 ^{+\infty}_{ -0.08 }$
\\
MKW~3S        &
$  2.43^{+  1.09}_{ -1.36}$&
$  0.67^{+  0.30}_{ -0.15}$&&
$  3.91^{+ 41.13}_{ -3.46}$&
$  0.54 ^{+\infty}_{ -0.34}$&&
$  2.43^{+  0.88}_{ -1.39}$&
$  0.67^{+  0.32}_{ -0.10 }$
\\
A2199         &
$ 11.86^{+ 11.86}_{ -5.32}$&
$  0.30^{+  0.11}_{ -0.29}$&&
$  2.21^{+ 17.33}_{ -1.90}$&
$  0.67 ^{+\infty}_{ -0.44}$&&
$  9.15^{+  4.96}_{ -3.92}$&
$  0.42^{+  0.12}_{ -0.11 }$
\\
Hercules~A    &
$  0.10^{+  0.11}_{ -0.04}$&
$+\infty$&&
$  3.20^{+ 33.98}_{ -2.84}$&
$  0.61 ^{+\infty}_{ -0.38}$&&
$ 0.017^{+ 0.012}_{-0.007}$&
  $+\infty$
\\
3C~388        &
$  6.30^{+ 12.22}_{ -4.18}$&
$  0.44^{+  0.28}_{ -0.23}$&&
$  4.22^{+ 44.75}_{ -3.69}$&
$  0.52 ^{+\infty}_{ -0.33}$&&
$  6.30^{+ 12.11}_{ -4.18}$&
$  0.44^{+  0.28}_{ -0.22 }$
\\
Cygnus~A      &
$  1.14^{+  0.16}_{ -0.10}$&
$  0.95^{+  0.04}_{ -0.06}$&&
$  3.07^{+ 28.94}_{ -2.60}$&
$  0.62 ^{+\infty}_{ -0.38}$&&
$  0.80^{+  0.21}_{ -0.15}$&
$  1.21^{+  0.21}_{ -0.21 }$
\\
Sersic~159/03 &
$  1.71^{+  1.13}_{ -0.88}$&
$  0.80^{+  0.66}_{ -0.18}$&&
$  5.16^{+ 58.11}_{ -4.58}$&
$  0.51^{+  4.42}_{ -0.32}$&&
$  0.50^{+  0.39}_{ -0.26}$&
$  2.23^{+  2.98}_{ -1.07 }$
\\
A2597         &
$ 76.82^{+132.16}_{-45.35}$&
$  0.12^{+  0.10}_{ -0.11}$&&
$  1.08^{+ 15.97}_{ -0.99}$&
$  0.96 ^{+\infty}_{ -0.67}$&&
$  5.38^{+  5.75}_{ -3.17}$&
$  0.54^{+  0.22}_{ -0.16 }$
\\
A4059         &
$ 44.18^{+ 30.25}_{-26.42}$&
$<  0.19 $&&
$  0.18^{+  1.20}_{ -0.12}$&
$>  0.87 $&&
$ 13.83^{+  7.75}_{ -8.10}$&
$  0.35^{+  0.09}_{ -0.34 }$
\\

     \hline
    \end{tabular}}\label{tab:jetII}
\end{table*}

\end{document}